\documentclass[11pt]{article}
\usepackage[matrix,arrow]{xy}
\usepackage{epsfig}
\usepackage[utf8x]{inputenc}
\usepackage{amssymb,amsthm}
\usepackage{amsmath}
\usepackage{amscd}
\usepackage{latexsym}
\usepackage{graphics}
\usepackage{color}
\usepackage{mathtools}
\usepackage{dsfont}
\usepackage{stmaryrd}
\usepackage{graphicx}
\usepackage{cite}
\usepackage{hyperref}
\input{xy}
\xyoption{all}

\def\beq{\begin{equation}}
\def\eeq{\end{equation}}
\def\bea{\begin{eqnarray}}
\def\eea{\end{eqnarray}}

\catcode`@=11
\renewcommand{\section}{\@startsection{section}{1}{0pt}{\medskipamount}
{\medskipamount}{\Large\bf}}
\numberwithin{equation}{section}
\catcode`@=12
\def\a{\alpha}
\def\b{\beta}

\def\1{{\bar 1}}
\def\2{{\bar 2}}
\def\3{{\bar 3}}
\def\4{{\bar 4}}

\newcommand{\K}{{\rm K}}
\newcommand{\Hh}{{\rm H}}
\newcommand{\G}{{\rm G}}

\newcommand{\su}{{{\rm SU}(2)}}
\newcommand{\suL}{{{\mathfrak{su}}(2)}}
\newcommand{\sutL}{{{\mathfrak{su}}(3)}}
\newcommand{\sut}{{{\rm SU}(3)}}
\newcommand{\uo}{{{\rm U}(1)}}
\newcommand{\uoL}{{{\mathfrak u}(1)}}

\newcommand{\glrm}{{{\rm GL}}}

\newcommand{\urm}{{{\rm U}}}
\newcommand{\urmL}{{{\mathfrak u}}}

\newcommand{\surmL}{{{\mathfrak{su}}}}

\newcommand{\utwo}{{{\rm U}(2)}}

\newcommand{\utwoL}{{{\mathfrak u}(2)}}

\newcommand{\sltcL}{{{\mathfrak{sl}}(3,\C)}}

\newcommand{\gfrak}{{\mathfrak{g}}}

\newcommand{\Hom}{{\rm Hom}}
\newcommand{\End}{{\rm End}}

\newcommand{\C}{\mathbb C}
\newcommand{\R}{\mathbb R}

\newcommand{\Z}{\mathbb Z}

\newcommand{\Zcal}{{\mathcal Z}}
\newcommand{\Acal}{{\mathcal A}}
\newcommand{\Ncal}{{\mathcal N}}
\newcommand{\Mcal}{{\mathcal M}}

\newcommand{\Fcal}{{\mathcal F}}

\newcommand{\Lcal}{{\mathcal L}}

\newcommand{\U}{{\mathcal U}}

\newcommand{\Gcal}{{\mathcal G}}
\newcommand{\Ocal}{{\mathcal O}}

\newcommand{\Qcal}{{\mathcal Q}}

\def\im{{\rm i}}
\def\e{{\,\rm e}\,}
\def\N2{$N{=}2$}

\def\diff{{\rm d}}
\def\Diff{{\rm D}}
\def\tr{{\rm tr}}

\def\>{\rangle}
\def\<{\langle}
\def\+{\dagger}
\def\={\ =\ }

\def\and{\quad\textrm{and}\quad}
\def\for{\qquad\textrm{for}\quad}
\def\with{\qquad\textrm{with}\quad}

  \def\rank{\mathrm{rk}}
  
  \def\dim{\mathrm{dim}}

  \newcommand{\betaphi}{\beta_{\varphi}}
  \newcommand{\barbetaphi}{\bar{\beta}_{\varphi}}
  \newcommand{\RepSu}[2]{\underline{C}^{#1,#2}}
  \newcommand{\RepSuHead}[2]{C^{#1,#2}}
  \newcommand{\Reps}[1]{\underline{#1}}

  \newcommand{\HomoPhii}[1]{\phi_{#1}}
  \newcommand{\HomoPhiAdji}[1]{\phi_{#1}^{\dagger}}
  \newcommand{\EndoPsi}[1]{\psi_{#1}}
  
 \newcommand{\zk}{\zeta_{q+1}}
 \newcommand{\ZK}{\mathbb{Z}_{q+1}}
 \newcommand{\betaZK}{\beta_{q+1}}
 \newcommand{\barbetaZK}{\bar{\beta}_{q+1}}
 
 \newcommand{\Bfield}[1]{B_{(#1)}}
 \newcommand{\UniEnd}[1]{\Pi_{(#1)}}
 \newcommand{\UniEndQui}[1]{\pi_{(#1)}}
 
 \newcommand{\CurvQui}[2]{(\mathcal{F})_{(#1),(#2)}}

 \newcommand{\BetaPlusi}[1]{{\beta}_{(#1)}^+}  
\newcommand{\BarBetaPlusi}[1]{{\bar{\beta}}_{(#1)}^+}
 \newcommand{\BetaMinusi}[1]{{\beta}_{(#1)}^-}
\newcommand{\BarBetaMinusi}[1]{{\bar{\beta}}_{(#1)}^-}
 \newcommand{\BetaPMi}[1]{{\beta}_{(#1)}^{\pm}}
 \newcommand{\BetaMPi}[1]{{\beta}_{(#1)}^{\mp}}
\newcommand{\BarBetaPMi}[1]{{\bar{\beta}}_{(#1)}^{\pm}}
\newcommand{\BarBetaMPi}[1]{{\bar{\beta}}_{(#1)}^{\mp}}
 \newcommand{\HomoPhiPlusi}[1]{{\phi}_{(#1)}^{+}}
 \newcommand{\HomoPhiMinusi}[1]{{\phi}_{(#1)}^{-}}
 \newcommand{\HomoPhiPMi}[1]{{\phi}_{(#1)}^{\pm}}
\newcommand{\HomoPhiAdjPMi}[1]{{(\phi^{\pm})}_{(#1)}^\dagger}
 \newcommand{\HomoPhiAdjPlusi}[1]{{(\phi^+)}_{(#1)}^{\dagger}}
\newcommand{\HomoPhiAdjMinusi}[1]{{(\phi^-)}_{(#1)}^{\dagger}}
 \newcommand{\EndPsiQui}[1]{{\psi}_{(#1)}}
 \newcommand{\ConnQui}[1]{A_{(#1)}}
 \newcommand{\FConnQui}[1]{F_{(#1)}}

 \newcommand{\Geni}[2]{I_{#1}^{#2}}
 \newcommand{\GeniAdj}[2]{\big(I_{#1}^{#2}\big)^\dagger}
  \newcommand{\bra}[3]{\left< \begin{matrix} #1 \\ #2  \end{matrix}\, #3  
\right|} % für Bra-Vektor
  \newcommand{\ket}[3]{\left| \begin{matrix} #1 \\ #2  \end{matrix} \,#3  
\right>} % für Ket-Vektor
  \newcommand{\mfd}[1]{M^{#1}}
  \newcommand{\starM}{\star_{M^d}} %
  \newcommand{\starCP}{\star_{\mathbb{C}P^2}}
  \newcommand{\starS}{\star_{S^5}}
  \newcommand{\CP}{\mathbb{C}P^2}
  \newcommand{\betavol}{\beta_{\mathrm{vol}}}
  
  \newcommand{\etavol}{\eta_{\mathrm{vol}}}
  \newcommand{\LaPlus}[1]{\Lambda_{k,l}^{+}(#1)}
  \newcommand{\LaMinus}[1]{\Lambda_{k,l}^{-}(#1)}
  \newcommand{\LaPM}[1]{\Lambda_{k,l}^{\pm}(#1)}
  \newcommand{\LaMP}[1]{\Lambda_{k,l}^{\mp}(#1)}
\allowdisplaybreaks

\begin{document}
\begin{titlepage}
\setcounter{page}{0}
\begin{flushright}
ITP--UH--11/15\\
EMPG--15--07
\end{flushright}

\vskip 2cm

\begin{center}

{\Large\bf Sasakian quiver gauge theories\\[2mm] and instantons on cones over lens 5-spaces 
}

\vspace{15mm}

{\large Olaf Lechtenfeld}${}^1$ , \ {\large Alexander D. Popov${}^{1}$} , \ 
{\large Marcus Sperling${}^{1}$} \ \ and \ \ {\large Richard J. Szabo${}^2$}
\\[5mm]
\noindent ${}^1${\em Institut f\"ur Theoretische Physik} and {\em Riemann Center for Geometry and Physics}\\
{\em Leibniz Universit\"at Hannover}\\
{\em Appelstra\ss e 2, 30167 Hannover, Germany}\\
Email: {\tt lechtenf@itp.uni-hannover.de , alexander.popov@itp.uni-hannover.de, 
marcus.sperling@itp.uni-hannover.de}
\\[5mm]
\noindent ${}^2${\em Department of Mathematics, Heriot-Watt University\\
Colin Maclaurin Building, Riccarton, Edinburgh EH14 4AS, U.K.}\\
and
{\em Maxwell Institute 
  for Mathematical Sciences, Edinburgh, U.K.}\\
and
{\em The Higgs Centre for Theoretical Physics, Edinburgh, U.K.}\\
{Email: {\tt R.J.Szabo@hw.ac.uk}}

\vspace{15mm}

\begin{abstract}
\noindent
We consider $\sut$-equivariant dimensional reduction of Yang-Mills
theory over certain cyclic orbifolds of the $5$-sphere which are
Sasaki-Einstein manifolds. We obtain new quiver gauge theories
extending those induced via reduction over the leaf spaces of the
characteristic foliation of the Sasaki-Einstein structure, which are
projective planes. We describe the Higgs branches of these quiver
gauge theories as moduli spaces of spherically symmetric instantons
which are $\sut$-equivariant solutions to the Hermitian Yang-Mills
equations on the associated Calabi-Yau cones, and further compare them
to moduli spaces of translationally-invariant instantons on the
cones. We provide an explicit unified construction of these moduli spaces as 
K\"ahler quotients and show that they have the same cyclic orbifold 
singularities as the cones over the lens 5-spaces.
\end{abstract}

\end{center}

\end{titlepage}

{\baselineskip=12pt
\tableofcontents
}

\bigskip

%%%%%%%%%%%%%%%%%%%%%%%%%%%%%%%%%%%%%%%%%%%%%%%%%%%%%%%%%%%%%%%%%%%%%%%%%%%%%%%%
 \bigskip \section{Introduction and summary}
\noindent
Sasaki-Einstein $5$-manifolds $M^5$ have played a prominent role in developments in string theory. For example, type~IIB string theory on $AdS_5\times M^5$ is conjecturally dual to the $4$-dimensional $\Ncal=1$ superconformal worldvolume field theory on a stack of D$3$-branes placed at the apex singularity of the $6$-dimensional Calabi-Yau cone $C(M^5)$ over $M^5$~\cite{Kachru:1998ys,Lawrence:1998ja,Kehagias:1998gn,Klebanov:1998hh,Morrison:1998cs,Martelli:2004wu}. They have moreover served as interesting testing grounds for the suggestion that maximally supersymmetric Yang-Mills theory in $5$~dimensions contains all degrees of freedom of the $6$-dimensional $(2, 0)$ superconformal theory compactified on a circle~\cite{Kallen:2012cs,Qiu:2013pta}. Metrics on the non-compact spaces $C(M^5)$ are also known explicitly~\cite{Gauntlett:2004yd,Gauntlett:2004hh,Cvetic:2005ft}, in contrast to the compact examples of Calabi-Yau string compactifications. 

In this paper we derive new quiver gauge theories via equivariant dimensional reduction over $M^5$ and describe their vacua in terms of moduli spaces of generalised instantons on the cones~$C(M^5)$;\footnote{The analogous instanton moduli spaces were studied by~\cite{Ivanova:2013mea} for the $3$-dimensional case and by~\cite{Sperling:2015sra} in arbitrary (odd) dimension.} such instantons play a central role in supersymmetric compactifications of heterotic string theory~\cite{Candelas:1985en}. This extends the constructions of~\cite{Lechtenfeld:2014fza} which dealt with the case of $3$-dimensional Sasaki-Einstein manifolds, wherein these field theories were dubbed as ``Sasakian'' quiver gauge theories. The only Sasaki-Einstein $3$-manifolds are the ADE orbifolds $S^3/\Gamma$ of the $3$-sphere by a discrete subgroup $\Gamma$ of $\su$. They have natural extensions as ADE orbifolds $M^5=S^5/\Gamma$ of the $5$-sphere which preserve $\Ncal=2$ supersymmetry~\cite{Lawrence:1998ja,Kachru:1998ys}. In the following we shall be interested in generalisations of these orbifolds to cases where $\Gamma$ is instead a finite subgroup of $\sut$. The corresponding affine cones $C(S^5/\Gamma)$ play a central role in the McKay correspondence for Calabi-Yau $3$-folds~\cite{IR,IN}. Moreover, the BPS configurations in the worldvolume gauge theories on D-branes located at points of Calabi-Yau manifolds which are resolutions of the orbifolds $\C^3/\Gamma$~\cite{DGM,Douglas:2000qw} are parameterised by moduli spaces of translationally-invariant solutions of Hermitian Yang-Mills equations on $\C^3/\Gamma$, which coincide with Calabi-Yau spaces that are partial resolutions of these orbifolds~\cite{SardoInfirri:1996ga,Cirafici:2010bd}.
Drawing from the situation in the $3$-dimensional case~\cite{Lechtenfeld:2014fza}, it is natural to expect the same sort of similarities between these moduli spaces and those of ``spherically symmetric'' instantons on cones over any Sasaki-Einstein $5$-manifold, where the generalised instanton equations can also be reduced to generalised Nahm equations of the form considered in~\cite{Ivanova:2012vz}.

On general grounds, any quasi-regular Sasaki-Einstein $5$-manifold $M^5$ is a $\uo$ V-bundle over a $4$-dimensional K\"ahler-Einstein orbifold $M^4$. In this paper we consider the special case where $M^5=S^5/\Gamma$ with $\Gamma=\ZK\subset\sut$ a cyclic group. Then $M^4=\CP$ and we can exploit the constructions from~\cite{Lechtenfeld:2008nh} which thoroughly studies $\sut$-equivariant dimensional reduction over the K\"ahler coset space $\CP\cong\sut/S(\utwo\times\uo)$. We shall construct the relevant quiver bundles and study the corresponding quiver gauge theories in detail; these quivers are new and we relate them explicitly to those arising from dimensional reduction over the leaf spaces $\CP$ of the characteristic foliation of $S^5/\ZK$. In particular, we will compare the moduli spaces of spherically symmetric and translationally-invariant instantons on the cones $C(S^5/\ZK)\cong \C^3/ \ZK $, and show that they contain the same orbifold singularities $\C^3/\Z_N$ (where $N$ is the rank of the gauge group) analogously to the cases of~\cite{Lechtenfeld:2014fza}. In analogy to the interpretations of~\cite{Lechtenfeld:2014fza}, our constructions thereby shed light on the interplay between the Higgs branches of the worldvolume quiver gauge theories on D$p$-branes which probe a set of D$(p+6)$-branes wrapping a (partial) resolution of $C(S^5/\ZK)$, and BPS states of the quiver gauge theories on pairs of D$(p+4)$-branes wrapping $C(S^5/\ZK)$ which transversally intersect D$(p+6)$-branes at the apex of the cone $C(S^5/\ZK) $. In this scenario, it is the codimensionality of the D-brane bound states which selects both the quiver type and the abelian category in which the quiver representation is realised; in particular, the arrows of the quivers keep track of the massless bifundamental transverse scalars stretching between constituent fractional D-branes at the vertices.

The outline of the remainder of this paper is as follows. In Section~\ref{sec:Geometry} we provide a detailed description of the geometry of the orbifold $S^5/\ZK$ using its realisation as both a coset space and as a Sasaki-Einstein manifold. In Section~\ref{sec:Quiver_bundles} we give a detailed description of the quiver gauge theory induced via $\sut$-equivariant dimensional reduction over $S^5/\ZK$, including explicit constructions of the quiver bundles and their connections as well as the form of the action functional. We then describe the Higgs branch vacuum states of quiver gauge theories on the cone $C(S^5/\ZK)$ as $\sut$-equivariant solutions to the Hermitian Yang-Mills equations in Section~\ref{sec:SU(3)-equivariant_cone} and as translationally-invariant solutions in Section~\ref{sec:TransInv_Instantons}. In Section~\ref{sec:Comparison_cone} we compare the two quiver gauge theories in some detail, including a contrasting of their quiver bundles and explicit universal constructions of their instanton moduli spaces as K\"ahler quotients. Four appendices at the end of the paper contain technical details and results which are employed in the main text.
%%%%%%%%%%%%%%%%%%%%%%%%%%%%%%%%%%%%%%%%%%%%%%%%%%%%%%%%%%%%%%%%%%%%%%%%%%%%%%%%
 \bigskip \section{Sasaki-Einstein geometry}
\label{sec:Geometry}
\noindent
In this section we shall introduce the basic geometrical constructions
that we shall need throughout this paper.
\subsection{Preliminaries}
\emph{Sasakian manifolds} $\mfd{2n+1}$ of dimension $2n+1$ are contact manifolds which form a natural
bridge between two different Kähler spaces $\mfd{2n}$ and $\mfd{2n+2}$ of 
dimensions $2n$ and $2n+2$, respectively. On the one hand, the metric cone over a 
Sasakian manifold 
$\mfd{2n+1}$ gives a Kähler space $\mfd{2n+2}= C(\mfd{2n+1})$. On the other 
hand, the Reeb vector field on $\mfd{2n+1}$ defines a foliation of $\mfd{2n+1}$ 
and the transverse space $\mfd{2n}$ is also Kähler. For further 
details, see for example~\cite{Boyer:2008}.

A Riemannian manifold is called \emph{Einstein} if its Ricci tensor is a 
scalar multiple of its metric. A Sasakian manifold which is also Einstein is 
called a \emph{Sasaki-Einstein manifold}~\cite{Boyer:2008}. Since the cone 
over an Einstein manifold is also an Einstein space, the metric cone over a 
Sasaki-Einstein manifold is a Calabi-Yau space and in this case the
transverse space $M^{2n}$ is K\"ahler-Einstein. Because of the $\R_{>0}$ scaling action on the cones we can write the Calabi-Yau metric as
\begin{equation}
\diff s^2_{C(\mfd{2n+1})}= \diff r^2 + r^2 \, \diff s^2_{\mfd{2n+1}} \ ,
\end{equation}
where $r\in\R_{\geq0}$ and the tensor $\diff s^2_{\mfd{2n+1}}$ defines a metric on the intersection $\mfd{2n+1}$ of the cone with the unit sphere in $\C^{n+2}$. 

Given a Riemannian manifold $M$ and a finite group $\Gamma$ acting isometrically on $M$, 
one can, loosely speaking, define the Riemannian space of $\Gamma$-orbits
$M/\Gamma$, 
which is called an \emph{orbifold} or sometimes \emph{V-manifold}, see for 
instance~\cite{Boyer:2008}. The notion of fibre bundle can be adapted to 
the category of orbifolds, and we follow~\cite{Boyer:2008} in calling them 
\emph{V-bundles}. Any quasi-regular Sasaki-Einstein manifold $M^{2n+1}$ is a
principal $\uo$ V-bundle over its transverse space $M^{2n}$. In this
case the Sasaki-Einstein metric can be expressed as
\begin{equation}
\diff s^2_{\mfd{2n+1}} = \diff s^2_{\mfd{2n}} +\eta\otimes\eta \ ,
\end{equation}
where $\diff s^2_{\mfd{2n}}$ is the (pullback of the)
K\"ahler-Einstein metric of $\mfd{2n}$, and $\eta$ is the contact
$1$-form which is a connection on the fibration
$M^{2n+1}\to M^{2n}$ of curvature $\diff\eta=-2\omega_{\mfd{2n}}$ with $\omega_{\mfd{2n}}$ the K\"ahler form of the base $\mfd{2n}$.
\subsection{Sphere $S^5$}
The $5$-dimensional sphere $S^5$ has two realisations: Firstly, as 
the coset space $S^5 = \sut / \su$ and, secondly, as a principal
$\uo$-bundle over the complex projective plane $\CP$. As such, we have the chain of principal bundles
\begin{equation}
 \sut \xrightarrow{\su} S^5 \xrightarrow{\uo} \CP \; .
\end{equation}
Our description of $S^5$ will be based on the 
principal $\uo$-bundle over $\CP$, and we will 
construct a flat connection on the 
principal $\su$-bundle over $S^5$ by employing 
this feature. 

\paragraph{Connections on $\CP$}
Let us consider a local section $U$ over a patch 
$\U_0$ of $\CP$ for the principal bundle $\sut \to \CP$. For this, let $\G= 
\sut$ and $\Hh=S(\utwo\times \uo) \subset \G$, and consider the principal bundle 
associated to the coset $\G/\Hh$ given by
\begin{equation}
 \G= \sut \xrightarrow{\Hh=S(\utwo\times \uo)} \G/\Hh \cong \CP \; . 
\label{eqn:bundle_SU3_CP2}
\end{equation}
By the definition of the complex projective plane
\begin{equation}
 \CP= \C^3 \, \big/ \, \sim \; = \big\{ [z^1:z^2:z^3] \in 
\C^3 \, : \;  [z^1 : z^2 : z^3] \sim [ \lambda \, z^1 :\lambda\, z^2 : \lambda \,
z^3] \; , \ \lambda \in \C^* \big\} \label{eqn:def_CP} \; ,
\end{equation}
one introduces on the patch $\U_0 = \{ [z^1:z^2:z^3] \in \CP \, : \; 
z^3 \neq 0 \}$ the coordinates 
\begin{equation}
 Y \coloneqq \begin{pmatrix} y^1 \\ y^2 \end{pmatrix} \sim \begin{pmatrix} 
z^1/z^3 \\ z^2/z^3 \end{pmatrix} \; .
\end{equation}
Define a local section on $\U_0$ of the principal 
bundle~\eqref{eqn:bundle_SU3_CP2} via~\cite{Lechtenfeld:2008nh}
\begin{equation}
\begin{split}
U : \mathcal{U}_0 &\longrightarrow \sut \\
 Y &\longmapsto U(Y) \coloneqq \frac{1}{\gamma} \, \begin{pmatrix} \bar{\Lambda} & 
\bar{Y} \\ 
- \bar{Y}^\dagger & 1  \end{pmatrix} \; ,
\end{split}
  \label{eqn:def_section_CP2}
\end{equation}
with the definitions
\begin{equation}
 \bar{\Lambda} \coloneqq \gamma \, \mathds{1}_2 - \frac{1}{\gamma + 1} \,\bar{Y} \,
\bar{Y}^\dagger \and
\gamma \coloneqq \sqrt{1 + Y^\dagger \,Y} \; .
\end{equation}
From these two definitions, one observes the properties
\begin{equation}%
\bar{\Lambda}^\dagger = \bar{\Lambda} \; , \quad 
\bar{\Lambda}^2 =  \gamma^2 \, \mathds{1}_2 - \bar{Y} \, \bar{Y}^\dagger \; , \quad 
 \bar{\Lambda} \bar{Y} =\bar{Y}   \and \bar{Y}^\dagger \bar{\Lambda} = 
\bar{Y}^\dagger \; . 
\label{eqn:properties_CP2}
\end{equation}
It is immediate from~\eqref{eqn:properties_CP2} that $U$ as defined 
in~\eqref{eqn:def_section_CP2} is $\sut$-valued. 

One can 
define a flat connection $A_0$ on the bundle~\eqref{eqn:bundle_SU3_CP2} via
\begin{equation}
 A_0 = U^\dagger \, \diff U \equiv \begin{pmatrix} B & \bar{\beta} \\ - \beta^\top & 
-a \end{pmatrix} \; ,
\end{equation}
with the definitions
\begin{subequations}
\label{eqn:definitions}
\begin{align}
 B &\coloneqq \frac{1}{\gamma^2} \, \Big(  \bar{\Lambda}\, \diff \bar{\Lambda} + 
\bar{Y} \,
\diff \bar{Y}^\dagger - \frac{1}{2 }\, \mathds{1}_2 \, \diff \big(
     Y^\dagger \, Y 
\big) \Big)  \; ,\\[4pt]
\bar{\beta} &\coloneqq \frac{1}{\gamma^2}\, \bar{\Lambda} \, \diff \bar{Y}  \and 
\beta^\top \coloneqq 
\frac{1}{\gamma^2} \, \diff \bar{Y}^\dagger \, \bar{\Lambda} \; , \\[4pt]
 a & \coloneqq  -\frac{1}{2 \gamma^2} \, \Big(    \bar{Y}^\dagger \, \diff \bar{Y} - 
 \diff \bar{Y}^\dagger  \, \bar{Y} \Big) = -\bar{a} \label{eqn:def_monopole_conn} 
\; .
\end{align}
\end{subequations}
That $U\in\sut$ directly implies the vanishing of the curvature $2$-form  $ 
F_0 = \diff A_0 + A_0 \wedge A_0 $, which is equivalent to the set of 
relations
\begin{subequations}
\label{eqn:identities}
 \begin{alignat}{3}
  \diff B + B \wedge B &= \bar{\beta} \wedge \beta^\top & &\and &
  \diff a = - \beta^\top \wedge \bar{\beta} &=  \beta^\dagger 
\wedge \beta \; ,\\[4pt]
 \diff \bar{\beta} + B \wedge \bar{\beta} &=  \bar{\beta} \wedge a & &\and &
 \diff \beta^\top + \beta^\top \wedge B &=  a \wedge \beta^\top \; .
 \end{alignat}
\end{subequations}
As elaborated in~\cite{Lechtenfeld:2008nh,Dolan:2009nz}, $B$ can be regarded
as a $\utwoL$-valued connection $1$-form and $a$ as a $\uoL 
$-valued connection. Consequently, one can
introduce an $\surmL(2)$-valued connection $B_{(1)}$ by removing the trace of 
$B$. An explicit parametrisation yields
\begin{align}
 B_{(1)} &\coloneqq B- \frac{1}{2} \, \tr(B) \, \mathds{1}_2 \equiv \begin{pmatrix} B_{11} 
& \bar{B}_{12} \\ -B_{12} & -B_{11}  
\end{pmatrix} \with  
\tr(B) = a \ , \;  B_{11}=-\bar{B}_{11}  \, . %
\label{eqn:def_B-one}
\end{align}
The geometry of $\CP$ including the properties of the
$\sut$-equivariant $1$-forms $\beta^i$, the 
instanton connection $B_{(1)}$ and the monopole connection $a$ are described 
in Appendix~\ref{subsec:geometry_CP}.
\paragraph{Connections on $S^5$}
Consider now the principal $\su$-bundle
\begin{equation}
 \G= \sut \xrightarrow{\K=\su} \G/\K = S^5 \; , \label{eqn:bundle_SU3_S5}
\end{equation}
where $\K \subset \G$. Then the section $U$ 
from~\eqref{eqn:def_section_CP2} can be modified as
\begin{equation}
\begin{split}
 \hat{U} : \U_0 \times [0,2\pi) &\longrightarrow \sut \\
 (Y,\varphi) &\longmapsto \hat{U}(Y,\varphi)\coloneqq U(Y) \, \begin{pmatrix} \e^{\im\, 
\varphi} \, \mathds{1}_2  & 0 \\ 0  & \e^{- 2\, \im\, \varphi}  \end{pmatrix} \equiv
U(Y) \, Z(\varphi) \; ,
\end{split}
 \end{equation}
which is a local section of the bundle~\eqref{eqn:bundle_SU3_S5} on the 
patch $\U_0 \times \left[0,2\pi \right)$ with coordinates $\{y^1,y^2,\varphi 
\}$. Note that $Z^{-1} = Z^\dagger = \bar{Z}$ and $\mathrm{det}(Z)=1$,
and furthermore $Z(\varphi) \, Z(\psi ) = Z(\psi) \, Z( \varphi ) = Z(\psi 
+ \varphi ) $, which implies that $Z$ realises the embedding $\uo 
\hookrightarrow \sut$; this also shows that $\hat{U}\in \sut$. The modified (flat) connection $\hat{A}$ 
on the bundle~\eqref{eqn:bundle_SU3_S5}
and the corresponding curvature $\hat{F}$ are given as
\begin{subequations}
 \begin{align}   
 \hat{A}&\coloneqq \hat{U}^\dagger\, \diff \hat{U} = \mathrm{Ad}(Z^{-1}) A_0 + 
Z^\dagger \,
\diff Z = \begin{pmatrix} B + \im \,\mathds{1}_2 \,\diff \varphi & \bar{\beta} 
\, \e^{-3\, \im \,\varphi} \\ - \beta^\top \,\e^{3\, \im\, \varphi} & -(a
+ 2 \,\im \, \diff \varphi) 
 \end{pmatrix} \ , \label{eqn:connection_S5_flat}\\[4pt]
\hat{F} &\coloneqq \diff \hat{A} + \hat{A} \wedge \hat{A}  = 
\mathrm{Ad}(Z^{-1}) F_0 \notag \\[4pt]
  &= \begin{pmatrix} \diff B + B \wedge B - \bar{\beta} \wedge \beta^\top & 
\left(\diff \bar{\beta} + B \wedge \bar{\beta} -  \bar{\beta} \wedge a \right) 
\, \e^{-3 \, \im \, \varphi}   \\ 
- \left( \diff \beta^\top + \beta^\top \wedge B -  a \wedge \beta^\top \right)
\, \e^{3\, \im\, 
\varphi} & - \diff a - \beta^\top \wedge \bar{\beta}  \end{pmatrix} =0 \ .
 \end{align}
\end{subequations}
Again the flatness of $\hat{A}$ yields the same set of identities~\eqref{eqn:identities}, because $\hat{F}$ differs from $F$ only by the 
adjoint action of $Z^{-1}$. 
\paragraph{Contact geometry of $S^5$}
By construction, the base space of~\eqref{eqn:bundle_SU3_S5} is a $5$-sphere. 
The aim now is to choose a basis of the cotangent bundle $T^* S^5$ over 
the patch $\U_0 \times [0,2\pi)$ such that one recovers the 
Sasaki-Einstein structure on $S^5$. For this, we start with the 
identifications
\begin{equation}
 \betaphi^1 \coloneqq  \beta^1 \, \e^{3\, \im\, \varphi}\equiv e^1 +
 \im \, e^2 \; , \quad 
\betaphi^2 \coloneqq \beta^2 \, \e^{3\, \im\, \varphi} \equiv e^3 + \im\, e^4 \and
\kappa \, e^5 \coloneqq \tfrac{1}{2}\, a + \im \, \diff \varphi  \; ,
\label{eqn:def_forms}
\end{equation}
where $\kappa \in \C$ is a constant to be determined and the
1-forms $\beta^i$ 
originate from the complex cotangent space $T_{(Y,\bar Y)}^* \CP$ at a point 
$(Y,\bar Y) \in \U_0\subset\CP$. Next we define the forms
\begin{equation}
 \omega_1 \coloneqq e^{14} + e^{23} \; , \quad \omega_2 \coloneqq e^{31} + 
e^{24} \; , \quad  \omega_3 \coloneqq e^{12} + e^{34} \and \eta \coloneqq 
e^5 \; ,
\end{equation}
where generally $e^{a_1\cdots a_k}=e^{a_1}\wedge\cdots\wedge e^{a_k}$.
In the basis~\eqref{eqn:def_forms}, one obtains
\begin{equation}
\begin{split}
 \omega_1 = \tfrac{1}{2 \,\im} \, \left( \betaphi^1 \wedge \betaphi^2 - 
\barbetaphi^1 
\wedge \barbetaphi^2  \right) \; ,\quad \omega_2 &= -\tfrac{1}{2} \, \left( 
\betaphi^1 \wedge \betaphi^2 + \barbetaphi^1 \wedge \barbetaphi^2  \right) \\
\and \omega_3 &= - \tfrac{1}{2\, \im} \, \left( \betaphi^1 \wedge \barbetaphi^1 +
\betaphi^2  \wedge \barbetaphi^2  \right) \; .
\end{split}
\end{equation}
Note that $\omega_3$ coincides (up to a normalisation factor) with the Kähler 
form on $\CP$, cf. Appendix~\ref{subsec:geometry_CP}. The 
exterior derivatives of $\betaphi^i$ and $\barbetaphi^i$ are given as 
\begin{equation}
 \diff \betaphi^i = \e^{3\, \im \, \varphi}  \,\diff \beta^i - 3\, \im
 \, \betaphi^i \wedge 
\diff \varphi 
\and 
\diff \barbetaphi^i = \e^{-3 \,\im \,\varphi}  \,
\diff \bar{\beta}^i + 3 \,\im \,\barbetaphi^i \wedge \diff \varphi \; .
\end{equation}
The distinguished $1$-form $\eta$ is taken to be the contact $1$-form dual to the Reeb vector field of the 
Sasaki-Einstein structure. At this stage, the choice of the quadruple
($\eta,\omega_1,\omega_2,\omega_3$) defines an $\su$-structure on the 
$5$-sphere~\cite{Conti:2005}. For it to be Sasaki-Einstein, one needs
the relations
\begin{equation}
 \diff \omega_1 = 3 \eta \wedge \omega_2 \; , \quad \diff \omega_2 = - 3 \eta 
\wedge \omega_1 \and \diff \eta = - 2 \omega_3 \label{eqn:def_Sasaki-Einstein}
\end{equation}
to hold~\cite{Fernandez:2006ux}. Employing~\eqref{eqn:identities} one 
arrives at
\begin{subequations}
\label{eqn:diffs_2-forms_S5}
 \begin{alignat}{3}
  \diff \omega_1 &= 6 \,\im \,\kappa \,\eta \wedge \omega_2 & &\and & 
  \diff \omega_2 &= -6 \,\im \,\kappa \, \eta \wedge \omega_1 \; , \\[4pt]
  \diff \eta &= \tfrac{\im}{\kappa}\, \omega_3 & &\and & \diff 
\omega_3 &=0 \; .
 \end{alignat}
\end{subequations}
Consequently, the coframe $\{ \eta , \betaphi^1,\betaphi^2 \}$ yields a 
Sasaki-Einstein structure on $S^5$ if and only if $\kappa = -\tfrac{\im}{2}$, and from 
now on this will be the case.
%
%%%%%%%%%%%%%%%%%%%%%%%%%%%%%%%%%%%%%%%%%%%%%%%%%%%%%%%%%%%%%%%%%%%%%%%%%%%%%%%%
%%%%%%%%%%%%%%%%%%%%%%%%%%%%%%%%%%%%%%%%%%%%%%%%%%%%%%%%%%%%%%%%%%%%%%%%%%%%%%%%
% 
\subsection{Orbifold $S^5/ \ZK $}
\label{subsec:geometry_orbifold}
Our aim is to now construct a principal V-bundle over the orbifold 
$S^5/\ZK $ by the following steps: Take the principal $\su$-bundle 
$\pi :\G=\sut \to \sut\slash \su \cong S^5$, which 
is 
$\su$-equivariant. Embed $\ZK \hookrightarrow \uo \subset \sut$ such that 
$\uo$ commutes with $\su \subset \sut$, and define a $\ZK$-action $\gamma$ on 
$S^5$. The action $\gamma : \ZK \times S^5 \to S^5$ can be lifted to
an action $ 
\widetilde{\gamma}: \ZK \times \G \to \G$ with 
an isomorphism on the $\su$ fibres induced by this action. The crucial 
point is that the fibre isomorphism is trivial as $\su$ commutes with $\ZK$ by 
construction. Hence one can consider the $\ZK$-projection of 
$\G$ to the principal $\su$ V-bundle $\widetilde{\G}$, which is schematically 
given as
\begin{equation}
 \xymatrix{ \G  \ar[r]^{\widetilde{\gamma}} \ar[d]_{\pi}  & \widetilde{\G} 
\ar[d]^{\widetilde{\pi}}  \\
 S^5  \ar[r]_{\!\!\!\!\!\!\!\!\! \gamma }  &  S^5 \slash \ZK } \label{eqn:bundle_SU3_S5_ZK} 
\end{equation}
With an abuse of notation, we will denote the V-bundles obtained via
$\ZK$-projection by the same symbols as the fibre bundles they originate 
from; only $\ZK$-equivariant field configurations survive this
orbifold projection.

A section $\widetilde{U}$ of the principal 
V-bundle~\eqref{eqn:bundle_SU3_S5_ZK} is obtained by a (further) 
modification of the section~\eqref{eqn:def_section_CP2} as 
\begin{equation}
\begin{split}
 \widetilde{U} : \U_0 \times \big[0, \tfrac{2 \pi}{q+1} \big) &\longrightarrow 
\sut \\
 \big( Y,\tfrac{\varphi}{q+1} \big) &\longmapsto 
\widetilde{U}\big(Y,\tfrac{\varphi}{q+1} \big)\coloneqq U(Y) \, \begin{pmatrix} 
\e^{\frac{\im\, \varphi}{q+1}} \, \mathds{1}_2 & 0 
\\ 0 & \e^{-2\,\frac{ \im\, \varphi}{q+1}}  \end{pmatrix} \equiv 
U(Y) \, Z_{q+1}(\varphi) \; .
\end{split}
\end{equation}
Here $\varphi \in [0,2 \pi)$ is again the local coordinate on the 
$S^1$-fibration $S^5 \xrightarrow{\uo} \C P^2$; hence $ \e^{\frac{\im\, 
\varphi}{q+1}} \in S^1/\ZK$. Analogously to the $q=0$ case of 
$S^5$ above, one can prove that $Z_{q+1}$ realises the embedding $ S^1 / 
\ZK 
\hookrightarrow 
\uo \subset \sut$, and $\widetilde{U}\in\sut$. 
As before, one computes the connection 1-form $\widetilde{A}$ and the curvature 
$\widetilde{F}$ of the flat connection on the 
V-bundle~\eqref{eqn:bundle_SU3_S5_ZK}. This yields
\begin{subequations}
\begin{align}
  \widetilde{A}&\coloneqq \widetilde{U}^\dagger \, \diff \widetilde{U} = 
\mathrm{Ad}(Z_{q+1}^{-1}) A_0 + Z_{q+1}^\dagger \, \diff Z_{q+1} = \begin{pmatrix} 
B +  \mathds{1}_2 \, \frac{ \im \, \diff \varphi}{q+1} & 
\bar{\beta} \,\e^{-3\, \frac{\im \, \varphi}{q+1}} \\ - \beta^\top \,
\e^{3 \, \frac{\im \,
\varphi}{q+1}} & - \big(a +  2\, \frac{\im \, \diff \varphi}{q+1} \big)     
     \end{pmatrix} \; , \label{eqn:flat_connection_orbifold} \\[4pt]
\widetilde{F} &\coloneqq \diff \widetilde{A} + \widetilde{A} \wedge 
\widetilde{A} =  \mathrm{Ad}(Z_{q+1}^{-1}) F_0 \notag \\[4pt]
  &= \begin{pmatrix} \diff B + B \wedge B - \bar{\beta} \wedge \beta^\top & 
\left(\diff \bar{\beta} + B \wedge \bar{\beta} -  \bar{\beta} \wedge a \right) 
\, \e^{-3 \,\frac{\im \, \varphi}{q+1}}   \\ 
- \left( \diff \beta^\top + \beta^\top \wedge B -  a \wedge \beta^\top \right)
\, \e^{3 \,
\frac{\im \, \varphi}{q+1}} & - \diff a - \beta^\top \wedge \bar{\beta}  
\end{pmatrix} =0 \; .
\end{align}
\end{subequations}
Again the flatness of the connection $\widetilde{A}$ yields the very 
same relations~\eqref{eqn:identities}.
\paragraph{Local coordinates}
Our description of the orbifold $S^5/\ZK$ follows~\cite{Lechtenfeld:2014fza}. 
The key idea is the embedding $S^5 = \sut / 
\su  \hookrightarrow \mathbb{R}^6 \cong \C^3$ via the relation
\begin{equation}
 r^2 = \delta_{\hat{\mu} \hat{\nu}} \, x^{\hat{\mu}}\, x^{\hat{\nu}} = |z^1|^2 + 
|z^2|^2 + |z^3|^2
\label{eqn:radialcoord}\end{equation}
where $x^{\hat{\mu}}$ ($\hat{\mu}=1,\ldots, 6$) are coordinates of 
$\mathbb{R}^6$ and $z^\alpha$ ($\alpha=1,2,3$) are coordinates of 
$\mathbb{C}^3$; here $r\in\R_{>0}$ gives the radius of the embedded 
$5$-sphere. In general, on the coordinates $z^\alpha$ the $\ZK$-action is 
realised linearly by a representation $h \mapsto (h^\alpha_{ \ \beta})$ such that 
\begin{equation}
 z^\alpha \longmapsto h^\alpha_{ \ \beta} \, z^\beta \and \bar{z}^\alpha 
\longmapsto \bar{h}^\alpha_{ \ \beta} \, \bar{z}^\beta =  (h^{-1})^\alpha_{ \ \beta} \,
\bar{z}^\beta \; ,
\end{equation}
where $h$ is the generator of the cyclic group $\ZK$.
In this paper the action of the finite group $\ZK$ is chosen to be 
realised by the embedding of $\ZK$ in the fundamental $3$-dimensional 
complex representation $\RepSu{1}{0}$ of $\sut$ given by
\begin{equation}
 (h^\alpha_{\ \beta}) = \begin{pmatrix} \zk & 0 & 0 \\ 0 & \zk & 0 \\ 0 & 
0 & \zk^{-2} \end{pmatrix} \in \sut \with \zk^l := \e^{\frac{2 \pi \,
\im}{q+1}\, l} \; . \label{eqn:ZK-action}
\end{equation}
Since $\CP$ is naturally defined via a quotient of $\C^3$, see~\eqref{eqn:def_CP}, one can deduce the 
$\ZK$-action on the local coordinates $(y^1,y^2)$ of the patch $\mathcal{U}_0$ 
to be
\begin{equation}
 y^\alpha \longmapsto \frac{\zk\, z^\alpha}{\zk^{-2}\, z^3} = \zk^3 \, y^\alpha \and \bar{y}^\alpha \longmapsto \frac{\zk^{-1}\, \bar{z}^\alpha}{\zk^{2}\, \bar{z}^3} 
= \zk^{-3} \, \bar{y}^\alpha \for \alpha = 1,2 \; . 
\label{eqn:trafo_coord_ZK}
\end{equation}
Next consider the action of $\ZK$ on the $S^1$ coordinate $\varphi$. 
By \eqref{eqn:ZK-action} one naturally has
\begin{equation}
\e^{\im \,\frac{ \varphi}{q+1}} \xrightarrow{\ZK} \e^{ \im \,(  \frac{ 
\varphi}{q+1} + \frac{2 \pi \, l }{q+1})} 
= \e^{\im \, \frac{ \varphi}{q+1}}  \, 
\zk^l \for l \in \{0,1,\ldots,q\} \; ,
\end{equation}
i.e. the transformed coordinate $\e^{ \im\, (  \frac{ \varphi}{q+1} + \frac{2 
\pi\, l }{q+1})}$  lies in the $\ZK$-orbit $  \big[ \e^{\im \,
\frac{ 
\varphi}{q+1}  } \big] $ of $ \e^{\im\, \frac{ \varphi}{q+1}}$. 
\paragraph{Lens spaces}
The spaces $S^5 / \ZK$ are known as lens spaces, see for instance~\cite{Boyer:2008}. For this, one usually embeds $S^5$ 
into $\C^3$ and chooses the action of $p\in \{0,1,\dots,q\}$ as
\begin{equation}
\begin{split}
\ZK \times \C^3 &\longrightarrow \C^3 \\
 \big(p\,,\, (z^1,z^2,z^3) \big) &\longmapsto p\cdot\big(z^1,z^2,z^3 \big) \coloneqq \big(\e^{\frac{2 \pi\, \im \,
p}{q+1}}\, z^1, \e^{\frac{2 \pi\, \im \, p}{q+1} \, r_1}\, z^2,
\e^{\frac{2 \pi\, \im \, p}{q+1} \,
r_2}\, z^3 \big) \; ,
\end{split}
\label{eqn:action_ZK_lens-space}
\end{equation}
where the integers $r_1$ and $r_2$ are chosen to be coprime to $q+1$. The
coprime condition is necessary for the $\ZK$-action to be free away from 
the origin of $\C^3$. The quotient space $S^5/\ZK$ with the 
action~\eqref{eqn:action_ZK_lens-space} is 
called the lens space $L(q+1,r_1,r_2)$ or $L^2(q+1,r_1,r_2)$. It is a 
$5$-dimensional orbifold with fundamental group $\ZK$.

We choose the $\ZK$-action to be given by~\eqref{eqn:ZK-action}, i.e. $r_1=1$ 
and $r_2=-2$. Then $r_1$ is always coprime to $q+1$, but $r_2$ is coprime to 
$q+1$ only if $q$ is even. Thus for $q+1\in 2\mathbb{Z}_{\geq0}+1$ the 
only singular point in $\C^3/\ZK$ is the origin, and its isotropy group is 
$\ZK$. However, for $q+1 \in 2\mathbb{Z}_{\geq0}$ there is a
singularity at the origin and also along the circle $\{z^1=z^2=0 \; , |z^3|=1\} 
\subset S^5$ of singularities with isotropy group $\big\{0,\tfrac{q+1}{2}\big\} \cong 
\mathbb{Z}_2\subset \ZK $. Hence for the chosen action~\eqref{eqn:ZK-action} we are forced to take 
$q\in 2\mathbb{Z}_{\geq0}$ in all considerations.
\paragraph{Differential forms}
Similarly to the previous case, one can construct locally a basis of differential 
forms. However, one has to work with a \emph{uniformising system} of local 
charts on the orbifold $S^5 \slash \ZK$ instead of local charts for the 
manifold $S^5$. Choosing the identifications
\begin{equation}
 \betaZK^1 := \beta^1 \, \e^{\frac{3 \,\im\, \varphi }{q+1}}  \equiv e^1 + \im \,
e^2 \; , \quad \betaZK^2 := 
\beta^2 \, \e^{\frac{3\, \im \, \varphi }{q+1}}  \equiv e^3 + \im \, e^4 \and 
 \eta := e^5 \equiv  \im \, a -  \tfrac{2\, \diff \varphi}{q+1}   %
\label{eqn:def_1-forms_ZK}
\end{equation}
and by means of the relations imposed by the flatness 
of~\eqref{eqn:flat_connection_orbifold}, one can study the geometry of 
$S^5 \slash \ZK$. 
Defining the three 2-forms 
\begin{equation}
\begin{split}
 \omega_1 := \tfrac{1}{2 \,\im}\, \left( \betaZK^1 \wedge \betaZK^2 - \barbetaZK^1 
\wedge \barbetaZK^2 \right) \; , \qquad
\omega_2 &:= - \tfrac{1}{2} \, \left( \betaZK^1 \wedge \betaZK^2 + \barbetaZK^1 
\wedge \barbetaZK^2 \right) \\
\and \omega_3 &:= -\tfrac{1}{2\, \im} \, \left( \betaZK^1 \wedge \barbetaZK^1 + 
\betaZK^2 \wedge \barbetaZK^2 \right) 
\end{split}
\end{equation}
and employing~\eqref{eqn:identities} implied by the flatness of 
$\widetilde{A}$, one obtains 
the correct Sasaki-Einstein relations~\eqref{eqn:def_Sasaki-Einstein}.
\paragraph{$\ZK$-action on 1-forms}
Consider the $\ZK$-action on the forms $\betaZK^i$, $\barbetaZK^i$, and 
$\eta$. Firstly, recall the definitions~\eqref{eqn:def_1-forms_ZK} and 
\eqref{eqn:diffs_1-forms}, from which one sees that
\begin{equation}
 \betaZK^i \xrightarrow{\ZK} \zk^3 \, \betaZK^i \and \barbetaZK^i 
\xrightarrow{\ZK} \zk^{-3}\, \barbetaZK^i \; . \label{eqn:ZK-action_betas}
\end{equation}
This follows directly from the transformation~\eqref{eqn:trafo_coord_ZK}. 
Moreover, it agrees with the monodromy of $\betaZK^i$ and $\barbetaZK^i$ 
along the $S^1$ fibres, i.e.
\begin{equation}
 \betaZK^i = \beta^i \, \e^{3\, \frac{\im \, \varphi }{q+1}} \xrightarrow{\varphi \mapsto 
\varphi + 2 \pi} \betaZK^i \, \zk^3 \; .
\end{equation}
Secondly, for the $1$-form $\eta$ from~\eqref{eqn:def_1-forms_ZK} we 
know that $a$ is a $\uo$ connection. As any $\uo $ connection is automatically $\uo $-invariant, due to the embedding $\ZK \hookrightarrow \uo $ one also has 
$\ZK$-invariance of $a$.\footnote{Alternatively, one can work out the transformation behaviour of $a$ directly from the 
explicit expression~\eqref{eqn:def_monopole_conn}.} We conclude that
\begin{equation}
 \eta \xrightarrow{\ZK} \eta \; .
\end{equation}

From the definition~\eqref{eqn:radialcoord} of the radial 
coordinate, one observes that $r$ is invariant under $\ZK$. The
same is true for the corresponding $1$-form, so that
\begin{equation}
 \diff r \xrightarrow{\ZK} \diff r \; .
\end{equation}
Following~\cite{Lechtenfeld:2014fza}, let $T$ be a $\ZK$-invariant 
1-form on the metric cone $C(S^5/\ZK)$ which is locally expressed as 
\begin{equation}
 T = T_\mu \, e^\mu + T_r \, \diff r \equiv W_i \, \betaZK^i + \overline{W}_i \,
\barbetaZK^i + W_5 \, e^5 + W_r \, \diff r 
\end{equation}
with $ i = 1,2 $ and $\mu = 
1,\ldots, 5$, where $W_1= \frac{1}{2}\,  (T_1 - \im \, T_2)$, $W_2=
\frac{1}{2} \, (T_3 - \im \, T_4)$, 
$W_5 =T_5$ and $W_r=T_r$.
This induces a representation $\pi$ of $\ZK$ in 
$\Omega^1\big(C(S^5)\big)$ as
\begin{subequations}
\label{eqn:transf_1forms_ZK}
\begin{alignat}{2}
 W_i &\longmapsto \pi(h)(W_i) = \zk^{-3} \, W_i \; , & \qquad 
\overline{W}_i &\longmapsto \pi(h)(\, \overline{W}_i) = \zk^{3} \, \overline{W}_i \; , \\[4pt]
W_5 &\longmapsto \pi(h)(W_5) = W_5 \; , & \qquad
W_r &\longmapsto \pi(h)(W_r) = W_r \; .
\end{alignat}
\end{subequations}

%%%%%%%%%%%%%%%%%%%%%%%%%%%%%%%%%%%%%%%%%%%%%%%%%%%%%%%%%%%%%%%%%%%%%%%%%%%%%%%%
  \bigskip \section{Quiver gauge theory}
\label{sec:Quiver_bundles}
\noindent
In this section we define quiver bundles over 
a $d$-dimensional manifold $M^d$ via equivariant dimensional reduction
over $\mfd{d}\times S^5 \slash \ZK$, and derive the generic form of a 
$\G$-equivariant connection. For this, we recall some aspects from the
representation theory of 
$\G= \sut$, and exemplify the relation between quiver 
representations and homogeneous bundles over $S^5 \slash \ZK$. Then we 
extend our constructions to $\G$-equivariant bundles over $\mfd{d} \times S^5 \slash 
\ZK$, which will furnish a quiver representation in the category of
vector bundles 
instead of vector spaces. We shall also derive the dimensional
reduction of the pure Yang-Mills action on 
$\mfd{d}\times S^5$ to obtain a Yang-Mills-Higgs theory on $\mfd{d}$
from our twisted reduction procedure (for the special case 
$q=0$).
\subsection{Preliminaries}
\paragraph{Cartan-Weyl basis of $\sutL$}
Our considerations are based on certain irreducible representations of the 
Lie group $\G=\sut$, which are decomposed into irreducible representations of 
the subgroup $\Hh=\su \times \uo \subset \sut$.
For this, we recall the root decomposition of the Lie algebra $\sutL$. 
There is a pair of simple roots $\alpha_1$ and $\alpha_2$, and the non-null 
roots are given by $\pm\, \alpha_1$, $\pm\, \alpha_2$, and $\pm\, (\alpha_1 + 
\alpha_2)$. The Lie algebra
$\sutL$ is $8$-dimensional and has a $2$-dimensional Cartan subalgebra spanned 
by $H_{\alpha_1}$ and $H_{\alpha_2}$. We distinguish one $\suL$ subalgebra, 
which is spanned by $H_{\alpha_1} $ and $ E_{\pm\, \alpha_1}$ with the
commutation relations  
\begin{subequations}
\label{eqn:def_Lie-algebra_SU(3)}
\begin{equation}
 \left[ H_{\alpha_1}, E_{\pm \, \alpha_{1}} \right] = \pm \, 2 E_{\pm\, \alpha_{1}} 
\and \left[ E_{\alpha_{1}},E_{-\alpha_{1}} \right] = 
H_{\alpha_1} \; .
\end{equation}
The element $H_{\alpha_2}$ generates a $\uoL$ subalgebra that commutes
with this $\suL$ subalgebra, i.e.
\begin{equation}
 \left[ H_{\alpha_2}, H_{\alpha_1} \right] = 0 
\and \left[ H_{\alpha_{2}},E_{\pm \, \alpha_{1}} \right] = 0 \; .
\end{equation}
In the \emph{Cartan-Weyl basis}, the remaining generators $E_{\pm \, \alpha_2}$ 
and $E_{\pm \, (\alpha_1 + \alpha_2)}$ satisfy non-vanishing commutation relations 
with the $\suL$ generators given by
\begin{alignat}{3}
 \left[ H_{\alpha_1} , E_{\pm \, \alpha_2} \right] &= \mp \, E_{\pm \,
   \alpha_2} &  
&\and &
 \left[ E_{\pm\, \alpha_1} , E_{\pm\, \alpha_2} \right] &= \pm\, E_{\pm\, (\alpha_1 + 
\alpha_2)}  \; , \\[4pt]
\left[ H_{\alpha_1} , E_{\pm\, (\alpha_1 + \alpha_2)} \right] &= \pm \,
E_{\pm\, (\alpha_1 + \alpha_2)}
 & &\and &
\left[ E_{\pm\, \alpha_1} , E_{\mp\, (\alpha_1 + \alpha_2)} \right] &= \mp\, E_{\mp \,
\alpha_2} \; ,
\end{alignat}
with the $\uoL$ generator given by
\begin{alignat}{3}
  \left[ H_{\alpha_2}  , E_{\pm \, \alpha_2} \right] &= \pm \, 3
  E_{\pm\,  \alpha_2} 
& &\and &
 \left[ H_{\alpha_2}, E_{\pm\, (\alpha_1 + \alpha_2)} \right] &= \pm\, 3 E_{\pm \,
(\alpha_1 + \alpha_2)} \; , 
\end{alignat}
and amongst each other given by
\begin{alignat}{3}
   \left[ E_{\alpha_2} , E_{-\alpha_2} \right] &= \tfrac{1}{2} \, \left( 
H_{\alpha_2} - H_{\alpha_1} \right) & &\and &
\left[ E_{\alpha_1 + \alpha_2}, 
E_{-\alpha_1 - \alpha_2} \right] &= \tfrac{1}{2}\, \left( H_{\alpha_1} + 
H_{\alpha_2} \right) \; ,\\[4pt]
 \left[ E_{\pm\, \alpha_2} , E_{\mp \, (\alpha_1 + \alpha_2)} \right]
 &= \pm \, E_{\mp \,
\alpha_1} \; .
\end{alignat}
\end{subequations}
\paragraph{Skew-Hermitian basis of $\sltcL$}
Equivalently, we introduce the complex basis given by
\begin{subequations}
\label{eqn:alternative_basis_SU(3)}
\begin{alignat}{2}
 I_1 &\coloneqq E_{\alpha_1 + \alpha_2}  -  E_{-\alpha_1 - \alpha_2} \; , 
&\qquad  I_2 &\coloneqq - \im \, \left( E_{\alpha_1 + \alpha_2}  + E_{-\alpha_1 - 
\alpha_2} \right) \; 
, \\[4pt]
 I_3 &\coloneqq E_{ \alpha_2}  -  E_{ - \alpha_2} \; , &\qquad  I_4 
&\coloneqq - \im\, \left( E_{ \alpha_2}  + E_{ - \alpha_2} \right) \; , \\[4pt]
 I_5 &\coloneqq  -\tfrac{\im}{2} \, H_{\alpha_2 } \; , & & \\[4pt]
I_6 &\coloneqq  E_{\alpha_1} - E_{-\alpha_1} \; , &\qquad I_7 &\coloneqq -\im \,
\left( E_{\alpha_1}+ E_{-\alpha_1} \right) \; , \\[4pt]
I_8 &\coloneqq \im \, H_{\alpha_1} \; , & &
\end{alignat}
\end{subequations}
which reflects the splitting $\sutL= \suL \oplus 
\mathfrak{m}$ in which
\begin{equation}
 I_i \in \suL \for i = 6,7,8 \and I_\mu \in \mathfrak{m} \for \mu=1, 
\ldots,5 \; .
\end{equation}
This representation of 
generators is skew-Hermitian, i.e. $I_\mu = - 
I_\mu^\dagger$ for $\mu = 1,\ldots, 5$ and $I_i = - I_i^\dagger$ for $i=6,7,8$, in contrast to the Cartan-Weyl basis. The chosen Cartan 
subalgebra is spanned by $I_5$ and $I_8$, and $\left[I_5,I_i \right]=0$.
From the commutation relations~\eqref{eqn:def_Lie-algebra_SU(3)} one can infer 
the non-vanishing structure constants of these generators as
\begin{subequations}
 \label{eqs:structure_constants}
\begin{align}
 f_{67}^{\ \ 8 }&= -2 \qquad \text{plus cyclic} \; , \\[4pt]
f_{63}^{\ \ 1}&= f_{64}^{\ \ 2} = 
f_{71}^{\ \ 4} = f_{73}^{\ \ 2} =  
f_{82}^{\ \ 1}=  f_{83}^{\ \ 4} = 1 \qquad \text{plus cyclic} \; , \\[4pt]
f_{12}^{\ \ 5}&= f_{34}^{\ \ 5} =2 \; , \\[4pt]
f_{25}^{\ \ 1} &= - f_{15}^{\ \ 2} = f_{45}^{\ \ 3} = - f_{35}^{\ \ 4} = 
\tfrac{3}{2}  \; .
\end{align}
\end{subequations}
The Killing form $K_{AB} \coloneqq f_{AC}^{\ \ D} f_{DB}^{\ \ C}$ (with $ 
A,B,\ldots=1,\ldots,8$) associated to this basis is diagonal but not 
proportional to the identity, and is given by
\begin{equation}
 K_{ab}=12\, \delta_{ab} \for a,b=1,2,3,4 \; , \quad K_{55}=9 \and K_{ij}= 12 \,
\delta_{ij} \for i,j=6,7,8 \; .
\end{equation}
Introducing the 't~Hooft tensors $\eta^\alpha_{ab} $ for $a,b=1,2,3,4$ and 
$\alpha=1,2,3$ one has
\begin{equation}
 f_{ab}^{\ \ 5} = 2\, \eta^3_{ab} \and f_{a5}^{\ \ b} = -\tfrac{3}{2}
 \, \eta^3_{ab}
\; .
\end{equation}
\paragraph{Biedenharn basis}
The irreducible $\sut $-representations $\RepSu{k}{l}$ are labelled by a pair of non-negative integers $(k,l)$ and have
(complex) dimension
\begin{equation}
p_0:= \dim\big(\, \RepSu{k}{l}\, \big)= \tfrac{1}{2}\, (k+l+2)\,
(k+1)\, (l+1) \ .
\end{equation}
We decompose $\RepSu{k}{l}$ with respect to the subgroup $\Hh= \su \times \uo \subset 
\G$, just as in~\cite{Lechtenfeld:2008nh}.
A particularly convenient choice of basis for the vector space $\RepSu{k}{l}$ is 
the \emph{Biedenharn 
basis}~\cite{Biedenharn:1962haa,Baird:1963wv,Mukunda:1965}, which is defined 
to be the eigenvector basis given by
\begin{equation}
  H_{\alpha_1} \ket{n}{q}{m} = q \, \ket{n}{q}{m} \; , \quad 
L^2 \ket{n}{q}{m} = n\, (n+2 )\, \ket{n}{q}{m} \; \and
H_{\alpha_2} \ket{n}{q}{m} = m \, \ket{n}{q}{m} \; , %
\label{eqn:def_Biederharn}
\end{equation}
where $L^2 := 2\left( E_{\alpha_{1}}\, E_{-\alpha_{1}} + E_{-\alpha_{1}} \,
E_{\alpha_{1}}  \right) + H_{\alpha_1}^2$ is the isospin operator of $\suL$. 
Define the representation space $\Reps{(n,m)}$ as the eigenspace with definite 
isospin $n\in \mathbb{Z}_{\geq0}$ and magnetic monopole charge $\tfrac{m}{2}$ for 
$m\in \Z$. 
Then the $\sut $-representation $\RepSu{k}{l}$ decomposes into irreducible 
$\su \times \uo $-representations $\Reps{(n,m)}$ as
\begin{equation}
 \RepSu{k}{l} = \bigoplus_{(n,m)\in Q_0(k,l)} \, \Reps{(n,m)} \; , 
\label{eqn:decomp_C(k,l)}
\end{equation}
where $Q_0(k,l)$ parameterises the set of all occurring representations 
$\underline{(n,m)}$. In 
Appendix~\ref{sec:Rep_Theory_SU3} we 
summarise the matrix elements of all generators in the Biedenharn basis.
\paragraph{Representations of $\ZK$}
As the cyclic group $\ZK$ is abelian, each of its irreducible representations is $1$-dimensional. 
There are exactly $q+1$ inequivalent irreducible unitary 
representations $\rho_l$ given by 
\begin{equation}
 \rho_l \, : \, \begin{matrix} \ZK &\longrightarrow \ S^1 \subset \mathbb{C}^* \\
p &\longmapsto \ \e^{\frac{2 \pi \, \im\, (p+l) }{q+1}} \end{matrix}  \for l=0,1,\ldots, 
q\; . %
\label{eqn:irred_reps_ZK}
\end{equation}
\subsection{Homogeneous bundles and quiver representations}
Consider the groups $\G=\sut $, 
$\Hh = \su \times \uo$ , $\K=\su $, $\widetilde{\K}=\su 
\times \ZK \subset \Hh$ and a finite-dimensional $\K$-representation 
$\Reps{R}$ which descends from a $\G$-representation. Associate to the 
principal bundle~\eqref{eqn:bundle_SU3_S5} the $\K$-equivariant vector 
bundle $\mathcal{V}_{\Reps{R}} \coloneqq \G \times_{\K} 
\Reps{R}$. Due to the embedding $\ZK \hookrightarrow \uo \subset \sut$ and 
the origin of $\underline{R}$ from a $\G$-module, it follows that 
$\Reps{R}$ is also a $\ZK$-module. Consequently, as in 
Section~\ref{subsec:geometry_orbifold}, the $\ZK$-action $\gamma : \ZK \times 
S^5 \to S^5$ can be lifted to a $\ZK$-action $\widetilde{\gamma} : \ZK  
\times \mathcal{V}_{\Reps{R}} \to \mathcal{V}_{\Reps{R}}$ wherein the linear 
$\ZK$-action on the fibres is trivial. Thus one can define 
the corresponding $\widetilde{\K}$-equivariant vector V-bundle 
$\widetilde{\mathcal{V}}_{\Reps{R}}$ by suitable $\ZK$-projection as\footnote{See also 
the treatment in~\cite{Lechtenfeld:2014fza}.}
\begin{equation}
 \xymatrix{ \mathcal{V}_{\Reps{R}} \ar[r]^{\widetilde{\gamma}} \ar[d]_{\pi} & 
\widetilde{\mathcal{V}}_{\Reps{R}} \ar[d]^{\widetilde{\pi}} \\
S^5 \ar[r]_{\!\!\!\!\!\!\!\!\!\gamma} & S^5 \slash \ZK } \label{eqn:homogeneous_bundle}
\end{equation}
and again we denote the vector V-bundle 
$\widetilde{\mathcal{V}}_{\Reps{R}}$ by the same symbol 
$\mathcal{V}_{\Reps{R}}$ whenever the context is clear.

It is known~\cite{AlvarezConsul:2001uk} that the category of such holomorphic 
homogeneous vector bundles $\mathcal{V}_{\Reps{R}}$ is equivalent to the 
category of finite-dimensional representations of certain quivers with 
relations.
We use this equivalence to associate quivers to homogeneous 
bundles related to an irreducible $\sut$-representation $\underline{R} 
=\RepSu{k}{l}$, which is evidently a finite-dimensional (and usually 
reducible) representation of $\su \times \ZK \hookrightarrow \su \times \uo$.
\paragraph{Flat connections}
Inspired by the structure of the flat 
connection~\eqref{eqn:flat_connection_orbifold} on the 
V-bundle~\eqref{eqn:bundle_SU3_S5_ZK}, one observes that it 
can be written as\footnote{Note that~\eqref{eqn:flat_connection_orbifold} 
implicitly uses the fundamental representation $\RepSu{1}{0}$ of $\sut$.}  
\begin{subequations}
\label{eqn:ansatz_flat_conn_all}
\begin{equation}
\begin{split}
 \Acal_0 = \big[B_{11} \, H_{\alpha_1}  + B_{12} \, E_{\alpha_{1}} - \left( 
B_{12} \, E_{ \alpha_{1}} \right)^\dagger \big] - \tfrac{\im}{2} \, \eta \,
H_{\alpha_2} &+ \barbetaZK^1 \, E_{ \alpha_{1}+ \alpha_2} + \barbetaZK^2\, E_{ 
\alpha_{2}}  \\
&- \betaZK^1\, E_{- \alpha_{1}- \alpha_2}- \betaZK^2\, E_{- \alpha_2}
\ ,
\end{split}
\label{eqn:ansatz_flat_connection_S5_ZK} 
\end{equation}
or equivalently
\begin{equation}
\mathcal{A}_0 =  \Gamma + I_\mu \, e^\mu 
\label{eqn:alternative_flat_conn} 
\end{equation}
\end{subequations}
with the coframe $\{e^\mu\}_{\mu=1,\ldots, 5}$ defined 
in~\eqref{eqn:def_1-forms_ZK} and the definition
\begin{equation}
 \Gamma \coloneqq \Gamma^i \, I_i \with
\Gamma^6 = \tfrac{\im}{2} \, \left(B_{12} - \bar{B}_{12} \right) \, , \quad 
\Gamma^7 = \tfrac{1}{2} \, \left(B_{12} + \bar{B}_{12} \right) \, , \quad 
\Gamma^8 = - \im\, B_{11} \, . %
\label{eqn:def_Gamma}
\end{equation}
Note that $\Gamma$ is an $\suL$-valued connection $1$-form. The flatness 
of $\Acal_0$, i.e. $\Fcal_0 = \diff \Acal_0 + \Acal_0 \wedge 
\Acal_0 =0$, is encoded in the relation
\begin{subequations}
 \label{eqn:flat_connection_alternative}
\begin{alignat}{2}
 & & \Fcal_0 &= F_{\Gamma} + I_{\mu} \, \diff e^{\mu} + \Gamma^i\, \left[I_i,I_\mu 
\right] \wedge e^{\mu}  + \tfrac{1}{2} \, \left[I_{\mu}, I_{\nu}
\right] \, e^{\mu \nu} 
 =0 \; , \\[4pt]
  &\Fcal_0\big|_{\suL} =0 \; : \qquad &   
F_{\Gamma} &= - \tfrac{1}{2}\, f_{\mu \nu}^{ \ \ i} \, I_i \,  e^{\mu \nu} \; ,\\[4pt]
  &\Fcal_0\big|_{\mathfrak{m}} =0 \; : &  
  \diff e^\mu &=-\Gamma^i \, f_{i \nu}^{ \ \ \mu} \wedge e^\nu - \tfrac{1}{2}\,  f_{\rho 
\sigma}^{ \ \ \mu}\,  e^{\rho \sigma} \; ,
\end{alignat}
\end{subequations}
where $F_\Gamma=\diff\Gamma+\Gamma\wedge\Gamma$.
The equivalent information can be cast in a set of relations 
starting from~\eqref{eqn:ansatz_flat_connection_S5_ZK} and using the Biedenharn 
basis; see Appendix~\ref{sec:flat_conn_appendix} for details.
\paragraph{$\ZK$-equivariance}
Consider the principal V-bundle~\eqref{eqn:bundle_SU3_S5_ZK}, 
where the $\ZK$-action is defined on $S^5$ as in 
Section~\ref{subsec:geometry_orbifold}. The 
connection~\eqref{eqn:ansatz_flat_conn_all} is $\sut$-equivariant by 
construction, but one can also check its
$\ZK$-equivariance explicitly. For this, one needs to specify an action of $\ZK$ on the 
fibre $\RepSu{k}{l}$, which 
decomposes as an $\su$-module via~\eqref{eqn:decomp_C(k,l)}. Demanding
that the $\ZK$-action commutes with the $\su$-action on $\RepSu{k}{l}$ forces it to act as a multiple of the identity on each irreducible $\su$-representation by Schur's lemma.
Hence we choose a representation $\gamma: \ZK \to \urm (p_0)$ of $\ZK$ on $\RepSu{k}{l}$ such that 
 $\ZK$ acts on $\Reps{(n,m)} $ as
  \begin{equation}
   \gamma(h)\big|_{\Reps{(n,m)}} = \zk^m \, \mathds{1}_{n+1} \in 
\uo \ . \label{eqn:rep_ZK_flat}
  \end{equation}
Consider the two parts of the 
connection~\eqref{eqn:ansatz_flat_conn_all}: The connection $\Gamma$ and the 
endomorphism-valued $1$-form $I_\mu\, e^\mu$. In terms of matrix elements, 
$\Gamma$ is completely determined by the $1$-forms $\Bfield{n,m} \in 
\Omega^1\left(\suL,\End(\, \Reps{n,m}\, ) \right)$ which are instanton connections on the $\widetilde{\K}$-equivariant vector 
V-bundle 
  \begin{equation}
    \widetilde{ \mathcal{V}}_{\Reps{(n,m)}} \xrightarrow{\Reps{(n,m)} 
\;} \G/\widetilde{\K} 
\cong S^5/\ZK  \with  \mathcal{V}_{\Reps{(n,m)}} := \G 
\times_{\K} \Reps{(n,m)} \; ,
  \label{eqn:def_K-bundle_over_S5}\end{equation}
simply because they are $\K$-equivariant by construction and $\ZK 
\hookrightarrow \uo \subset \sut$ commutes with this particular $\su$
subgroup (see 
also Appendix~\ref{subsec:geometry_CP}). More 
explicitly, taking~\eqref{eqn:rep_ZK_flat} one observes that $\ZK$ acts 
trivially on the endomorphism part,
\begin{equation}
 \gamma(h) \, \Bfield{n,m} \, \gamma(h)^{-1} = \Bfield{n,m} \; ,
\label{eqn:ZK-equiv_Bfields}
\end{equation}
as well as on the $1$-form parts $\Gamma_i$ because they are horizontal 
in the V-bundle~\eqref{eqn:bundle_SU3_S5_ZK}.  
For $\ZK$-equivariance of the second term $I_\mu \, e^\mu$, from~\eqref{eqn:ansatz_flat_connection_S5_ZK} and the 
representation $\pi$ defined in~\eqref{eqn:transf_1forms_ZK} one demands the 
conditions
\begin{subequations}
\label{eqn:ZK-equivariance_flat_conn}
 \begin{align}
\gamma(h) \, E_{ \mathfrak{w}} \, \gamma(h)^{-1} &= \pi(h)^{-1} (E_{ 
\mathfrak{w}}) = \zk^3 \, E_{ 
\mathfrak{w}} \for \mathfrak{w}=\alpha_2 , \alpha_1 + \alpha_2 \; , \\[4pt]
  \gamma(h) \, E_{ -\mathfrak{w}}  \, \gamma(h)^{-1} &= \pi(h)^{-1} 
(E_{ -\mathfrak{w}} )= \zk^{-3}\, E_{- 
\mathfrak{w}} \for 
\mathfrak{w}=\alpha_2 , \alpha_1 + \alpha_2 \; ,\\[4pt]
  \gamma(h) \, H_{\alpha_2} \, \gamma(h)^{-1} &=\pi(h)^{-1} 
(H_{\alpha_2})=  H_{\alpha_2} \; .
 \end{align}
\end{subequations}
One can check that these conditions are satisfied by our 
choice of representation~\eqref{eqn:rep_ZK_flat}, due to the explicit 
components of the generators~\eqref{eqn:matrix_elements_E1_E1+2}. We
conclude that, 
due to our ansatz for the connection~\eqref{eqn:ansatz_flat_conn_all} on 
the principal V-bundle~\eqref{eqn:bundle_SU3_S5_ZK} and the 
embedding $\ZK \hookrightarrow \uo  \subset \sut $, the 1-form $\Acal_0$ 
is indeed $\ZK$-equivariant.
\paragraph{Quiver representations}
Recall from~\cite{Lechtenfeld:2008nh} that one can 
interpret the decomposition~\eqref{eqn:decomp_C(k,l)} and the structure of the 
connection~\eqref{eqn:ansatz_flat_conn_all} as a quiver associated to 
$\RepSu{k}{l}$ as follows: The appearing $\Hh$-representations 
$\Reps{(n,m)}$ form a set $Q_0(k,l)$ of vertices, whereas the actions of the 
generators $E_{\alpha_2}$ and $E_{\alpha_1 + \alpha_2}$ 
intertwine the $\Hh$-modules. These $\Hh$-morphisms, together with $H_{\alpha_2}$,
constitute a set $Q_1(k,l)$ of arrows $(n,m)\to(n',m'\,)$ between the vertices. The quiver 
$\mathcal{Q}^{k,l}$ is then given by the pair 
$\mathcal{Q}^{k,l}=\big(Q_0(k,l)\,,\, 
Q_1(k,l) \big)$; the underlying graph of this quiver is obtained from the weight diagram of the representation $\RepSu{k}{l}$ by collapsing all horizontal edges to vertices, cf.~\cite{Lechtenfeld:2008nh}. See Appendix~\ref{sec:examples_quiver} 
for an explicit treatment of the examples $\RepSu{1}{0}$, $\RepSu{2}{0}$ and~$\RepSu{1}{1}$.
% 
%%%%%%%%%%%%%%%%%%%%%%%%%%%%%%%%%%%%%%%%%%%%%%%%%%%%%%%%%%%%%%%%%%%%%%%%%%%%%%%%
%%%%%%%%%%%%%%%%%%%%%%%%%%%%%%%%%%%%%%%%%%%%%%%%%%%%%%%%%%%%%%%%%%%%%%%%%%%%%%%%
% 
\subsection{Quiver bundles and connections}
 \label{subsec:construction_quiver}
In the following we will consider representations of
quivers not in the category of vector spaces, but rather in the category of 
vector bundles.
 We shall construct a $\G$-equivariant gauge theory on 
the product space
\begin{equation}
 \mfd{d} \times_{\widetilde{\K}} \G := \mfd{d} \times 
\G \slash \widetilde{\K} = \mfd{d} \times S^5 \slash \ZK
\end{equation}
where $\G$ and all of its subgroups act trivially on a $d$-dimensional
Riemannian manifold $\mfd{d}$. The 
equivariant dimensional reduction compensates isometries on ${\G \slash \widetilde{\K}}$ 
with gauge transformations, thus leading to quiver gauge theories on the manifold $\mfd{d}$. 

Roughly speaking, the reduction is achieved by extending the
homogeneous V-bundles~\eqref{eqn:homogeneous_bundle} by 
$\widetilde{\K}$-equivariant bundles $E \to \mfd{d}$, which furnish a 
representation of the corresponding quiver in the category of 
complex vector bundles over $\mfd{d}$. Such a representation is called a
\emph{quiver bundle} and it originates from the one-to-one correspondence between 
$\G$-equivariant Hermitian vector V-bundles over $\mfd{d} \times 
\G \slash \widetilde{\K} $ and $\widetilde{\K}$-equivariant Hermitian vector 
bundles over $\mfd{d}$, where $\widetilde{\K}$ acts trivially on the 
base space $\mfd{d}$~\cite{AlvarezConsul:2001uk}.
\paragraph{Equivariant bundles}
For each irreducible 
$\Hh$-representation $\Reps{(n,m)}$ in the decomposition of $\RepSu{k}{l}$, construct the 
(trivial) vector bundle
  \begin{equation}
   \Reps{(n,m)}_{\, \mfd{d}}\coloneqq \mfd{d} \times_{\widetilde{\K}} \Reps{(n,m)} 
 \xrightarrow{ \Reps{(n,m)} 
\; } \mfd{d} 
  \end{equation}
of rank $n+1$, which is $ \widetilde{\K}$-equivariant due to the trivial $\widetilde{\K}$-action on $\mfd{d}$ and the linear 
action on the fibres. For each module $\Reps{(n,m)}$ introduce also a Hermitian 
vector bundle
  \begin{equation}
   E_{p_{(n,m)}} \xrightarrow{\C^{p_{(n,m)}}} 
\mfd{d} \with \rank(E_{p_{(n,m)}}) = p_{(n,m)} %
  \label{eqn:def_hermitian_bundles}
  \end{equation}
  with structure group $\urm (p_{(n,m)})$ and a $\urmL(p_{(n,m)})$-valued 
connection $\ConnQui{n,m}$, and with trivial $\widetilde{\K}$-action. Denote the identity endomorphism 
on the fibres of $E_{p_{(n,m)}}$ by $\UniEndQui{n,m}$. With these data one 
constructs a $\widetilde{\K}$-equivariant bundle 
  \begin{equation}
   E^{k,l} \cong \bigoplus_{(n,m)\in Q_0(k,l)} \, E_{p_{(n,m)}} \otimes 
\Reps{(n,m)}_{\, \mfd{d}} \xrightarrow{ \ {\C}^{p} \ } \mfd{d} 
\label{eqn:decomp_E(k,l)}
  \end{equation}
  whose rank $p$ is given by 
  \begin{equation}
   p= \sum_{(n,m)\in Q_0(k,l)}\, p_{(n,m)} \ \dim \, \Reps{(n,m)} = 
\sum_{(n,m)\in Q_0(k,l)}\, p_{(n,m)} \, (n+1) \; .
  \end{equation}
Following~\cite{Lechtenfeld:2008nh}, the bundle $E^{k,l}$ is the 
$\widetilde{\K}$-equivariant vector bundle of rank $p$ associated to 
the representation $\RepSu{k}{l} \big|_{\widetilde{\K}}$ of 
$\widetilde{\K}$, and~\eqref{eqn:decomp_E(k,l)} is its 
isotopical decomposition. This construction breaks the structure group $\urm(p)$ of $E^{k,l}$ via the Higgs effect to the subgroup
\begin{equation}
\Gcal^{k,l}:= \prod_{(n,m)\in Q_0(k,l)}\, \urm\big(p_{(n,m)}\big)^{n+1}
\label{eqn:gauge_group_fibre}\end{equation}
which commutes with the $\su$-action on the fibres of~\eqref{eqn:decomp_E(k,l)}.

On the other hand, one can introduce 
$\widetilde{\K}$-equivariant V-bundles over $S^5 \slash \ZK$
by~(\ref{eqn:def_K-bundle_over_S5}). On $\mathcal{V}_{\Reps{(n,m)}}$ one has the 
$\mathfrak{su}(2)$-valued $1$-instanton connection $\Bfield{n,m}$ in the $(n+1)$-dimensional 
irreducible representation.
  The aim is to establish a $\G$-equivariant V-bundle 
$\mathcal{E}^{k,l}$ over $\mfd{d} \times S^5 \slash \ZK$ as an extension of the 
$\widetilde{\K}$-equivariant bundle $E^{k,l}$. By the results 
of~\cite{AlvarezConsul:2001uk} such a V-bundle $\mathcal{E}^{k,l}$ exists and
according to~\cite{Lechtenfeld:2008nh} it is realised as
\begin{equation}
\mathcal{E}^{k,l} \coloneqq \G \times_{\widetilde{\K}} E^{k,l}  
= \bigoplus_{(n,m)\in Q_0(k,l)}\, E_{p_{(n,m)}} 
\boxtimes \mathcal{V}_{\Reps{(n,m)}} \xrightarrow{ \ V^{k,l} \ } \mfd{d} \times S^5 \slash \ZK \; ,
\label{eqn:equiv_quiver_bundle}
\end{equation}
where
\begin{equation}
V^{k,l} =
\bigoplus_{(n,m)\in Q_0(k,l) } \,
\C^{p_{(n,m)}} \otimes \Reps{(n,m)} 
\label{eqn:typical_fibre}\end{equation}
is the typical fibre of (\ref{eqn:equiv_quiver_bundle}).
 \paragraph{Generic $\G$-equivariant connection}
The task now is to determine the generic form of a $\G$-equivariant 
connection on~\eqref{eqn:equiv_quiver_bundle}. Since the space of connections 
on $\mathcal{E}^{k,l}$ is an affine space modelled over $\Omega^1( 
\End(\mathcal{E}^{k,l}))^\G$, one has to study the 
$\G$-representations on this vector space. Recall 
from~\cite{Lechtenfeld:2008nh} that the decomposition of $\Omega^1( 
\End(\mathcal{E}^{k,l}))^\G$ with respect to $\G$ yields 
a ``diagonal'' subspace which accommodates the connections $\ConnQui{n,m}$ 
on~\eqref{eqn:def_hermitian_bundles} twisted by 
$\G$-equivariant connections on~\eqref{eqn:def_K-bundle_over_S5}, and 
an ``off-diagonal'' subspace which gives rise to bundle morphisms.

In other words, $\K$-equivariance alone introduces only the connections 
$\ConnQui{n,m}$ on each bundle~\eqref{eqn:def_hermitian_bundles} as well as the 
$\su$-connections $\Bfield{n,m}$ on the 
V-bundles~\eqref{eqn:def_K-bundle_over_S5}. On the other hand,
$\G$-equivariance additionally requires one to introduce a set of 
bundle morphisms
\begin{subequations}
\begin{align}
 \HomoPhiPMi{n,m} \in 
\Hom\big(E_{p_{(n,m)}},E_{p_{(n\pm1,m+3)}}\big)
\end{align}
and their adjoint maps
\begin{align} 
\HomoPhiAdjPMi{n,m} &\in 
\Hom\big(E_{p_{(n\pm1,m+3)}},E_{p_{(n,m)}}\big) \; , 
\end{align}
for all $(n,m)\in Q_0(k,l)$; one further introduces the bundle endomorphisms
\begin{equation}
 \EndPsiQui{n,m} \in \End\big(E_{p_{(n,m)}}\big)
\end{equation}
\end{subequations}
at each vertex $(n,m)\in Q_0(k,l)$ with $m\neq0$.
The morphisms $\HomoPhiPMi{n,m}$ and $\EndPsiQui{n,m}$ are collectively 
called \emph{Higgs fields}, and they realise the $\G$-action in the 
same way that the generators $I_\mu$ (or more precisely the $1$-forms
$\BarBetaPMi{n,m}$ and $\frac{\im\, m}2\, \eta\,\Pi_{(n,m)}$) 
do 
in the case of the flat connection~\eqref{eqn:ansatz_flat_conn_all}.
The ``new'' Higgs fields $\EndPsiQui{n,m}$ implementing the
vertical connection components on the (orbifold of the) Hopf bundle $S^5\to\CP$ must be Hermitian, i.e. 
$\EndPsiQui{n,m}= \EndPsiQui{n,m}^\dagger$, by construction in order
for the 
connection to be $\urmL(p)$-valued. 

\paragraph{Ansatz for connection}
The ansatz for a $\G$-equivariant connection on the equivariant
V-bundle \eqref{eqn:equiv_quiver_bundle} is given by
\begin{equation}
 \Acal= \widehat{A} + \widehat{\Gamma} + X_\mu \, e^\mu 
\label{eqn:connection_quiver}
\end{equation}
wherein the $\urmL(p_{(n,m)})$-valued connections $\ConnQui{n,m}$ and the 
$\suL$-valued connection $\Gamma$ are extended as
\begin{equation}
 \widehat{A} \coloneqq \bigoplus_{(n,m)}\, \ConnQui{n,m} \otimes \Pi_{(n,m)} 
\equiv A \otimes \mathds{1} \and
 \widehat{\Gamma} \coloneqq \bigoplus_{(n,m)}\, \pi_{(n,m)} \otimes \Gamma^i \,
I_i^{(n,m)} = \Gamma^i \, \widehat{I}_i \equiv \mathds{1} \otimes \Gamma \; 
,
\end{equation}
together with $\widehat{I}_i = \bigoplus_{  (n,m)}\, \pi_{(n,m)} \otimes 
I_i^{ (n,m)}$.
The matrices $X_\mu$ are required to satisfy the \emph{equivariance 
condition}~\cite{Ivanova:2012vz,Bunk:2014coa}
\begin{equation}
 \big[\, \widehat{I}_i , X_\mu \big] = f_{i \mu}^{\ \ \nu} \, X_\nu  \for i=6,7,8 \and 
\mu=1,\ldots,5 \; .%
 \label{eqn:equivariance}
\end{equation}
As explained in~\cite{Bunk:2014coa}, the equivariance condition 
ensures that $X_\mu$ are frame-independently defined endomorphisms that 
are the components of an endomorphism-valued $1$-form, which is here given as 
the difference $\Acal -(\widehat{A} + \widehat{\Gamma}\, )$.

The general solution to~\eqref{eqn:equivariance} expresses $X_\mu$ in terms of 
Higgs fields and generators as
\begin{subequations}
\label{eqn:explicit_form_X-matrices}
\begin{align}
\tfrac{1}{2}\, ( X_1 + \im \, X_2 ) &= \bigoplus_{\pm,(n,m)}  \,
\HomoPhiPMi{n,m} \otimes   E_{\alpha_1 + 
\alpha_2}^{\pm\,(n,m)} \ , \quad
\tfrac{1}{2} \, (X_1 - \im \, X_2) =  - \bigoplus_{\pm,(n,m)}  \,
\HomoPhiAdjPMi{n,m} \otimes E_{-\alpha_1 - 
\alpha_2}^{\pm\,(n,m)}  \; ,  \\[4pt]
\tfrac{1}{2}\,( X_3 + \im\, X_4) &= \bigoplus_{\pm,(n,m)}  \,
\HomoPhiPMi{n,m} \otimes 
E_{\alpha_2}^{\pm\, (n,m)} \ , \quad
\tfrac{1}{2}\, ( X_3 - \im X_4)  = -  \bigoplus_{\pm,(n,m)} 
\HomoPhiAdjPMi{n,m} \otimes E_{ - 
\alpha_2}^{\pm\, (n,m)} \; ,  \\[4pt]
X_5 &= -\frac{\im}{2} \, \bigoplus_{(n,m)}\, \EndPsiQui{n,m} \otimes 
H_{\alpha_2}^{(n,m)} \; .
\end{align}
\end{subequations}
Altogether the $\G$-equivariant connection takes the form
\begin{align}
 \Acal =  & \bigoplus_{(n,m)\in Q_0(k,l)} \, \Big( \ConnQui{n,m} \otimes 
\Pi_{(n,m)} + 
\pi_{(n,m)} \otimes \Bfield{n,m} - \EndPsiQui{n,m} \otimes \tfrac{\im\, m}{2} \,
\eta \, \Pi_{(n,m)} %
\label{eqn:connection_quiver_detail}\\*
&\qquad\qquad + \, \HomoPhiPlusi{n,m} \otimes \BarBetaPlusi{n,m} + 
\HomoPhiMinusi{n,m} \otimes \BarBetaMinusi{n,m} - \HomoPhiAdjPlusi{n,m} \otimes \BetaPlusi{n,m} - 
\HomoPhiAdjMinusi{n,m} \otimes \BetaMinusi{n,m} \Big)
  \ . \notag
\end{align}
\paragraph{$\ZK$-equivariance}
One needs 
to extend the $\ZK$-re\-pre\-sen\-tation $\gamma$ of~\eqref{eqn:rep_ZK_flat} 
to act 
on the fibres~\eqref{eqn:typical_fibre} of the equivariant V-bundle~\eqref{eqn:equiv_quiver_bundle}. Since by construction $\widetilde{\K}=\su\times\Z_{q+1}$ acts trivially on the fibres of the 
bundles~\eqref{eqn:def_hermitian_bundles}, one ends up with the representation $\gamma : \ZK \to \urm (p)$ given by
\begin{equation}
 \gamma(h) = \bigoplus_{(n,m)\in Q_0(k,l)} \,
\mathds{1}_{p_{(n,m)}} \otimes \gamma(h)\big|_{\Reps{(n,m)}} = \bigoplus_{(n,m)\in Q_0(k,l)} \, \mathds{1}_{p_{(n,m)}} \otimes 
\zk^m \,
\mathds{1}_{n+1} \ . %
\label{eqn:rep_ZK_quiver}
\end{equation}

To prove $\ZK$-equivariance of~\eqref{eqn:connection_quiver} one
again needs to show two things. Firstly, the connections $A \otimes
\mathds{1} $ and $\mathds{1} \otimes \Gamma$ have to be $\ZK$-equivariant. This can be seen as 
follows: For $A \otimes \mathds{1}$ the representation $\gamma$ 
of~\eqref{eqn:rep_ZK_quiver} acts trivially on each bundle $E_{p_{(n,m)}}$, and thus
\begin{equation}
 \gamma(h) \, \left( A \otimes \mathds{1} \right) \, \gamma(h)^{-1} = A \otimes 
\mathds{1} \; .
\end{equation}
Furthermore, $\mathds{1} \otimes \Gamma$ is $\ZK$-equivariant because $\Gamma$ is by~\eqref{eqn:ZK-equiv_Bfields}, and hence the connection $A \otimes \mathds{1} + \mathds{1} \otimes 
\Gamma$ satisfies the equivariance conditions.

Secondly, the endomorphism-valued $1$-form $X_\mu \, e^\mu  =  \Acal- 
\widehat{A} - \widehat{\Gamma} $ needs to be $\ZK$-equivariant as well. Due to 
its structure, one needs to consider a combination of the adjoint action of 
$\gamma$ from~\eqref{eqn:rep_ZK_quiver} and the $\ZK$-action on 
forms from~\eqref{eqn:transf_1forms_ZK}. As $\gamma$ acts 
trivially on each bundle $E_{p_{(n,m)}} $, the $\ZK$-equivariance 
conditions 
 \begin{equation}
  \gamma(h) \, X_\mu \, \gamma(h)^{-1} = \pi(h)^{-1} (X_\mu) \for \mu=1,\ldots,5
  \label{eqn:ZK-equivariance_quiver_conn}
 \end{equation}
hold also for the quiver connection $\Acal$ just as they hold for the
flat connection 
$\Acal_0$ by~\eqref{eqn:ZK-equivariance_flat_conn}.

Thus the chosen representations~\eqref{eqn:transf_1forms_ZK} and 
\eqref{eqn:rep_ZK_quiver} render the quiver 
connection~\eqref{eqn:connection_quiver} equivariant with respect to the action of $\ZK$. On 
each irreducible representation $\Reps{(n,m)}$ the generator $h$ of $\ZK$ is 
represented by $\zk^m\, \mathds{1}_{n+1}$ which depends on the $\uo $
monopole charge but not on the $\su $ isospin. This 
comes about as follows: The bundle morphisms associated to $\betaZK^i$ map 
between bundles $E_{p_{(n,m)}}  \otimes  \Reps{(n,m)}_{\, \mfd{d}}$ that differ in $m$ by 
$-3$ 
(from source to target vertex), but differ in $n$ by either $+1$ or $-1$. Thus
the representation $\gamma$ should only be sensitive to $m$ and not to
$n$. We shall elucidate this point further in Section~\ref{subsec:quiver_graphs}.
\paragraph{Curvature}
The curvature $\Fcal=\diff\Acal+\Acal\wedge \Acal$ of the 
connection~\eqref{eqn:connection_quiver} is given by
\begin{subequations}
\begin{equation}
 \Fcal= F_A \otimes \mathds{1} + \mathds{1} \otimes F_{\Gamma} + \big( 
\diff X_\mu + \big[\, \widehat{A}, X_\mu \big] \big) \wedge e^\mu + X_\mu \,
\diff 
e^\mu + \big[\, \widehat{\Gamma},X_\mu \big] \wedge e^\mu + \tfrac{1}{2}\, \left[ 
X_\mu , X_\nu \right] \, e^{\mu \nu} \; ,
\end{equation}
where $F_A=\diff A+A\wedge A$. Employing the relations~\eqref{eqn:flat_connection_alternative} then yields
\begin{equation}
\begin{split}
 \mathcal{F}= F_A \otimes \mathds{1} +  \big( \diff X_\mu + \big[\, 
\widehat{A}, 
X_\mu \big] \big) \wedge e^\mu &+ \Gamma^i \,\big( \big[\, \widehat{I}_i , 
X_\mu 
\big] - f_{i \mu}^{\ \ \nu} \, X_\nu \big) \wedge e^\mu \\
&+ \tfrac{1}{2} \, \big( \left[ X_\mu , X_\nu \right] - f_{\mu \nu}^{\
  \ \rho} \, X_\rho 
- f_{\mu \nu}^{\ \ i} \, \widehat{I}_i \, \big)\, e^{\mu \nu } \; .
\end{split}
\end{equation}
Since the matrices $X_\mu$ satisfy the equivariance 
relation~\eqref{eqn:equivariance}, the final form 
of the curvature reads
\begin{equation}
 \mathcal{F}= F_A \otimes \mathds{1} +  (DX)_\mu\, \wedge e^\mu + \tfrac{1}{2}\, \big( 
\left[ X_\mu , X_\nu \right] - f_{\mu \nu}^{ \ \ \rho} \, X_\rho - f_{\mu \nu}^{\ \ i} \,
\widehat{I}_i \, \big)\, e^{\mu \nu } \; , \label{eqn:curv_quiver_matrix}
\end{equation}
where we defined the bifundamental covariant derivatives as 
\begin{equation}
 (D X)_\mu := \diff X_\mu + \big[\, \widehat{A},X_\mu \big] \; .
\end{equation}
\end{subequations}
Inserting the explicit form~\eqref{eqn:explicit_form_X-matrices} for
the scalar fields $X_\mu$ 
leads to the curvature components in the Biedenharn basis; the 
detailed expressions are summarised in Appendix~\ref{sec:Quiver_conn_appendix}.
\paragraph{Quiver bundles}
Let us now exemplify and clarify how the equivariant bundle
$E^{k,l}\to M^d$ from (\ref{eqn:decomp_E(k,l)}) realises a quiver
bundle from our constructions above.
Recall that the quiver consists of the pair $\Qcal^{k,l}= \big( Q_0(k,l)\,,\,Q_1(k,l) \big) $, with 
vertices $(n,m)\in Q_0(k,l)$ and arrows $(n,m) \to (n',m'\,) \in 
Q_1(k,l)$ between certain pairs of vertices which are here determined 
by the decomposition~\eqref{eqn:decomp_C(k,l)}. We consider a 
representation $\widetilde{\Qcal}\,^{k,l}= \big(\, \widetilde{Q}_0(k,l)\,,\,\widetilde{Q}_1(k,l)\, \big) $ of this
quiver in the category 
of complex vector bundles. The set of vertices is  
\begin{equation}
 \widetilde{Q}_0(k,l) = \big\{ E_{p_{(n,m)}} \longrightarrow \mfd{d} \, , \quad (n,m)\in 
Q_0(k,l) \big\} \; ,
\end{equation}
i.e. the set of Hermitian vector bundles each equipped with a unitary connection 
$\ConnQui{n,m}$. The set of arrows is
\begin{align}
 \widetilde{Q}_1(k,l) = & \ \Big\{ \HomoPhiPMi{n,m} \in 
\Hom\big(E_{p_{(n,m)}},E_{p_{(n\pm1,m+3)}}\big)\, , \quad
(n,m)\in Q_0(k,l) \Big\} \notag \\*
& \qquad \cup \ \Big\{ \EndPsiQui{n,m} \in \End\big(E_{p_{(n,m)}}\big)\, ,
  \quad (n,m) \in Q_0(k,l) \ , \ m\neq0 \Big\}\; ,
\end{align}
which is precisely the set of bundle morphisms, i.e. the Higgs fields. These quivers differ from those
considered in~\cite{Lechtenfeld:2008nh} by the appearance of vertex loops 
corresponding to the endomorphisms $\EndPsiQui{n,m}$. See 
Appendix~\ref{sec:examples_quiver} for details of the quiver bundles based on the representations $\RepSu{1}{0}$, $\RepSu{2}{0}$ and 
$\RepSu{1}{1}$.

These constructions yield representations of quivers 
without any relations. We will see later on that relations can arise by minimising 
the scalar potential of the quiver gauge theory (see
Section~\ref{sec:Reduction_YM}) or by imposing a generalised
instanton equation on the connection $\Acal$ (see 
Section~\ref{sec:SU(3)-equivariant_cone}).

\subsection{Dimensional reduction of the Yang-Mills action}
\label{sec:Reduction_YM}
Consider the reduction of the 
pure Yang-Mills action from $\mfd{d} \times S^5$ to $\mfd{d}$. On $S^5$ we take as basis of coframes $\{\betaphi^j, 
\barbetaphi^j\}_{j=1,2}$ and $e^5 = \eta$, and as metric
 \begin{equation}
  \diff s^2_{S^5} = R^2 \, \left( \betaphi^1 \otimes \barbetaphi^1 + 
\barbetaphi^1 \otimes \betaphi^1 + \betaphi^2 \otimes \barbetaphi^2 
+\barbetaphi^2 \otimes \betaphi^2 \right) + r^2 \, \eta \otimes \eta \; . 
\label{eqn:metric_S5}
 \end{equation}
The Yang-Mills action is given by
\begin{equation}
 S= 
-\frac{1}{4 \tilde{g}^2} \, \int_{\mfd{d} \times S^5} \, \tr \, \Fcal \wedge \star \, \Fcal \; , 
\label{eqn:YM-action_MxS5}
\end{equation}
with coupling constant $\tilde{g}$ and $\star$ the Hodge duality
operator corresponding to the metric on $\mfd{d} \times S^5 $ given by
\begin{equation}
 \diff s^2 = 
\diff s_{M^d}^2 + \diff s^2_{S^5} \ .
\end{equation}
We denote the Hodge operator corresponding to the metric $\diff s_{M^d}^2$ on $M^d$ by $\starM$. 
The reduction of~\eqref{eqn:YM-action_MxS5} proceeds by inserting the 
curvature~\eqref{eqn:curv_quiver_matrix} and performing the integrals over 
$S^5$, which can be evaluated by using
\eqref{eqn:metric_S5} and the identities of Appendix~\ref{subsec:reduction_S5}.
One finally obtains for the reduced action
\begin{align}
 S= -\frac{2\pi^3\, r\, R^4}{\tilde{g}^2}\, &\Big( \,
\int_{\mfd{d}} \tr \, \big(F_A \wedge \starM F_A \big) \otimes 
\mathds{1} \notag \\* 
&\qquad + \frac1{2R^2}\, \int_{\mfd{d}} \ \sum_{a=1}^4 \,\tr \,
(DX)_a \wedge \starM (DX)_a
+ \frac{1}{r^2}\, \int_{\mfd{d}} \, \tr \, (DX)_5 \wedge \starM (DX)_5 \notag  \\
&\qquad +\frac{1}{8 R^4} \, \int_{\mfd{d}} \, \starM \ \sum_{a,b=1}^4 \, \tr\big( 
\left[X_a,X_b \right] - f_{ab}^{\ \ 5} \, X_5 - f_{ab}^{\ \ i} \, \widehat{I}_i \,
\big)^2 \notag \\
&\qquad +\frac{1}{8R^2\, r^2}\, \int_{\mfd{d}}\, \starM \ \sum_{a=1}^4 \, \tr\big( 
\left[X_a,X_5 \right] - f_{a5}^{\ \ b} \, X_b\big)^2 \, \Big) \; .
\label{eqn:YM-action_matrix}\end{align}
Here the explicit structure 
constants~\eqref{eqs:structure_constants}, i.e. $f_{ab}^{\ \ c} =f_{a5}^{\ \ 5} =f_{a5}^{\ \ i} =0$, have been used. One may detail this action further by inserting the $\G$-equivariant 
solution~\eqref{eqn:explicit_form_X-matrices} for the scalar fields $X_\mu$ in the 
Biedenharn basis, which allows one to perform the trace over the $\su \times 
\uo$-representations $\Reps{(n,m)}$. The explicit but lengthy formulas are 
given in Appendix~\ref{sec:details_red_S5_appendix}.
\paragraph{Higgs branch} 
On the Higgs branch of the quiver gauge theory where all connections $\ConnQui{n,m}$ are 
trivial and the Higgs fields are constant, the vacuum is solely determined by 
the vanishing locus of the scalar potential. The vanishing of the potential 
gives rise to holomorphic F-term constraints as well as non-holomorphic D-term 
constraints which read as
\begin{equation}
\left[X_a,X_b \right] = f_{ab}^{\ \ 5} \, X_5 + f_{ab}^{\ \ i} \,
\widehat{I}_i \and \left[X_a,X_5 \right] = f_{a5}^{\ \ b} \,
X_b \ ,
\label{eqn:BPSequations}\end{equation}
for $a,b=1,2,3,4$. The equivariance condition
\eqref{eqn:equivariance} implies that $X_\mu$ lie in a representation
of the $\suL$ Lie algebra. Hence the BPS configurations of the gauge
theory $X_\mu$, together with $\widehat{I}_i $, furnish a
representation of the Lie algebra $\sutL$ in the representation space of the quiver in
$\urmL(p)$. These constraints respectively give rise to a set of relations and a set of stability conditions for the 
corresponding quiver representation.
The details can be read off from the explicit expressions
in Appendix~\ref{sec:details_red_S5_appendix}.
% 
%%%%%%%%%%%%%%%%%%%%%%%%%%%%%%%%%%%%%%%%%%%%%%%%%%%%%%%%%%%%%%%%%%%%%%%%%%%%%%%%
  \bigskip \section{Spherically symmetric instantons}
\label{sec:SU(3)-equivariant_cone}
\noindent
In this section we specialise to the case where the Riemannian manifold $\mfd{d}=\mfd{1}$ is 
$1$-dimensional. We investigate the 
Hermitian Yang-Mills equations on the product $\mfd{1}\times S^5 / \ZK$ for the generic form of $\G$-equivariant connections 
derived in Section~\ref{subsec:construction_quiver}.
\subsection{Preliminaries}
Consider the product manifold $\mfd{1} \times S^5\slash \ZK$ with $\mfd{1}= \mathbb{R}$ such that 
$\mfd{1} \times S^5\slash \ZK \cong C(S^5\slash \ZK)$ is the metric cone over 
the Sasaki-Einstein space $S^5\slash \ZK$, which is an orbifold of the 
Calabi-Yau manifold $C(S^5)$.
The Calabi-Yau space $C(S^5)$ is conformally equivalent to the cylinder $\R\times S^5$ with the metric
\begin{equation}
 \diff s^2_{C(S^5)} = \diff r^2 + r^2 \, \diff s^2_{S^5} = r^2 \,
 \left( \diff \tau^2 + \diff s^2_{S^5} 
\right) = \e^{2 \tau} \, \left( \diff \tau^2  + \delta_{\mu \nu} \, e^\mu \otimes 
e^\nu 
\right)
\end{equation}
where $\tau=\log r$. The Kähler 2-form is given by 
\begin{equation}
   \omega_{C(S^5)} = \e^{2 \tau} \, \left( \omega_3 + \eta \wedge \diff \tau  
\right)\; .
\end{equation}

\paragraph{Connections}
As $\R$ is contractible, each bundle $E_{p_{(n,m)}} \to \R$ is necessarily trivial and hence one can gauge away the (global) connection $1$-forms $\ConnQui{n,m}=\ConnQui{n,m}(\tau) \, \diff \tau$; explicitly, there is a gauge transformation $g:\R\to\Gcal^{k,l}$ such that
\begin{equation}
 \tilde{A}_{(n,m)} = \mathrm{Ad}(g^{-1}) A_{(n,m)} + g^{-1} \, \frac{\mathrm{d} g}{\mathrm{d} 
\tau} =0 \with g = \exp\Big(-\int\, A_{(n,m)}(\tau) \, \diff 
\tau \Big) \; .
\end{equation}
The ansatz 
for the connection on the equivariant V-bundle then reads
\begin{equation}
 \mathcal{A}= \mathds{1}\otimes \Gamma + X_\mu\, e^\mu \; , 
\label{eqn:ansatz_HYM_cone}
\end{equation}
where the Higgs fields $\HomoPhiPMi{n,m}$ and $\EndPsiQui{n,m}$ depend only on the 
cone coordinate $\tau$ (compare also with~\cite[Section 4.1]{Bunk:2014coa}). 
The curvature of this connection can be read off from~\eqref{eqn:curv_quiver_matrix} and is evaluated to
\begin{equation}
\mathcal{F}= \frac{\diff X_\mu}{\diff\tau} \, \diff \tau \wedge e^\mu  + 
\frac{1}{2}\, \Big( \left[ X_\mu , X_\nu \right] - f_{\mu \nu}^{\ \ \rho}\, X_\rho -  f_{\mu \nu}^{\ \ i}\, \widehat{I}_i
\, \Big)\, e^{\mu \nu} \; .%
\label{eqn:curvature_C(S5)}
\end{equation}
\subsection{Generalised instanton equations}
\label{subsec:instanton_SU(3)}
The ansatz~\eqref{eqn:ansatz_HYM_cone} restricts the space of all 
connections on the $\sut$-equivariant vector V-bundle over $C(S^5\slash \ZK)$ to $\sut $-equivariant and 
$\ZK$-equivariant connections.
\paragraph{Quiver relations}
On this subspace of connections one can further restrict to holomorphic 
connections, i.e. 
connections which allow for a holomorphic structure.\footnote{For a Hermitian 
connection $\Acal$ on a complex vector bundle, the requirement for it to induce a 
holomorphic structure is equivalent to the $(0,1)$-part $\Acal^{0,1}$ of 
$\Acal$ being integrable, i.e. 
the corresponding curvature $\Fcal$ is of type~$(1,1)$.} For 
this, one 
requires the holomorphicity condition $\mathcal{F}^{0,2}= 0 = 
\mathcal{F}^{2,0}$ which for the connection~\eqref{eqn:ansatz_HYM_cone} is equivalent to
\begin{subequations}
  \label{eqn:holomorphicity_SU(3)}
  \begin{alignat}{3}
   \mathcal{F}_{14} + \mathcal{F}_{23} &=0 \; , & \qquad \mathcal{F}_{1\tau} + 
\mathcal{F}_{25} &=0 \; , & \qquad \mathcal{F}_{3\tau} + \mathcal{F}_{45} &=0 \; , 
\\[4pt]
   \mathcal{F}_{13} - \mathcal{F}_{24} &=0 \; , & \qquad \mathcal{F}_{15} - 
\mathcal{F}_{2\tau} &=0 \; , & \qquad \mathcal{F}_{35} - \mathcal{F}_{4\tau}  &=0 \; .
  \end{alignat}
\end{subequations}
Substituting the explicit components of the curvature~\eqref{eqn:curvature_C(S5)}, 
one finds relations for the endomorphisms $X_\mu$ given by
% 
%\begin{subequations}
\begin{align}\label{eqn:holo_matrices_SU(3)}
  \left[X_1 , X_4 \right] + \left[X_2 , X_3 \right] =0=
\left[X_1 , X_3 \right] - \left[X_2 , X_4 \right] \and
\left[X_a , X_5 \right] = f_{a5}^{\ \ b} \, \Big( \, X_b+ \frac{2}{3}
                          \, \frac{\diff 
X_b}{\diff\tau}\, \Big)
\end{align}
%\end{subequations}
for $a=1,2,3,4$.
\paragraph{Stability conditions}
By well-known theorems from algebraic geometry~\cite{Donaldson:1985,Uhlenbeck:1986,Donaldson:1987}, a 
holomorphic vector bundle admits solutions to the Hermitian Yang-Mills equations 
if and only if it is stable. This condition can be 
translated into a condition on the remaining $(1,1)$-part of the 
curvature $\mathcal{F}$: One demands that $\mathcal{F}$ is a 
primitive $(1,1)$-form, i.e. $\omega_{C(S^5)} \, \lrcorner \, \mathcal{F}=0$, or in components
\begin{equation}
 \mathcal{F}_{12} + \mathcal{F}_{34} + \mathcal{F}_{5\tau}=0 \; .
\end{equation}
Using the explicit components~\eqref{eqn:curvature_C(S5)} 
one can deduce the matrix differential equation for $X_\mu$ given by
\begin{equation}
\label{eqn:stab_matrices_SU(3)}
\left[X_1 , X_2 \right] + \left[X_3 , X_4 \right] = 4 X_5 + \frac{\diff X_5 }{\diff\tau}
\; .
\end{equation} 

One can also regard the stability condition in terms of a moment map 
$\mu$ from the space of holomorphic connections to the dual of the Lie algebra 
of the gauge group~\cite{Atiyah:1982}. The dual $\mu^*$ then acts on a 
connection $\Acal$ via $\mu^*(\Acal) = \omega_{C(S^5)} \, \lrcorner \, \Fcal$, 
which 
is well-defined as the curvature $\Fcal$ is a Lie algebra-valued $2$-form. Then the 
stability conditions correspond to the level set of zeroes
$\mu^{*-1}(0)$; we shall return to this interpretation in 
Section~\ref{subsec:su3-equi_case}.
% 
%%%%%%%%%%%%%%%%%%%%%%%%%%%%%%%%%%%%%%%%%%%%%%%%%%%%%%%%%%%%%%%%%%%%%%%%%%%%%%%
%
% 
\subsection{Examples}
We shall now apply these considerations to the three simplest 
examples: 
The quivers based on the representations $\RepSu{1}{0}$, $\RepSu{2}{0}$ and $\RepSu{1}{1}$. For 
each example we explicitly provide the representation of the generators and 
the form of the matrices $X_\mu$, followed by the quiver relations and the stability 
conditions.
\subsubsection{$\RepSuHead{1}{0} $-quiver}
\label{subsec:HYM_Cone_C(1,0)}
The generators in the fundamental representation $\RepSu{1}{0}$, which splits as 
in~\eqref{eqn:split_C(1,0)}, are given as
\begin{subequations}
\begin{equation}
 I_a = \begin{pmatrix} 0_2 & I_a^{(0,-2)} \\ 
-\big(I_a^{(0,-2)}\big)^\dag & 0 \end{pmatrix} \and
I_5 = \begin{pmatrix} I_5^{(1,1)} & 0 \\ 0 & 
I_5^{(0,-2)} \end{pmatrix} 
\end{equation}
for $a=1,2,3,4$, with components
\begin{align}
I_1^{(0,-2)} = \begin{pmatrix} 1\\0 \end{pmatrix} =\im\, I_2^{(0,-2)} \and &
I_3^{(0,-2)}= \begin{pmatrix} 0\\1 \end{pmatrix}  = \im \, I_4^{(0,-2)} \; , \\[4pt]
I_5^{(0,-2)} = \im \, \mathds{1}_{2} \and &
I_5^{(1,1)} = -\tfrac{\im}{2} \; .
\end{align}
\end{subequations}
The endomorphisms $X_\mu$ read as 
\begin{equation}
 X_a = \begin{pmatrix} 0_2 & \phi \otimes I_a^{(0,-2)} \\ 
-\phi^\dagger \otimes \big(I_a^{(0,-2)}\big)^\dag & 0 \end{pmatrix} \and
X_5 = \begin{pmatrix} \psi_1 \otimes I_5^{(1,1)} & 0 \\ 0 & 
\psi_0 
\otimes I_5^{(0,-2)} \end{pmatrix}
\end{equation}
where the Higgs fields from Appendix~\ref{sec:examples_quiver} give a representation of the quiver
\begin{equation}
 \xymatrix{*[F.]{(0,-2)} \ar@(ur,ul)[]_{\psi_0}  \ar[rr]^{\phi}  & 
 & *[F.]{(1,1)} \ar@(ur,ul)[]_{\psi_1}
 } 
\end{equation}
The $\ZK$-representation~\eqref{eqn:rep_ZK_quiver} reads
\begin{equation}
 \gamma : h \longmapsto \begin{pmatrix} \mathds{1}_{p_{(1,1)}} \otimes 
\mathds{1}_{2} \ \zk  & 0 \\ 0&  \mathds{1}_{p_{(0,-2)}} 
\otimes \zk^{-2}    \end{pmatrix} \; ,
\end{equation}
where $h$ is the generator of the cyclic group $\ZK$.
\paragraph{Quiver relations}
The first two equations from~\eqref{eqn:holo_matrices_SU(3)} are trivially satisfied without any further 
constraints. The second set of equations all have the 
same non-trivial off-diagonal component (and its adjoint) 
which yields
\begin{equation}
   2 \, \frac{\diff\phi }{\diff \tau}= - 3\, \phi +2 \, \phi \, \psi_0 + \psi_1 \, \phi 
 \; .
\label{eqn:C10-holomorphic}\end{equation}
Thus for the $\RepSuHead{1}{0} $-quiver there are no purely algebraic
quiver relations.
\paragraph{Stability conditions}
From~\eqref{eqn:stab_matrices_SU(3)} we read off the two non-trivial diagonal components which yield  
  \begin{equation}
   \frac{1}{4}\, \frac{\diff\EndoPsi{0}}{\diff \tau} =- \EndoPsi{0} +
\phi^\dagger \, \phi\and
    \frac{1}{4}\, \frac{\diff\EndoPsi{1}}{\diff \tau} =- \EndoPsi{1} +\phi\,
\phi^\dagger 
\; .
 \label{eqn:C10-stability} \end{equation}
By taking $\psi_0$ and $\psi_1$ to be identity endomorphisms, we
recover the Higgs branch BPS equations from equivariant dimensional reduction over
$\CP$: In this limit (\ref{eqn:C10-holomorphic}) implies that the
scalar field $\phi$ is independent of $\tau$, while (\ref{eqn:C10-stability})
correctly reproduces the D-term constraints of the quiver gauge theory
for constant matrices~\cite{Lechtenfeld:2008nh,Dolan:2009nz}.
%
% 
%%%%%%%%%%%%%%%%%%%%%%%%%%%%%%%%%%%%%%%%%%%%%%%%%%%%%%%%%%%%%%%%%%%%%%%%%%%%%%%%
%%%%%%%%%%%%%%%%%%%%%%%%%%%%%%%%%%%%%%%%%%%%%%%%%%%%%%%%%%%%%%%%%%%%%%%%%%%%%%%%
% 
\subsubsection{$\RepSuHead{2}{0} $-quiver}
\label{subsec:HYM_Cone_C(2,0)}
The generators in the 6-dimensional representation $\RepSu{2}{0} $, 
which splits as in~\eqref{eqn:split_C(2,0)}, are given by
\begin{subequations}
\begin{align}
 I_a = \begin{pmatrix} 0_{3} & \Geni{a}{(1,-1)} & 0 \\
 -\GeniAdj{a}{(1,-1)} & 0_{2} & \Geni{a}{(0,-4)} \\
 0 & -\GeniAdj{a}{(0,-4)} & 0  \end{pmatrix} \and
 I_5 = \begin{pmatrix} \Geni{5}{(2,2)} & 0 & 0 \\
 0 & \Geni{5}{(1,-1)} & 0 \\
 0 & 0 & \Geni{5}{(0,-4)}  \end{pmatrix} 
 \end{align}
for $a=1,2,3,4$, with components
\begin{alignat}{2}
  \Geni{1}{(1,-1)} &= \begin{pmatrix} \sqrt{2} & 0 \\ 0 & 1 \\ 0 & 0 
\end{pmatrix} = \im\, \Geni{2}{(1,-1)} \and & \Geni{1}{(0,-4)} &= \begin{pmatrix} \sqrt{2} \, \\ 0  \end{pmatrix} =\im\, \Geni{2}{(0,-4)} \; , \\[4pt]
  \Geni{3}{(1,-1)} &= \begin{pmatrix} 0 & 0 \\ 1 & 0 \\ 0 & \sqrt{2} 
\end{pmatrix} = \im\, \Geni{4}{(1,-1)} \and &
\Geni{3}{(0,-4)} &= \begin{pmatrix} 0 \\ \sqrt{2} \end{pmatrix} =\im\, \Geni{4}{(0,-4)} \; , \\[4pt]
\Geni{5}{(2,2)} &= -\im \, \mathds{1}_{3} \; , \quad 
\Geni{5}{(1,-1)} = 
\tfrac{\im}{2} \, \mathds{1}_{2} \and & \Geni{5}{(0,-4)} &= 2 \,
\im \; .
\end{alignat}
\end{subequations}
The endomorphisms $X_{\mu}$ read
\begin{equation}
\begin{split}
 X_a &= \begin{pmatrix} 0_{3} & \HomoPhii{1} \otimes 
\Geni{a}{(1,-1)} & 0 
\\
 - \HomoPhiAdji{1} \otimes \GeniAdj{a}{(1,-1)} & 0_{2} & 
\HomoPhii{0} 
\otimes \Geni{a}{(0,-4)} \\
0 & -\HomoPhiAdji{0} \otimes \GeniAdj{a}{(0,-4)} & 0
       \end{pmatrix} \; , \\[4pt]
       X_5 &= \begin{pmatrix} \EndoPsi{2} \otimes 
\Geni{5}{(2,2)} & 0 & 0 \\ 0 & \EndoPsi{1} \otimes 
\Geni{5}{(1,-1)} & 0 \\ 0 & 0 
& \EndoPsi{0} \otimes \Geni{5}{(0,-4)} \end{pmatrix}  \; ,
\end{split}
\end{equation}
with the Higgs field content from Appendix~\ref{sec:examples_quiver}
that furnishes a representation of the quiver
\begin{equation}
  \xymatrix{*[F.]{(0,-4)} \ar@(ur,ul)[]_{\EndoPsi{0}}  \ar[rr]^{\HomoPhii{0}} 
   &  &
  *[F.]{(1,-1)} \ar@(ur,ul)[]_{\EndoPsi{1}} \ar[rr]^{\HomoPhii{1}}  
 & &
 *[F.]{(2,2)} \ar@(ur,ul)[]_{\EndoPsi{2}}
 }
\end{equation}
The representation~\eqref{eqn:rep_ZK_quiver} in this case reads
\begin{equation}
 \gamma : h \longmapsto \begin{pmatrix} \, \mathds{1}_{p_{(2,2)}} \otimes 
\mathds{1}_{3} \ \zk^2  & 0 &0 \\
0& \mathds{1}_{p_{(1,-1)}} \otimes 
\mathds{1}_{2} \ \zk^{-1}  & 0 \\
0& 0&  \mathds{1}_{p_{(0,-4)}} 
\otimes \zk^{-4} \, \end{pmatrix} \; .
\end{equation}
\paragraph{Quiver relations}
Again the first two equations of~\eqref{eqn:holo_matrices_SU(3)} turn out to be trivial, while the second set of equations
have two non-vanishing off-diagonal components (plus their conjugates) which yield
  \begin{equation}
       2\, \frac{\diff\HomoPhii{0}}{\diff \tau} = - 3 \, \HomoPhii{0} 
   - \EndoPsi{1} \, \HomoPhii{0} + 4\, \HomoPhii{0} \, \EndoPsi{0}
 \and 2 \, \frac{\diff\HomoPhii{1}}{\diff  \tau} = - 3 \, \HomoPhii{1}
      + \HomoPhii{1} \, \EndoPsi{1} + 2\, \EndoPsi{2} \, \HomoPhii{1}
\label{eqn:C20-holomorphic}  \end{equation}
and the $C^{2,0}$-quiver has no purely algebraic quiver relations 
either.
\paragraph{Stability conditions}
From~\eqref{eqn:stab_matrices_SU(3)} one obtains three non-trivial diagonal components that yield
\begin{subequations}  
  \begin{align}
    \frac{1}{4} \,\frac{\diff\EndoPsi{0}}{\diff \tau} &= - \EndoPsi{0} 
     +\HomoPhiAdji{0} \, \HomoPhii{0} \; ,
\\[4pt]
   \frac{1}{4}\,  \frac{\diff\EndoPsi{1}}{\diff \tau} &= - 
\EndoPsi{1} - 2 \, \HomoPhii{0} \, \HomoPhiAdji{0} + 3 \, \HomoPhiAdji{1} \,
\HomoPhii{1} \; ,
  \\[4pt]
 \frac{1}{4} \, \frac{\diff\EndoPsi{2}}{\diff \tau} &= - \EndoPsi{2}
   + \HomoPhii{1} \, \HomoPhiAdji{1} \; .
  \end{align}
\label{eqn:C20-stability}\end{subequations}
Taking $\psi_0$, $\psi_1$ and $\psi_2$ again to be identity morphisms,
from (\ref{eqn:C20-holomorphic}) we obtain constant matrices $\phi_0$
and $\phi_1$ which by (\ref{eqn:C20-stability}) obey the expected
D-term constraints from equivariant dimensional reduction over
$\CP$~\cite{Lechtenfeld:2008nh,Dolan:2009nz}.
% 
%%%%%%%%%%%%%%%%%%%%%%%%%%%%%%%%%%%%%%%%%%%%%%%%%%%%%%%%%%%%%%%%%%%%%%%%%%%%%%%%
%%%%%%%%%%%%%%%%%%%%%%%%%%%%%%%%%%%%%%%%%%%%%%%%%%%%%%%%%%%%%%%%%%%%%%%%%%%%%%%%
% 
\subsubsection{$\RepSuHead{1}{1} $-quiver}
\label{subsec:HYM_Cone_C(1,1)}
The decomposition of the adjoint representation $\RepSu{1}{1}$, which splits as given 
in~\eqref{eqn:split_C(1,1)}, yields
\begin{subequations}
\begin{align}
 I_a &= \begin{pmatrix} 0_{2} & \Geni{a}{(0,0)} & 
\Geni{a}{(2,0)} & 0 
\\
-\GeniAdj{a}{(0,0)} & 0 & 0 & \Geni{a}{-\,(1,-3)} \\
-\GeniAdj{a}{(2,0)} & 0 & 0_{3} & \Geni{a}{+\,(1,-3)} \\
0 & -\GeniAdj{a}{-\,(1,-3)} & - \GeniAdj{a}{+\,(1,-3)} & 0_{2} 
\end{pmatrix} \; 
, \\[4pt]
I_5 &= 
\begin{pmatrix} \Geni{5}{(1,3)} & 0 & 0 & 0 \\
0 & \Geni{5}{(0,0)} & 0 & 0 \\
0 & 0 & \Geni{5}{(2,0)} & 0 \\
0 & 0 & 0 & \Geni{5}{(1,-3)} \end{pmatrix}
\end{align}
for $a=1,2,3,4$, with components
\begin{align}
\Geni{1}{(0,0)} = \begin{pmatrix} \sqrt{\frac{3}{2}}\, \\ 0   
\end{pmatrix} = \im\,\Geni{2}{(0,0)} \and &
\Geni{1}{(2,0)} = \begin{pmatrix} 0 & 
-\sqrt{\frac{1}{2}} & 0 \\ 0 & 0 & -1 \end{pmatrix} =\im\,\Geni{2}{(2,0)}  \; , \\[4pt] 
% % 
% % 
\Geni{1}{-\,(1,-3)} = \begin{pmatrix} 0 & 
-\sqrt{\frac{3}{2}}\, \end{pmatrix} =\im\, \Geni{2}{-\,(1,-3)}  \and &
\Geni{1}{+\,(1,-3)} =  \begin{pmatrix}
1& 0  \\ 0 & \sqrt{\frac{1}{2}} \\ 0 & 0 \end{pmatrix} =\im\,\Geni{2}{+\,(1,-3)} \; , \\[4pt]
% %
% % 
\Geni{3}{(0,0)} = \begin{pmatrix}0 \\ \sqrt{\frac{3}{2}} 
\end{pmatrix} = \im\, \Geni{4}{(0,0)} \and &
\Geni{3}{(2,0)} = \begin{pmatrix} 1 & 0 & 0 \\ 0 & 
\sqrt{\frac{1}{2}} & 0 \end{pmatrix} = \im\, \Geni{4}{(2,0)} \; , \\[4pt]
% % 
% % 
\Geni{3}{-\,(1,-3)} = \begin{pmatrix} 
\sqrt{\frac{3}{2}} & 0 \end{pmatrix} =\im\, \Geni{4}{-\,(1,-3)} \and &
\Geni{3}{+\,(1,-3)} =  \begin{pmatrix}
0 & 0  \\ \sqrt{\frac{1}{2}} & 0 \\ 0 & 1 \end{pmatrix} = \im\,\Geni{4}{+\,(1,-3)} \; , \\[4pt]
% % 
% % 
\Geni{5}{(1,3)} = - \tfrac{3 \, \im}{2} \, \mathds{1}_{2} \; , \quad 
\Geni{5}{(0,0)} = 0 \; , \quad
% % 
% % 
\Geni{5}{(2,0)} = 0_{3} \and &
\Geni{5}{(1,-3)} = \tfrac{3 \, \im}{2} \, \mathds{1}_{2} \; .
\end{align}
\end{subequations}
The matrices $X_\mu$ are given by 
\begin{subequations}
\begin{align}
 X_a &= \begin{pmatrix} 0_{2} & \phi^+_0
\otimes \Geni{a}{(0,0)} & \phi_0^- \otimes 
\Geni{a}{(2,0)}
& 0 \\
-(\phi_0^+)^\dag \otimes \GeniAdj{a}{(0,0)} & 0 & 0 & 
\phi_1^- \otimes \Geni{a}{-\,(1,-3)} \\
- (\phi_0^-)^\dag \otimes \GeniAdj{a}{(2,0)} & 0 & 0_{3} 
& 
\phi_1^+ \otimes \Geni{a}{+\,(1,-3)} \\
0 & - (\phi_1^-)^\dag \otimes \GeniAdj{a}{-\,(1,-3)} & - 
(\phi_1^+)^\dag 
\otimes \GeniAdj{a}{+\,(1,-3)} & 0_{2} \end{pmatrix} \; , \\[4pt]
X_5 &= 
\begin{pmatrix} \psi^+ \otimes \Geni{5}{(1,3)} & 0 & 0 & 0 
\\
0 & 0 & 0 & 0 \\
0 & 0 & 0_{3} & 0 \\
0 & 0 & 0 & \psi^- \otimes \Geni{5}{(1,-3)} \end{pmatrix} 
\; .
\end{align}
\end{subequations}
This example involves the collection of Higgs fields from Appendix~\ref{sec:examples_quiver}
which furnish a representation of the quiver
\begin{equation}
 \xymatrix{ &   *[F.]{(1,+3)}  \ar@(ur,ul)[]_{\psi^+}
   &  &\\
  *[F.]{(0,0)}  \ar@/^/[ur]^{\phi_0^+}
&  & 
*[F.]{(2,0)}  \ar@/_/[ul]_{\phi_0^-}  & \\
 & *[F.]{(1,-3)} \ar@/^/[ul]^{\phi_1^-} 
\ar@/_/[ur]_{\phi_1^+} 
\ar@(dr,dl)[]^{\psi^-}
&  &
 } 
\end{equation}
In this case the $\ZK$-representation~\eqref{eqn:rep_ZK_quiver} has the form
\begin{equation}
  \gamma : h \longmapsto \begin{pmatrix} \mathds{1}_{p_{(1,3)}} \otimes 
\mathds{1}_{2} \ \zk^3  & 0 &0 & 0  \\
0& \mathds{1}_{p_{(0,0)}} \otimes 1 & 0 & 0 \\
0& 0&  \mathds{1}_{p_{(2,0)}} \otimes \mathds{1}_{3}  &0 \\
0& 0& 0& \mathds{1}_{p_{(1,-3)}} \otimes 
\mathds{1}_{2} \ \zk^{-3 }
\end{pmatrix} \; .
\end{equation}
\paragraph{Quiver relations}
For this $8$-dimensional example, one finds that the first two equations of~\eqref{eqn:holo_matrices_SU(3)} have the same single non-trivial off-diagonal component (plus its adjoint) which yields 
\begin{equation}
\phi_0^+\, \phi_1^- = \phi_0^-\, \phi_1^+ \; .
\end{equation}
This equation is precisely the anticipated algebraic relation for the
$C^{1,1}$-quiver expressing equality of paths between the vertices $(1,\pm\,3)$, cf.~\cite{Lechtenfeld:2008nh}. 
The second set of equations have four non-trivial off-diagonal components (plus their conjugates) which yield
\begin{equation}
\frac{2}{3} \, \frac{\diff\phi_0^\pm}{\diff \tau} = - \phi_0^\pm + \psi^+\, \phi_0^\pm \and
\frac{2}{3} \, \frac{\diff\phi_1^\pm}{\diff \tau} = - \phi_1^\pm + \phi_1^\pm\, \psi^- \ .
\end{equation}
\paragraph{Stability conditions}
From~\eqref{eqn:stab_matrices_SU(3)} one 
computes four non-vanishing diagonal components that yield
\begin{subequations}
\begin{align}
(\phi_0^\pm)^\dag\, \phi_0^\pm &= \phi_1^\mp\,(\phi_1^\mp)^\dag \ , \\[4pt]
\frac{1}{4} \, \frac{\diff\psi^+}{\diff \tau} &= - \psi^+
 + \frac{1}{2} \,\left( \phi_0^+\, (\phi_0^+)^\dag +\phi_0^-\,(\phi_0^-)^\dag \right) \ , \\[4pt]
\frac{1}{4} \, \frac{\diff\psi^-}{\diff \tau} &= - \psi^-
 + \frac{1}{2} \,\left( (\phi_1^-)^\dag\, \phi_1^-+(\phi_1^+)^\dag\, \phi_1^+ \right) \ .
\end{align}
\end{subequations}
We thus obtain two non-holomorphic purely algebraic conditions, which coincide with D-term constraints of the quiver gauge theory for the $C^{1,1}$-quiver, and two 
further differential equations which for identity endomorphisms
$\psi^\pm$ reproduce the remaining stability equations for constant
matrices $\phi_0^\pm$ and $\phi_1^\pm$ in equivariant dimensional reduction over $\CP$~\cite{Lechtenfeld:2008nh,Dolan:2009nz}.
%%%%%%%%%%%%%%%%%%%%%%%%%%%%%%%%%%%%%%%%%%%%%%%%%%%%%%%%%%%%%%%%%%%%%%%%%%%%%%%%
 \bigskip \section{Translationally-invariant instantons}
\label{sec:TransInv_Instantons}
\noindent
In this section we study translationally-invariant instantons on a trivial 
vector V-bundle over the orbifold  $\C^3 /\ZK$. In contrast to the $\G$-equivariant Hermitian Yang-Mills 
instantons of Section~\ref{sec:SU(3)-equivariant_cone}, the 
generic form of a translationally-invariant connection is determined by 
$\ZK$-equivariance alone and is associated with a different quiver.
\subsection{Preliminaries}
Consider the cone $C(S^5)/\ZK \cong \C^3 /\ZK$, with the $\ZK$-action 
given by~\eqref{eqn:ZK-action}, and the (trivial) vector V-bundle 
\begin{equation}
 \mathfrak{E}^{k,l} \xrightarrow{ \ V^{k,l} \ } \C^3 / \ZK 
\label{eqn:bundle_TransInv}
\end{equation}
of rank $p$ which is obtained by suitable $\ZK$-projection from the trivial vector bundle 
$\C^3 \times V^{k,l} \to \C^3 $. The fibres of~\eqref{eqn:bundle_TransInv} 
can be regarded as representation spaces
\begin{equation}
 V^{k,l} = \bigoplus_{(n,m)\in Q_0(k,l)} \, \C^{p_{(n,m)}}\otimes \Reps{(n,m)} \cong \bigoplus_{(n,m)\in Q_0(k,l)} \,
\big(\C^{p_{(n,m)}} \otimes\C^{n+1}\big) \otimes V_{m} \; . \label{eqn:fibres_trans-inv}
\end{equation}
Here $V_m$ is the $[m]$-th irreducible representation
$\rho_{[m]}$ of $\ZK$ (cf.~\eqref{eqn:irred_reps_ZK}), with $[m]\in\{0,1,\dots,q\}$ the congruence class of $m\in\Z$ modulo $q+1$, and the vector space $\C^{p_{(n,m)}}\otimes \C^{n+1}$ serves 
as the multiplicity space of this representation. The structure group of the bundle $\mathfrak{E}^{k,l}$ is
\begin{equation}
\mathfrak{G}^{k,l}:= 
\prod_{(n,m)\in Q_0(k,l)}\, \urm\big(p_{(n,m)}\, (n+1) \big) \ ,
\label{eqn:gauge_group_transinv}\end{equation}
because the fibres are isomorphic 
to~\eqref{eqn:fibres_trans-inv} and hence it carries a natural complex structure $J$; this 
complex structure is simply multiplication with $\im$ on each factor $V_m$.
Consequently, the structure group is reduced to the stabiliser of $J$. 

On the base the canonical Kähler form of $\C^3$ is given by
\begin{equation}
 \omega_{\C^3} = \tfrac{\im}{2} \, \delta_{\alpha\beta} \, \diff z^\alpha \wedge \diff 
\bar{z}^{{\beta}} \; .
\end{equation}
This Kähler form is compatible with the standard metric $\diff s^2_{\C^3}= \tfrac{1}{2} \,
\delta_{\alpha\beta} \, (\diff z^\alpha \otimes \diff \bar{z}^{{\beta}} + \diff 
\bar{z}^{{\alpha}} \otimes \diff z^\beta)$ and the complex structure $ J 
(\diff z^\alpha) = \im \, \diff z^\alpha$, $J (\diff \bar{z}^{{\alpha}}) = - 
\im \,
\diff 
\bar{z}^{{\alpha}}$.
\paragraph{Connections}
Consider a connection $1$-form
\begin{equation}
 \mathcal{A}= W_\alpha \, \diff z^\alpha + 
 \overline{W}_{{\alpha}} \, \diff 
\bar{z}^{{\alpha}} \label{eqn:ansatz_trans_inv}
\end{equation}
on $\mathfrak{E}^{k,l}$, and impose translational invariance along the space 
$\C^3$. For the coordinate basis $\{ \diff z^\alpha, \diff 
\bar{z}^{{\alpha}} \}$ of $T_{(z,\bar z)}^* \C^3$ at any point $(z,\bar z)\in\C^3$, this 
translates into the condition 
\begin{equation}
 \diff W_\alpha =0 = \diff \overline{W}_{{\alpha}} \for \alpha=1,2,3
\; . \label{eqn:trans_inv_condition_matrices}
\end{equation}
Thus the curvature $\mathcal{F}=\diff \mathcal{A} + \mathcal{A} \wedge 
\mathcal{A}$ simplifies to
\begin{equation}
 \mathcal{F} =  \mathcal{A} \wedge \mathcal{A} = 
 \tfrac{1}{2}\, \big[W_\alpha, W_\beta \big] \, \diff z^\alpha \wedge \diff z^\beta
+ \big[W_\alpha , \overline{W}_{{\beta}} \big] \, \diff z^\alpha \wedge \diff 
\bar{z}^{{\beta}}
+\tfrac{1}{2}\, \big[\, \overline{W}_{{\alpha}}, \overline{W}_{{\beta}} \big] \, \diff 
\bar{z}^{{\alpha}} \wedge \diff \bar{z}^{{\beta}} \; .
\end{equation}

\paragraph{$\ZK$-action}
As before one demands $\ZK$-invariance due to the projection from the trivial 
vector bundle $\C^3 \times V^{k,l} \to \C^3$ to the 
trivial V-bundle $\mathfrak{E}^{k,l} \to \C^3/\ZK$. Again
one needs to choose a representation of $\ZK$ on the 
fibres~\eqref{eqn:fibres_trans-inv}. For reasons that will become 
clear later on (see Section~\ref{subsec:quiver_graphs}), this time one chooses
\begin{equation}
 \gamma(h) = \bigoplus_{(n,m)\in Q_0(k,l)}\, \mathds{1}_{p_{(n,m)}} \otimes \zk^n \, \mathds{1}_{n+1} \ \in \
\mathrm{Center}\big( \mathfrak{G}^{k,l} \big) \; .%
 \label{eqn:ZK-action_trans-inv}
\end{equation}
One immediately sees that all elements of $\mathfrak{G}^{k,l}$ commute 
with the action of $\ZK$ given by~\eqref{eqn:ZK-action_trans-inv}, i.e. 
$\gamma(\ZK) \subset \mathrm{Center}\left( \mathfrak{G}^{k,l}\right)$.
The action of $\ZK$ on the coordinates $z^\alpha$ defined 
in~\eqref{eqn:ZK-action} induces a representation $\pi$ of $\ZK$ in $\Omega^1(\C^3)$, which on the generator 
$h$ of $\ZK$ is given by
\begin{equation}
 \pi(h)( W_\alpha) = \left\{ \begin{matrix} \zk^{-1}\, W_i \ , & 
i =1,2 \\[4pt] \zk^{2} \, W_3 &    \end{matrix} \right. 
\and 
\pi(h)(\, \overline{W}_{{\alpha}}) = \left\{ \begin{matrix} \zk\, 
W_{{i}} \ , & i=1,2 \\[4pt] \zk^{-2} \, W_{{3}} &    
\end{matrix} 
\right. \; .
\end{equation}
The requirement of $\ZK$-equivariance of the connection $\mathcal{A}$ reduces to conditions 
similar to~\eqref{eqn:ZK-equivariance_quiver_conn}, i.e. the equivariance conditions 
read 
as 
 \begin{align}
  \gamma(h) \, W_\alpha \, \gamma(h)^{-1}  = \pi(h)^{-1}
( W_\alpha ) \and \gamma(h)\, \overline{W}_{{\alpha}} \, \gamma(h)^{-1} = 
\pi(h)^{-1}(\, \overline{W}_{{\alpha}}) 
 \end{align}
for $\alpha =1,2,3$, but this time with
different $\ZK$-actions $\gamma$ and $\pi$. 
\paragraph{Quiver representations} 
For a decomposition of the endomorphisms
\begin{equation}
 W_\alpha = \bigoplus_{(n,m),(n',m')}\, ( W_\alpha 
)_{(n,m),(n',m')} \ , \quad ( W_\alpha 
)_{(n,m),(n',m')} \in \Hom\big(\C^{p_{(n,m)}}\otimes \Reps{(n,m)}
\,,\, \C^{p_{(n',m')}}\otimes \Reps{(n',m'\,)} \, \big)
\end{equation}
as before, the equivariance conditions imply that the allowed non-vanishing 
components are given by
\begin{subequations}
\label{eqn:form_W_TransInv}
\begin{align}
\Phi^i_{(n,m)}:= \left( W_i  \right)_{(n,m),(n',m')} 
&\neq 0  \for  n' -n ={ 1  {\pmod {q+1}}} \; , \label{eqn:form_Wi_TransInv} \\[4pt]
\Psi_{(n,m)}:=\left( W_3  \right)_{(n,m),(n',m')} &\neq 0  
\for  n' -n ={ -2  {\pmod {q+1}}} \; , \label{eqn:form_W3_TransInv}
\end{align}
\end{subequations}
for $i=1,2$, together with the analogous conjugate decomposition for
$\overline{W}_\alpha$; in each instance $m'$ is implicitly determined by $n$ and $m$
via the requirement $(n',m'\,)\in Q_0(k,l)$. The structure of these
endomorphisms thus determines a representation of another quiver
$\mathfrak{Q}^{k,l}$ with the same vertex set $Q_0(k,l)$ as before for
the quiver $\Qcal^{k,l}$ but with new arrow set consisting of allowed components $(n,m)\to (n',m'\,)$.
\subsection{Generalised instanton equations}
Similarly to Section~\ref{subsec:instanton_SU(3)}, the Hermitian Yang-Mills equations on the 
complex $3$-space $\C^3\slash \ZK$ can be regarded in terms of holomorphicity and stability conditions.
\paragraph{Quiver relations} 
The condition that the connection $\Acal$ defines an integrable holomorphic 
structure on the bundle~\eqref{eqn:bundle_TransInv} is, as 
before, equivalent to the vanishing of the $(2,0)$- and $(0,2)$-parts
of the curvature $\Fcal$, i.e. $\Fcal^{0,2}=0=\Fcal^{2,0}$, which in
the present case is equivalent to
  \begin{equation}
\big[W_\alpha, W_\beta \big] =0  \and \big[\,\overline{W}_{{\alpha}}, 
\overline{W}_{{\beta}} \big] =0 \; . \label{eqn:gen_instanton_1}
  \end{equation}
The general solutions~\eqref{eqn:form_W_TransInv}
to the equivariance conditions allow for a decomposition of the
generalised instanton equations~\eqref{eqn:gen_instanton_1} into
components given by
\begin{subequations}
  \begin{align}
   ( W_1)_{(n,m),(n+1,m')}  \,
   ( W_2)_{(n-1,m''),(n,m)} &= ( W_2)_{(n,m),(n+1,m')}\, 
  ( W_1)_{(n-1,m''),(n,m)} \; , \label{eqn:holo_W1W2} \\[4pt]
  ( W_i)_{(n,m),(n+1,m')}  \,
   ( W_3)_{(n+2,m''),(n,m)} &=0= ( W_3)_{(n,m),(n-2,m')}\,
  ( W_i)_{(n-1,m''),(n,m)}  
\ , \label{eqn:holo_WiW3}
  \end{align}
  \end{subequations}
for $(n,m)\in Q_0(k,l)$ and $i=1,2$, together with their conjugate equations. Note that in~\eqref{eqn:holo_W1W2} both
combinations are
morphisms between the same representation spaces and
hence the commutation relation $[W_1,W_2]=0$ requires only that their
difference vanish. On the other hand, in~\eqref{eqn:holo_WiW3} the two terms are
morphisms between different 
spaces and so the relation $[W_i,W_3]=0$ implies that they each vanish
individually; in particular, in the generic case the solution has $W_3=0$. 
\paragraph{Stability conditions} 
For invariant connections there is a peculiarity involved in
formulating stability of a holomorphic vector bundle, see for example~\cite{SardoInfirri:1996ga}. On a 
$2n$-dimensional Kähler manifold with Kähler form $\omega$, the stability 
condition is usually formulated through the identity
\begin{equation}
\mathcal{F} \wedge \omega^{n-1} = (\omega \,
\lrcorner \, \mathcal{F}) \, \omega^{n}
\label{eqn:stab_identity}\end{equation}
by demanding that $\omega \,
\lrcorner \, \mathcal{F} \in \mathrm{Center} (\mathfrak{g})$,
where $\mathfrak{g}$ is the Lie algebra of the structure group. For generic 
connections the center of $\mathfrak{g}$ is trivial and the 
usual stability condition $\omega \, \lrcorner \, \mathcal{F}=0$ follows. 
However, for invariant connections the structure group is smaller and the center
can be non-trivial. This implies that there are \emph{several} moduli spaces 
of translationally-invariant (and $\ZK$-equivariant) instantons depending on 
a choice of element in $\mathrm{Center} (\mathfrak{g})$.

Analogously to Section~\ref{subsec:instanton_SU(3)}, the stability condition is 
associated to the moment map on the space of translationally-invariant and 
$\ZK$-equivariant connections as we elaborate on in
Section~\ref{subsec:trans-inv_case}. In this case one can use any
gauge-invariant element 
\begin{equation}
\Xi \coloneqq \bigoplus_{(n,m)\in Q_0(k,l)}\, \mathds{1}_{p_{(n,m)}} \otimes \im\, \xi_{(n,m)} \,
\mathds{1}_{n+1} \ \in \ \mathrm{Center}\big(\mathfrak{g}^{k,l} \big)
\label{eqn:levels}\end{equation}
from the 
center of the Lie algebra
\begin{equation}
\mathfrak{g}^{k,l} := \bigoplus_{(n,m)\in Q_0(k,l)}\, \urmL\big(p_{(n,m)}\, (n+1)\big) \ ,
\label{eqn:gauge_alg_transinv}\end{equation}
where $\xi_{(n,m)} \in\R$ are called \emph{Fayet-Iliopoulos parameters}. Thus the 
remaining instanton equations $\omega_{\C^3} \, \lrcorner \, \mathcal{F}=
-\im \, \Xi$ read
\begin{equation}
\big[W_1 
, \overline{W}_{{1}} \big] + \big[W_2 , \overline{W}_{{2}} \big] + \big[W_3 , 
\overline{W}_{{3}} \big] = -\im \, \Xi  \; . \label{eqn:gen_instanton_2}
\end{equation}
Again by substituting the general solutions~\eqref{eqn:form_Wi_TransInv} 
and~\eqref{eqn:form_W3_TransInv} to the equivariance conditions we can
decompose the generalised instanton
equation~\eqref{eqn:gen_instanton_2} explicitly into component equations
\begin{align}
&\sum_{i=1}^2 \, \Big(
( W_i)_{(n,m),(n+1,m')}  \,
   ( \, \overline{W}_{{i}})_{(n+1,m'),(n,m)}
  - ( \, \overline{W}_{{i}})_{(n,m),(n-1,m')} \,
  ( W_i)_{(n-1,m'),(n,m)} 
\Big) \notag\\*
& \qquad \qquad \qquad + ( W_3 
)_{(n,m),(n-2,m')}  \,
   ( \, \overline{W}_{{3}})_{(n-2,m'),(n,m)}
  - ( \, \overline{W}_{{3}})_{(n,m),(n+2,m')} \,
  ( W_3)_{(n+2,m'),(n,m)} \notag \\* 
& \qquad \qquad \qquad \qquad \qquad \qquad \ = \ \mathds{1}_{p_{(n,m)}}\otimes \mathds{1}_{n+1} \ \xi_{(n,m)} 
\end{align}
for $(n,m)\in Q_0(k,l)$.
\subsection{Examples}
We shall now elucidate this general construction for the three examples  
$\RepSu{1}{0}$, $\RepSu{2}{0}$ and $\RepSu{1}{1}$. In each case we highlight 
the non-vanishing components of the matrices $W_\alpha$ and the 
representation~\eqref{eqn:ZK-action_trans-inv}.
\subsubsection{$\RepSuHead{1}{0}$-quiver}
The decomposition of the fundamental representation $\RepSu{1}{0}$ into irreducible 
$\su$-representations is given by~\eqref{eqn:split_C(1,0)}. The non-vanishing 
components can be read off to be $( W_i)_{(0,-2),(1,1)}$ and their
adjoints $(\, \overline{W}_{{i}})_{(1,1),(0,-2)}$. Thus there are 
two complex Higgs fields 
\begin{equation}
 \Phi_i \coloneqq \left( W_i \right)_{(0,-2),(1,1)}   \for i = 1,2 \; ,
\end{equation}
which determine a representation of the $2$-Kronecker quiver
\begin{equation}
 \xymatrix{*[F.]{(0,-2)} \ar@/^/[rr]^{\Phi_1} \ar@/_/[rr]_{\Phi_2} &  & 
*[F.]{(1,1)} } \label{eqn:trans_inv_quiver-graph_C(1,0)}
\end{equation}
By~\eqref{eqn:ZK-action_trans-inv} the representation of the 
generator $h$ is given by
\begin{equation}
 \gamma : h \longmapsto \begin{pmatrix} \mathds{1}_{p_{(1,1)}}\otimes \mathds{1}_{2} \ \zk  & 0 \\
                     0 & \mathds{1}_{p_{(0,-2)}}\otimes 1 \end{pmatrix} \; .
\end{equation}
\paragraph{Quiver relations}
The mutual commutativity of the matrices $W_\alpha$ is trivial in this case, and thus there are no quiver relations among the arrows of~\eqref{eqn:trans_inv_quiver-graph_C(1,0)}.
\paragraph{Stability conditions}
Choosing Fayet-Iliopoulos parameters $\xi_0,\xi_1\in\R$, the requirement of a stable quiver bundle yields non-holomorphic 
matrix equations given by
% 
%\begin{subequations}
\begin{align}
\Phi_1 \, \Phi_1^\dagger + \Phi_2 \, \Phi_2^\dagger = \mathds{1}_{p_{(1,1)}}\otimes \xi_0 \and
\Phi_1^\dagger \, \Phi_1 + \Phi_2^\dagger \, \Phi_2 &= \mathds{1}_{p_{(0,-2)}}\otimes 
\mathds{1}_{2} \ \xi_1\; .
\end{align}
%\end{subequations}
% 
% 
\subsubsection{$\RepSuHead{2}{0}$-quiver}
The representation $\RepSu{2}{0}$ is decomposed according 
to~\eqref{eqn:split_C(2,0)}. The non-vanishing components can be determined as 
before to be 
$\left( W_i\right)_{(0,-4),(1,-1)}$, 
$\left( W_i\right)_{(1,-1),(2,2)}$ and 
$\left( W_3 \right)_{(2,2),(0,-4)}$, together with their adjoints
$( \, \overline{W}_{{i}} )_{(1,-1),(0,-4)}$, 
$( \, \overline{W}_{{i}} )_{(2,2),(1,-1)}$ and
$( \,\overline{W}_{{3}})_{(0,-4),(2,2)}$. Thus there are five complex Higgs fields
\begin{subequations}
\begin{align}
 \Phi_i \coloneqq \left( W_i \right)_{(0,-4),(1,-1)} \; ,\quad
\Phi_{i+2} \coloneqq \left( W_i \right)_{(1,-1),(2,2)} \and
\Psi &\coloneqq \left( W_3 \right)_{(2,2),(0,-4)} \; ,
\end{align}
\end{subequations}
for $i=1,2$, 
which can be encoded in a representation of the quiver
\begin{equation}
 \xymatrix{ *[F.]{(0,-4)}  \ar@/^/[rr]^{\Phi_1} \ar@/_/[rr]_{\Phi_2} 
  &  & *[F.]{(1,-1)}  \ar@/^/[rr]^{\Phi_3} 
\ar@/_/[rr]_{\Phi_4} & &*[F.]{(2,2)} \ar@/_2pc/[llll]_{\Psi}
} \label{eqn:trans_inv_quiver-graph_C(2,0)}
\end{equation}
As before the representation~\eqref{eqn:ZK-action_trans-inv} for this 
example is
\begin{equation}
  \gamma : h \longmapsto \begin{pmatrix} \mathds{1}_{p_{(2,2)}}\otimes \mathds{1}_{3} \ \zk^2 &  0 & 0 \\
   0 & \mathds{1}_{p_{(1,-1)}}\otimes \mathds{1}_{2} \ \zk &  0  \\
                    0& 0 & \mathds{1}_{p_{(0,-4)}}\otimes 1 \end{pmatrix} \; .
\end{equation}
\paragraph{Quiver relations}
The holomorphicity condition yields
 \begin{align}
\Phi_i \, \Psi   =0 \ , \quad \Psi \, \Phi_{i+2}  =0 \and
\Phi_3 \, \Phi_2 = \Phi_4 \, \Phi_1 \label{eqn:holo_W_C(2,0)}
 \end{align}
for $i=1,2$, plus the conjugate equations. The first two sets of
quiver relations of~\eqref{eqn:holo_W_C(2,0)} each describe the vanishing of a 
path of the quiver~\eqref{eqn:trans_inv_quiver-graph_C(2,0)}; an obvious trivial solution of these equations is $\Psi=0$. The last 
relation expresses equality of two paths with source vertex $(0,-4)$ and target vertex $(2,2)$. 
\paragraph{Stability conditions}
Choosing Fayet-Iliopoulos parameters $\xi_0,\xi_1,\xi_2\in\R$, the stability conditions yield
\begin{subequations}
\begin{align}
\Phi_1^\dagger \, \Phi_1 + 
\Phi_2^\dagger \, \Phi_2 - \Psi \, \Psi^\dagger&= \mathds{1}_{p_{(0,-4)}}\otimes \xi_0 \; 
, \\[4pt]
\Phi_1 \, \Phi_1^\dagger + \Phi_2 \, \Phi_2^\dagger - 
\Phi_{3}^\dagger \, \Phi_{3} - \Phi_{4}^\dagger \, \Phi_{4}  &= \mathds{1}_{p_{(1,-1)}}\otimes \mathds{1}_{2} \ \xi_1 \; ,\\[4pt]
\Phi_3 \, \Phi_3^\dagger +\Phi_4 \, \Phi_4^\dagger - 
\Psi^\dagger \, \Psi &= \mathds{1}_{p_{(2,2)}}\otimes \mathds{1}_{3} \ \xi_2 \; .
\end{align}
\end{subequations}
\subsubsection{$\RepSuHead{1}{1}$-quiver}
The decomposition of the adjoint representation $\RepSu{1}{1}$ is given 
by~\eqref{eqn:split_C(1,1)}. The non-vanishing components are 
$\left( W_i \right)_{(0,0),(1,3)}$, 
$\left( W_i \right)_{(0,0),(1,-3)}$, 
$\left( W_i \right)_{(1,3),(2,0)}$, 
$\left( W_i \right)_{(1,-3),(2,0)}$ and
$\left( W_3 \right)_{(2,0),(0,0)}$, together with their adjoint maps $(\, \overline{W}_{{i}} )_{(1,3),(0,0)}$, 
$( \, \overline{W}_{{i}})_{(1,-3),(0,0)}$, 
$(\, \overline{W}_{{i}} )_{(2,0),(1,3)}$, 
$( \, \overline{W}_{{i}} )_{(2,0),(1,-3)}$ and 
$(\, \overline{W}_{{3}} )_{(0,0),(2,0)}$. Thus there are nine complex Higgs fields 
\begin{align}
 \Phi_i^{\pm} \coloneqq \left( W_i \right)_{(0,0),(1,\pm\, 3)} \ , \quad
\Phi_{i+2}^{\pm} \coloneqq \left( W_i \right)_{(1,\pm\, 3),(2,0)} \and
\Psi \coloneqq \left( W_3 \right)_{(2,0),(0,0)} \; ,
\end{align}
for $i=1,2$, which can be assembled into a representation of the quiver
\begin{equation}
 \xymatrix{ &   *[F.]{(1,+3)}  
\ar@/_/[dr] \ar@/^/[dr]^{\Phi_{3}^{+},\Phi_{4}^{+}}   & & \\
  *[F.]{(0,0)}  \ar@/_/[ur] \ar@/^/[ur]^{\Phi_1^{+},\Phi_2^{+}} 
  \ar@/_/[dr]_{\Phi_1^-,\Phi_2^-} \ar@/^/[dr]
&  & 
*[F.]{(2,0)} \ar[ll]_\Psi & \\
 & *[F.]{(1,-3)} 
\ar@/_/[ur]_{\Phi_{3}^-,\Phi_{4}^-} \ar@/^/[ur] &  & 
 } \label{eqn:trans_inv_quiver-graph_C(1,1)}
\end{equation}
In this example the generator $h$ of $\ZK$ has the representation
\begin{equation}
  \gamma : h \longmapsto \begin{pmatrix} \mathds{1}_{p_{(1,3)}}\otimes \mathds{1}_{2} \ \zk & 0 &0 
& 0  \\
0& \mathds{1}_{p_{(0,0)}}\otimes 1  & 0 & 0 \\
0& 0& \mathds{1}_{p_{(2,0)}}\otimes \mathds{1}_{3} \ \zk^2 &0 \\
0& 0& 0& \mathds{1}_{p_{(1,-3)}}\otimes \mathds{1}_{2} \ \zk 
\end{pmatrix} \; .
\end{equation}
\paragraph{Quiver relations}
In this case the holomorphicity condition yields the relations
 \begin{align}
 \Phi_i^\pm \, \Psi =0 \ , \quad
 \Psi \, \Phi_{i+2}^\pm =0 \and
\Phi_3^+ \, \Phi_2^+ + \Phi_3^- \,\Phi_2^- = \Phi_4^+ \, \Phi_1^+ +
\Phi_4^- \, \Phi_1^- \label{eqn:holo_W_C(1,1)}
 \end{align}
for $i=1,2$. Again the first two sets of relations of~\eqref{eqn:holo_W_C(1,1)} each describe the 
vanishing of a path in the associated quiver~\eqref{eqn:trans_inv_quiver-graph_C(1,1)} (with the obvious trivial solution $\Psi=0$), while 
the last relation equates two sums of paths.
\paragraph{Stability conditions}
Introducing Fayet-Iliopoulos parameters $\xi_1^\pm,\xi_2,\xi_3\in\R$, from the stability conditions one obtains
\begin{subequations}
 \begin{align}
  (\Phi_1^+)^\dagger \, \Phi_1^+ + (\Phi_2^+)^\dagger \,
\Phi_2^+ + (\Phi_1^-)^\dagger \, \Phi_1^- +  (\Phi_2^-)^\dagger \, \Phi_2^- - \Psi\, \Psi^\dagger &= 
 \mathds{1}_{p_{(0,0)}}\otimes \xi_0 \; , \\[4pt]
 \Phi_1^\pm \, (\Phi_1^\pm)^\dagger + \Phi_2^\pm \,
(\Phi_2^\pm)^\dagger  -( \Phi_{3}^\pm )^\dagger \, \Phi_{3}^\pm -( \Phi_{4}^\pm )^\dagger \,
\Phi_{4}^\pm &= \mathds{1}_{p_{(1,\pm 3)}}\otimes \mathds{1}_{2} \ \xi_1^\pm \; ,\\[4pt]
 \Phi_{3}^+ \,(\Phi_{3}^+)^\dagger + \Phi_{4}^+ \,
(\Phi_{4}^+)^\dagger+ \Phi_{3}^-\, (\Phi_{3}^-)^\dagger + \Phi_{4}^- \,
(\Phi_{4}^-)^\dagger - \Psi^\dagger \,\Psi &= \mathds{1}_{p_{(2,0)}}\otimes \mathds{1}_{3} \ \xi_2
   \; .
 \end{align}
\end{subequations}
%
%%%%%%%%%%%%%%%%%%%%%%%%%%%%%%%%%%%%%%%%%%%%%%%%%%%%%%%%%%%%%%%%%%%%%%%%%%%%%%%%
  \bigskip \section{Quiver gauge theories on cones: Comparison}
\label{sec:Comparison_cone}
\noindent
In Sections~\ref{sec:SU(3)-equivariant_cone} and~\ref{sec:TransInv_Instantons} 
we defined Higgs branch moduli spaces of vacua of two distinct quiver gauge theories on the Calabi-Yau
cone over the orbifold $S^5/\ZK$. In this section we shall explore their constructions in more detail, and describe their 
similarities and differences.

\subsection{Quiver bundles\label{subsec:quiver_graphs}}
\paragraph{$\sut$-equivariance}
Consider the quiver bundle $\mathcal{E}^{k,l}$ over $\R \times S^5 \slash 
\ZK$ (as a special case of~\eqref{eqn:equiv_quiver_bundle}). By 
construction the space of all connections is restricted to those which are both
$\sut$-equivariant and $\ZK$-equivariant. For holomorphic quiver
bundles, one additionally imposes the holomorphicity condition on the allowed 
connections. The general solution to these constraints (up to gauge equivalence) is given by the 
ansatz~\eqref{eqn:ansatz_HYM_cone}, where the matrices $X_\mu$ satisfy 
the equivariance conditions~\eqref{eqn:equivariance} and~\eqref{eqn:ZK-equivariance_quiver_conn} as well as the quiver relations~\eqref{eqn:holo_matrices_SU(3)}. The induced quiver bundles have 
the following structure:
\begin{itemize}
 \item A single morphism (arrow) $\HomoPhiPMi{n,m}$ between two 
Hermitian bundles (vertices) $E_{p_{(n,m)}}$ and $E_{p_{(n',m')}}$ if $n-n'=\pm \,
1$ and $m-m'=\pm\,3$.
  \item An endomorphism (vertex loop) $\EndoPsi{(n,m)}$ at each Hermitian 
bundle (vertex) $E_{p_{(n,m)}}$ with non-trivial monopole charge $\frac{m}{2}$.
\end{itemize}
The reason why there is precisely one arrow between any two adjacent vertices is
$\sut$-equivariance, which forces the horizontal component matrices $X_a$ for $a=1,2,3,4$ to have exactly the same Higgs fields 
$\HomoPhiPMi{n,m}$, i.e. $\sut$-equivariance intertwines the horizontal 
components. The vertical component $X_5$ can be 
chosen independently as it originates from the Hopf fibration $S^5 \to 
\CP$. No further constraints arise from $\ZK$-equivariance as we embed $\ZK \hookrightarrow \uo \subset \sut$. 
These quivers are a simple extension of the quivers 
obtained by~\cite{Lechtenfeld:2008nh,Dolan:2009nz} from dimensional reduction over $\CP$, because 
the additional vertical components only contribute loops on vertices with $m\neq0$. 
This structure is reminescent of that of the quivers of~\cite{Lechtenfeld:2014fza} which arise from reduction over
$3$-dimensional Sasaki-Einstein manifolds.

The Hermitian Yang-Mills equations can be considered as the intersection of the
holomorphicity condition~\eqref{eqn:holo_matrices_SU(3)} and the
stability condition~\eqref{eqn:stab_matrices_SU(3)}. In this way their form can be recognised
as Nahm-type equations of the sort considered in~\cite{Ivanova:2012vz}. We will come 
back to this point in Section~\ref{subsec:su3-equi_case}.
\paragraph{$\C^3$-invariance}
Consider the V-bundle $\mathfrak{E}^{k,l}$ over $\C^3 / \ZK$ from~\eqref{eqn:bundle_TransInv}. Recall that $C(S^5) \cong \C^3$. In 
contrast to the former case, we now impose invariance under the translation group 
$\C^3$ acting on the base as well as $\ZK$-equivariance. We demand that 
these invariant connections also induce a holomorphic structure as previously. The 
general solution to these constraints is given by the 
ansatz~\eqref{eqn:ansatz_trans_inv} where the matrices $W_\alpha$ are constant
along the base by~\eqref{eqn:trans_inv_condition_matrices}, they commute with each 
other, and they solve the $\ZK$-equivariance conditions~\eqref{eqn:form_W_TransInv}. 
The induced quiver representations have the following characteristic structure:
\begin{itemize}
 \item Two morphisms (arrows) $\Phi_{(n,m)}^i$ ($i=1,2$) between each pair of
$\ZK$-representations (vertices) $\C^{p_{(n,m)}}\otimes \Reps{(n,m)}$ and $\C^{p_{(n',m')}}\otimes\Reps{(n',m'\, )}$ if $n - n' = \pm\, 1$ in $\ZK$.
 \item One homomorphism (arrow) $\Psi_{(n,m)}$ between each pair of
$\ZK$-representations (vertices) $\C^{p_{(n,m)}}\otimes \Reps{(n,m)}$ and $\C^{p_{(n',m')}}\otimes\Reps{(n',m'\,)}$ if $n - n' = \pm\, 2$ in $\ZK$.
\end{itemize}
The reason why there are exactly two arrows between adjacent vertices is that the 
chosen representation~\eqref{eqn:ZK-action_trans-inv} does not
intertwine $W_1$, $W_2$ and acts in the same way on both of them. Thus
both endomorphisms have the same 
allowed non-vanishing components independently of one another, which gives rise to two independent sets of Higgs fields. 
The next novelty, compared to the former case, is the additional arrow 
associated to $W_3$; its existence is again due to the chosen $\ZK$-action. Translational invariance plus $\ZK$-equivariance are 
weaker constraints than $\sut$-equivariance plus $\ZK$-equivariance, and consequently the allowed number 
of Higgs fields is larger. On the other hand, holomorphicity 
seems to impose the constraint $W_3=0$ for generic non-trivial
endomorphisms $W_1$ and $W_2$ as discussed in 
Section~\ref{sec:TransInv_Instantons}. Hence there are two arrows 
between adjacent vertices, i.e. with $n - n'=\pm\, 1$, but no vertex loops as in the former 
case.

It follows that the generalised instanton 
equations~\eqref{eqn:gen_instanton_1} and~\eqref{eqn:gen_instanton_2} give rise 
to non-linear matrix equations similar to those considered 
in~\cite{SardoInfirri:1996ga} for moduli spaces of Hermitian Yang-Mills-type generalised instantons and 
in~\cite{Lechtenfeld:2014fza} for instantons on cones over $3$-dimensional Sasaki-Einstein orbifolds. We will analyse
these equations further in Section~\ref{subsec:trans-inv_case}.
\paragraph{Fibrewise $\ZK$-actions}
We shall now explain the origin of the difference between the
choices 
of $\ZK$-representations~\eqref{eqn:rep_ZK_quiver} and~\eqref{eqn:ZK-action_trans-inv}. Consider the generic linear $\ZK$-action on $\C^3$: Letting $h$ denote the generator of the cyclic group $\ZK$, and choosing
$(\theta^\alpha) = (\theta^1,\theta^2,\theta^3) \in \Z^3$ and $(z^\alpha)=(z^1,z^2,z^3)\in\C^3$, one has
\begin{equation}
 h\cdot (z^\alpha) = \big(h^\alpha_{\ \beta} \, z^\beta \big) \with (h^\alpha_{\ 
\beta})=
 \begin{pmatrix} \zk^{\theta^1}  & 0 & 0 \\
                   0 & \zk^{\theta^2} & 0 \\
                    0 & 0 & \zk^{\theta^3}  \end{pmatrix}  \; .%
   \label{eqn:generic_ZK-action}
\end{equation}
This defines an embedding of $\ZK$ into $\sut$ if and only if $\theta^1 + \theta^2 + \theta^3 = 
0 \bmod{q+1}$.

However, we also have to account for the representation $\gamma$ of $\ZK$ in 
the fibres of the bundles~\eqref{eqn:equiv_quiver_bundle} and 
\eqref{eqn:bundle_TransInv}. These bundles are explicitly constructed 
from $\sut$-representations $\RepSu{k}{l}$ which decompose under $\su 
\times \uo$ into a sum of irreducible representations $\Reps{(n,m)}$ from~\eqref{eqn:decomp_C(k,l)}. If $\Reps{(n,m)}$ and $\Reps{(n',m'\,)}$ 
both appear in the decomposition~\eqref{eqn:decomp_C(k,l)}, then there 
exists $(r,s) \in \mathbb{Z}_{\geq0}^2$ such that $n-n' = \pm \, r$ and $m-m' = \pm \,3 
s$. 
\paragraph{\it $\sut$-equivariance} The $1$-forms $\betaZK^i$ transform 
under the generic $\ZK$-action~\eqref{eqn:generic_ZK-action} as 
    \begin{equation}
     \betaZK^i \longmapsto \zk^{\theta^i - \theta^3} \, \betaZK^i \for i=1,2  \;,
    \end{equation}
while $\eta$ and $\diff \tau$ are invariant. Thus
the equivariance condition for the connection~\eqref{eqn:connection_quiver} 
becomes
\begin{subequations}
\begin{alignat}{2}
 \gamma(h) \, \left(X_{2i -1} - \im \, X_{2i} \right) \, \gamma(h)^{-1} &= 
\zk^{-\theta^i + \theta^3} \, \left(X_{2i -1} - \im \, X_{2i} \right) &  &\for i= 1,2 
\; ,\\[4pt]
 \gamma(h)\, \left(X_{2i -1} + \im \, X_{2i} \right) \, \gamma(h)^{-1} &= 
\zk^{\theta^i - \theta^3} \, \left(X_{2i -1} + \im \, X_{2i} \right)  
 &  &\for i=1,2 \; ,\\[4pt]
 \gamma(h) \, X_5 \, \gamma(h)^{-1} &= X_5  \; .  & &
\end{alignat}
\end{subequations}
In this case the aim is to embed $\ZK$ in such a way that the entire quiver 
decomposition~\eqref{eqn:equiv_quiver_bundle} is 
automatically $\ZK$-equivariant; hence the 
non-vanishing components of the matrices $X_a$ and $X_5$ are already prescribed 
by $\sut$-equivariance. For generic $\theta^\alpha$ it seems quite difficult 
to realise this embedding, because if one assumes a diagonal $\ZK$-action on 
the fibre of the form
\begin{equation}
 \gamma(h) = \bigoplus_{(n,m)\in 
Q_0(k,l)} \, \mathds{1}_{p_{(n,m)}} \otimes \zk^{\gamma(n,m)} \,
\mathds{1}_{n+1} \with \gamma(n,m) \in \Z \; , 
\label{eqn:ZK-action_test_equiv}
\end{equation}
then these equivariance conditions translate into
\begin{equation}
 \gamma(n\pm1,m+3) - \gamma(n,m) = \theta^i - \theta^3 \bmod{q+1} \for i=1,2 \label{eqn:gen_equi_cond_SU(3)}
\end{equation}
on the non-vanishing components of $X_a$, $a=1,2,3,4$.

In this paper we specialise to the weights $(\theta^\alpha)=(1,1,-2)$ and 
obtain~\eqref{eqn:ZK-action_betas} for the $\ZK$-action on $\sut$-equivariant $1$-forms. 
From this action we naturally obtain factors $\zk^{\pm\, 3}$ for the induced 
representation $\pi(h)$. This justifies the choice of $\gamma$ in~\eqref{eqn:rep_ZK_quiver}, as $m$ changes by integer multiples of $3$ while
$n$ in~\eqref{eqn:gen_equi_cond_SU(3)} does not have such uniform behaviour.
\paragraph{\it $\C^3$-invariance} The modified equivariance condition 
under~\eqref{eqn:generic_ZK-action} is readily read off to be
  \begin{align}
 \gamma(h)\, W_\alpha \, \gamma(h)^{-1} = \zk^{\theta^\alpha}\, W_\alpha \for \alpha=1,2,3 \; .
\end{align}
In contrast to the $\sut$-equivariant case above, no 
particular form of the matrices $W_\alpha$ is fixed yet, i.e. here the choice of 
realisation of the $\ZK$-action on the fibres determines the field content.
By the same argument as above, a representation of $\ZK$ on the fibres of the 
form~\eqref{eqn:ZK-action_test_equiv}
allows the component $(W_\alpha)_{(n,m),(n',m')}$ to be 
non-trivial if and only if
\begin{equation}
 \gamma(n',m'\,)-\gamma(n,m)= \theta^\alpha  \bmod{q+1} \for \alpha=1,2,3 \; .
\end{equation}
For the weights $(\theta^\alpha)=(1,1,-2)$ we then pick up 
factors of $\zk^{\pm\, 1}$ or $\zk^{\pm\, 2}$, which excludes the 
choice~\eqref{eqn:rep_ZK_quiver}. However, the modification 
to~\eqref{eqn:ZK-action_trans-inv} is allowed as $n$ changes in integer 
increments.
\paragraph{McKay quiver}
In~\cite{SardoInfirri:1996gb,Lechtenfeld:2014fza} the correspondence 
between the Hermitian Yang-Mills moduli space for translationally-invariant and 
$\ZK$-equivariant connections and the representation moduli of the McKay 
quiver is employed. The McKay quiver associated to the orbifold singularity $\C^3/\Z_{q+1}$ and the weights $(\theta^\alpha)=(1,1,-2)$
is constructed in exactly the same way as the $C^{k,l}$-quivers from Section~\ref{sec:TransInv_Instantons}, except that it is based on the 
regular representation of $\ZK$ rather than the 
representations $\RepSu{k}{l}$ considered here. It is a cyclic quiver
with $q+1$ vertices labelled by the irreducible representations of
$\ZK$, whose underlying graph is the affine extended Dynkin diagram of
type $\widehat{A}_q$, and whose arrow set coincides with those of the $C^{k,l}$-quivers. See~\cite{Cirafici:2010bd,Cirafici:2011cd,Cirafici:2012qc} for explicit constructions of instanton moduli on $\C^3/\Z_{q+1}$ in this context.
% 
%%%%%%%%%%%%%%%%%%%%%%%%%%%%%%%%%%%%%%%%%%%%%%%%%%%%%%%%%%%%%%%%%%%%%%%%%%%%%%%%
%%%%%%%%%%%%%%%%%%%%%%%%%%%%%%%%%%%%%%%%%%%%%%%%%%%%%%%%%%%%%%%%%%%%%%%%%%%%%%%%
% 
\subsection{Moduli spaces}
We shall now formalise the treatment of the instanton moduli spaces. 
We will first present an account of the general construction 
following~\cite{Atiyah:1982,Deser:2014}, and then discuss the individual 
scenarios.
\subsubsection{K\"ahler quotient construction}
\label{subsec:moduli_generic}
Let $M$ be a Kähler manifold of complex dimension $n $ and $\Gcal$ a 
compact Lie group with Lie algebra~$\gfrak$. Assume that $\Gcal$ acts in the cotangent bundle $T^*M$ preserving the complex structure $J$ and the metric 
$g$; hence $\Gcal$ also preserves the Kähler form $\omega$. Let $P=P(M,\Gcal)$ be a 
principal $\Gcal$-bundle over $M$, $\Acal$ a connection 1-form and 
$\Fcal=\Fcal_\Acal= \diff \Acal + \Acal \wedge \Acal$ 
its curvature. 

Let $\mathrm{Ad}(P)\coloneqq P \times_\Gcal \Gcal $ be the group adjoint bundle (where 
$\Gcal$ 
acts on itself via the adjoint action, i.e. by the inner automorphism $h \mapsto g \, h \, g^{-1}$), and let 
$\mathrm{ad}(P)\coloneqq P \times_{\Gcal}\gfrak$ be the algebra adjoint bundle 
(where $\Gcal$ acts on $\gfrak$ via the adjoint action, i.e. by
$X\mapsto \mathrm{Ad}(g)X=g\, X\, g^{-1}$). Let $E\coloneqq P 
\times_{\Gcal}F$ be the complex vector bundle associated to a $\Gcal$-representation $F$.

Denote the space of all connections $\Acal$ on $P$ by $\mathbb{A}= \mathbb{A}(P)$ and note that all 
associated bundles $E$ inherit their space of connections $\mathbb{A}(E)$ from 
$P$. On $\mathbb{A}(P)$ there is a natural action of the gauge group 
$\widehat{\Gcal}$, i.e. the group of automorphisms of $P$ which are trivial on 
the base $M$. One can identify the gauge group with the space of global sections
\begin{equation}
 \widehat{\Gcal} = \Omega^0(M,\mathrm{Ad}(P))  
\end{equation}
of the group adjoint bundle, and the 
action is realised via the gauge transformations
\begin{equation}
 \Acal \longmapsto g\cdot \Acal = \mathrm{Ad}(g) \Acal + g^{-1} \, \diff g  \for g \in 
\Omega^0(M,\mathrm{Ad}(P)) \; .
\end{equation}
The Lie algebra of the gauge group can then be identified with the space of sections
\begin{equation}
 \widehat{\gfrak} = \Omega^0(M,\mathrm{ad}(P))
\end{equation}
of the algebra adjoint bundle, and the infinitesimal gauge transformations are given by
\begin{equation}
 \Acal \longmapsto \delta_\chi \Acal = \diff_\Acal \chi \coloneqq \diff \chi + \left[ 
\Acal, \chi \right] \for \chi \in \Omega^0(M,\mathrm{ad}(P)) \; . 
\label{eqn:inf_gauge_transf}
\end{equation}

Since $\mathbb{A}(P)$ is an affine space, its tangent space $T_{\Acal} 
\mathbb{A}$ at any point $\Acal \in \mathbb{A}$ can be canonically identified with 
$\Omega^1(M,\mathrm{ad}(P))$. If the structure group is a matrix Lie group, i.e. there is an embedding $\Gcal \hookrightarrow 
\urm(N)$ for some $N\in \mathbb{Z}_{>0}$, then $\gfrak$ is a matrix 
Lie algebra and the trace defines an $\mathrm{Ad}(\Gcal)$-invariant inner product on~$\gfrak$. The 
induced invariant inner product on $\Omega^1(M,\mathrm{ad}(P))$ is
\begin{subequations}
 \label{eqn:metric_connections}
\begin{equation}
 \langle X_1 , X_2 \rangle \coloneqq \int_M \, \tr \left( X_1 \wedge \star\, X_2 
\right) \for X_1,X_2 \in \Omega^1(M,\mathrm{ad}(P)) \; , 
\end{equation}
which gives rise to a gauge-invariant metric on $\mathbb{A}(P)$  via the pointwise 
definition
\begin{equation}
 \boldsymbol{g}_{|\Acal}(X_1 , X_2 ) := \langle X_1 , X_2 \rangle_{|\Acal} 
\for X_1,X_2 \in T_\Acal \mathbb{A} \; .
\end{equation}
\end{subequations}
The space $\mathbb{A}(P)$ moreover carries a gauge-invariant symplectic structure defined by
\begin{equation}
 \boldsymbol{\omega}_{|\Acal} (X_1,X_2) = \int_M \, \tr \left( X_1 \wedge X_2 
\right) \wedge \omega^{n-1}  \for X_1,X_2 \in T_\Acal \mathbb{A}\; . 
\label{eqn:symplectic_form_connections}
\end{equation}
Note that the $2$-form $\boldsymbol{\omega}$ is completely independent
of the base point $\Acal\in \mathbb{A}$. Let $\Diff$ denote the exterior derivative acting on forms on 
$\mathbb{A}$. Then by computing
\begin{equation}
\begin{split}
 \Diff \boldsymbol{\omega}_{|\Acal} (X_0,X_1,X_2) =  
& \, X_0 \big(\boldsymbol{\omega}_{|\Acal} (X_1,X_2) \big)
- X_1 \big(\boldsymbol{\omega}_{|\Acal} (X_0,X_2) \big)
+X_2 \big(\boldsymbol{\omega}_{|\Acal} (X_0,X_1)\big) \\
&-\boldsymbol{\omega}_{|\Acal} ([X_0,X_1],X_2) 
+\boldsymbol{\omega}_{|\Acal} ([X_0,X_2],X_1)
-\boldsymbol{\omega}_{|\Acal} ([X_1,X_2],X_0) \; ,
\end{split}
\end{equation}
one observes that $\Diff \boldsymbol{\omega}=0$ as $X_i \big(
\boldsymbol{\omega}_{|\Acal} (X_j,X_k) \big) =0 $ due to base point independence and
\begin{equation}
 \boldsymbol{\omega}_{|\Acal} ([X_i,X_j],X_k) = \int_M \, \tr \left( [X_i,X_j] 
\wedge X_k \right) \wedge \omega^{n-1} =0 
\end{equation}
as $\tr \left( 
[X_i,X_j] \wedge X_k \right)\in \Omega^3(M)$ which renders the integrand into a form of 
degree larger than the top degree. It follows that $\boldsymbol{\omega}$ is a 
symplectic form, which promotes $\mathbb{A}$
to an infinite-dimensional 
Riemannian symplectic manifold $(\mathbb{A},\boldsymbol{g},\boldsymbol{\omega})$ equipped with a compatible 
$\widehat{\Gcal}$-action.
\paragraph{Holomorphic structure}
Consider now the restriction to connections on $E \rightarrow
M$ which are generalised instanton connections. Recall that one part of the Hermitian Yang-Mills equations can be 
interpreted as holomorphicity conditions, and the corresponding subspace is
\begin{equation}
 \mathbb{A}^{1,1} = \big\{ \Acal \in \mathbb{A}(E) \,:\, 
\Fcal_\Acal^{0,2} = - \big( \Fcal_\Acal^{2,0} \big)^\dagger =0  
\big\} \subset \mathbb{A}(E) \; .
\end{equation}
This definition employs the underlying complex structure on $M$. As before, this condition is equivalent to the existence of a 
holomorphic structure on $E$, i.e. a Cauchy-Riemann operator 
$\overline{\partial}_E := \overline{\partial} + A^{0,1}$ that satisfies the Leibniz rule 
as well as $\overline{\partial}_E \circ \overline{\partial}_E =0$. Thus a 
$\Gcal$-bundle with only holomorphic connections induces a 
$\Gcal^\C$-bundle where $\Gcal^\C=\Gcal\otimes\C$. One can show that $ \mathbb{A}^{1,1}$ is an 
infinite-dimensional Kähler manifold, i.e. the metric $ \boldsymbol{g}$ is Hermitian 
and the symplectic form $\boldsymbol{\omega}$ is Kähler. These tensor fields 
descend from $\mathbb{A}$ to $\mathbb{A}^{1,1}$ simply by 
restriction.
\paragraph{Moment map}
The space $\mathbb{A}^{1,1}$ inherits a $\widehat{\Gcal}$-action from 
$\mathbb{A}$ and since it has a $\widehat{\Gcal}$-invariant symplectic form, i.e.~the Kähler form $\boldsymbol{\omega}$, one can 
introduce a moment map
\begin{equation}
 \begin{split}
 \mu : \mathbb{A}^{1,1} &\longrightarrow \widehat{\gfrak}\,^* 
\cong \Omega^{2n}(M,\mathrm{ad}(P)) \\
 \Acal &\longmapsto \mathcal{F}_\Acal \wedge \omega^{n-1} \; .
 \end{split} 
 \label{eqn:moment_map_aux}
\end{equation}
For this to be a moment map of the $\widehat{\Gcal}$-action one needs to 
verify the defining properties, generalising the arguments 
presented in~\cite{Atiyah:1982}. 
For this, note that $\mu$ is obviously $\widehat{\Gcal}$-equivariant. Next let 
$\phi \in \Omega^0(M, \mathrm{ad}(P))$ be an element of the gauge algebra, 
$\phi^\natural$ the corresponding vector field on $ \mathbb{A}^{1,1}$ and 
$\psi \in \Omega^1(M,\mathrm{ad}(P))$ a tangent vector at the base point 
$\Acal$. Then the condition to verify is
\begin{equation}
 (\phi,\Diff \mu_{| \Acal})(\psi) = \iota_{\phi^\natural}  
\boldsymbol{\omega}_{|\Acal} (\psi) \; , \label{eqn:aux_moment_map}
\end{equation}
wherein $\iota$ denotes contraction and the dual pairing $(\cdot, \cdot)$ of $\widehat{\gfrak}$ with $\widehat{\gfrak}\,^*$
is defined via integration over $M$ of the invariant inner product on $\gfrak$. 
Firstly, in the definition of $\mu$ only $\Fcal_\Acal$ is base point dependent, and a 
standard computation gives $\Fcal_{\Acal + t \, \psi} = \Fcal_\Acal + t\, \diff_\Acal \psi + 
\tfrac{1}{2}\, t^2 \, \psi\wedge\psi$ so that $\Diff \Fcal_{|\Acal} = 
\big( \frac{\diff}{\diff t} \Fcal_{\Acal + t \, \psi} \big)_{|{t=0}} 
=\diff_\Acal \psi $. Thus the left-hand side of~\eqref{eqn:aux_moment_map} is  
$(\phi,\Diff \mu_{| \Acal})(\psi)= \int_M \, \tr \big( (\diff_\Acal 
\psi) \wedge \phi \big) \wedge \omega^{n-1} $. Secondly, the vector field $\phi^\natural$ can be read off 
from~\eqref{eqn:inf_gauge_transf} to be $\phi^\natural_{|\Acal} = \diff_\Acal 
\phi \in \Omega^1(M , \mathrm{ad}(P))$. Hence the right-hand side is
$\iota_{\phi^\natural}  
\boldsymbol{\omega}_{|\Acal} (\psi)= \int_M\, \tr \big( (\diff_\Acal \phi) 
\wedge \psi \big) \wedge \omega^{n-1} $. But from $\int_M \, \diff 
\left( \tr \left( \psi \wedge \phi \right) \wedge \omega^{n-1} \right) =0$ 
and $\diff \omega =0$ one has $\int_M\, \tr \big( (\diff_\Acal \psi) 
\wedge \phi\big) \wedge \omega^{n-1} = - \int_M \tr\, \big(  
\psi \wedge (\diff_\Acal \phi )\big) \wedge \omega^{n-1} $, and therefore the
relation~\eqref{eqn:aux_moment_map} holds, i.e. $\mu$ is a moment map of the 
$\widehat{\Gcal}$-action on $\mathbb{A}^{1,1}$.

We will use the dual moment map defined by 
\begin{equation}
 \begin{split}
 \mu^* : \mathbb{A}^{1,1}  &\longrightarrow  \widehat{\gfrak} = \Omega^0(M,\mathrm{ad}(P)) 
\\
 \Acal &\longmapsto \omega\, \lrcorner \, \mathcal{F}_\Acal \; ,
 \end{split} 
  \label{eqn:moment_map_generic}
\end{equation}
 which is equivalent to the definition~\eqref{eqn:moment_map_aux} due to the  
identification $\widehat{\gfrak} \cong 
 \widehat{\gfrak}\,^*$ given by~\eqref{eqn:stab_identity} (generically by a choice of metric). Thus we will no longer explicitly distinguish between the moment map $\mu$ and its dual $\mu^*$.

For \emph{regular} elements $\Xi \in \widehat{\mathfrak{g}}$, the
centraliser of $\Xi$ in $\widehat{\Gcal}$ is the maximal torus and
$\mu^{-1}(\Xi) \subset \mathbb{A}^{1,1} $ defines a submanifold which
carries a $\widehat{\Gcal}$-action. The quotient of the level sets\footnote{One must in
  fact take $\Xi \in \mathrm{Center}(\, \widehat{\mathfrak{g}}\, )$ for a well-defined 
quotient.}
\begin{equation}
\mathbb{A}^{1,1} \sslash^{\phantom{\dag}}_\Xi \, \widehat{\Gcal} := \mu^{-1}(\Xi) \, \big\slash \,
\widehat{\Gcal}
\end{equation}
is well-defined, and moreover it defines a Kähler space as the Kähler form and the
complex structure descend from $\mathbb{A}^{1,1} $ by gauge-invariance. The level 
set of zeroes is precisely the Hermitian Yang-Mills moduli space.
\paragraph{Complex group action}
 As the $\Gcal$-action in $T^*M$ preserves the Kähler 
structure, one can extend it to a $\Gcal^\C$-action in $T^*M$. The same is true for the extension to the complexification of the
$\widehat{\Gcal}$-action on $\mathbb{A}^{1,1}$, i.e. the 
holomorphicity conditions 
$\Fcal_\Acal^{0,2}=0$ are invariant under the action of the complex gauge 
group
\begin{equation}
 \widehat{\Gcal}\,^\C =\widehat{\Gcal} \otimes \C \; .
\end{equation}
For $\Acal \in  \mathbb{A}^{1,1}$ the orbit $\widehat{\Gcal}_\Acal^{\,\C}$ 
of the $\widehat{\Gcal}\,^\C$-action is given by
\begin{equation}
 \widehat{\Gcal}_\Acal^{\,\C} = \big\{ \Acal' \in \mathbb{A}^{1,1} \, 
: \, \Acal'= g\cdot \Acal \ , \ g \in \widehat{\Gcal}\,^\C \,
\big\} \ .
\end{equation}
A point $\Acal \in  \mathbb{A}^{1,1}$ is called \emph{stable} if 
$\widehat{\Gcal}_\Acal^{\,\C} \cap \mu^{-1}(\Xi) \neq \emptyset$. Denote 
by $\mathbb{A}^{1,1}_{\rm st}(\Xi) \subset\mathbb{A}^{1,1}$ the set of all 
stable points (for a given regular element $\Xi$). Then the K\"ahler quotient can be identified with the GIT quotient (see for 
instance~\cite{Thomas:2006})
\begin{equation}
  \mathbb{A}^{1,1} \sslash^{\phantom{\dag}}_\Xi \, \widehat{\Gcal} \cong  \mathbb{A}^{1,1}_{\rm st}(\Xi) \, \big\slash \,
\widehat{\Gcal}\,^\C \; .
\end{equation}

In the following we discuss applications of this K\"ahler quotient construction to
$\sut$-equivariant and $\ZK$-equivariant instantons on the Calabi-Yau
cone $M=C(S^5/\ZK)$, as well as to the $\C^3$-invariant and $\ZK$-equivariant 
case. These vacuum moduli spaces are special cases of those constructed above, as we do not 
consider generic connections but rather equivariant connections. For
instance, equivariance reduces the gauge groups.
% 
%%%%%%%%%%%%%%%%%%%%%%%%%%%%%%%%%%%%%%%%%%%%%%%%%%%%%%%%%%%%%%%%%%%%%%%%%%%%%%%%
%%%%%%%%%%%%%%%%%%%%%%%%%%%%%%%%%%%%%%%%%%%%%%%%%%%%%%%%%%%%%%%%%%%%%%%%%%%%%%%%
% 
\subsubsection{$\sut$-equivariance}
\label{subsec:su3-equi_case}
Consider the space of $\sut$-equivariant connections 
$\mathbb{A}(\mathcal{E}^{k,l})$ on the bundle~\eqref{eqn:equiv_quiver_bundle} (for $d=1$), 
which is an affine space modelled on $\Omega^1\big( C(S^5/\ZK), 
\End_\uo(V^{k,l}) 
\big)$. The structure group $\Gcal^{k,l}$ of the bundle~\eqref{eqn:equiv_quiver_bundle} is given by~\eqref{eqn:gauge_group_fibre}. An element 
$X 
\in\Omega^1\big( C(S^5/\ZK), \End_\uo(V^{k,l}) \big)$ can be expressed as 
\begin{equation}
 X = X_\mu \, e^\mu + X_\tau \, \diff \tau \equiv X_j \, \theta^j 
+ \overline{X}_{{j}} \, \overline{\theta}{}^{{j}} \; ,
\end{equation}
once one has chosen the coframe $\{e^\mu ,\diff \tau\}$ of the conformally 
equivalent cylinder $\R\times S^5/\ZK$ with $r=\e^\tau$. One can equivalently use the 
complex basis $\theta^j = e^{2j-1} + \im \, e^{2j}$ for $j=1,2,3$, where 
$e^6:= \diff \tau$; then $(X_j)^\dagger = -\overline{X}_{{j}}$. Thus once one has 
fixed a choice of coframe on the Calabi-Yau cone $C(S^5/\ZK)$, the tangent space to
$\mathbb{A}(\mathcal{E}^{k,l})$ at a point $\Acal$ is described by all 
6-tuples $(\{ X_\mu \},X_\tau)$ or equivalently $(\{X_j\}, \{\,\overline{X}_{{j}} \})$. Here $X_\mu$ and
$X_\tau$ depend only on the cone coordinate $\tau$ by $\sut$-equivariance.\footnote{Recall that the equivariance 
condition~\eqref{eqn:equivariance} makes the endomorphisms $X_\mu$ base point independent on 
$S^5/\ZK$; hence it is consistent to have solely $\tau$-dependent matrices
$X_\mu$ in any 
coframe.}

\paragraph{Instanton equations}
One can eliminate the linear terms in~\eqref{eqn:holo_matrices_SU(3)} 
and~\eqref{eqn:stab_matrices_SU(3)} via the redefinitions
\begin{equation}
 X_a = \e^{-\frac{3}{2}\, \tau} \, \mathcal{X}_a \for a=1,2,3,4 \and X_5 = \e^{-4 
\tau} \,
\mathcal{X}_5 \; , \quad X_\tau = \e^{-4 \tau} \, \mathcal{X}_6 \; .  
\label{eqn:rescale_X-matrices}
\end{equation}
Using 't~Hooft tensors the matrix equations read
\begin{subequations}
\label{eqn:flow_rescaled_X-matrices}
\begin{align}
 \eta^1_{ab} \left[ \mathcal{X}_a , \mathcal{X}_b\right] &= 0 \and 
  \eta^2_{ab} \left[ \mathcal{X}_a , \mathcal{X}_b\right] = 0 \; , \\[4pt]
\frac{\diff\mathcal{X}_a}{\diff s} &= -\eta^3_{ab} \,\left[\mathcal{X}_b, \mathcal{X}_5 \right]-
\left[\mathcal{X}_a, \mathcal{X}_6 \right] \; , \\[4pt]
\frac{\diff\mathcal{X}_5}{\diff s} &= -\lambda(s) \, \big( \left[\mathcal{X}_1, \mathcal{X}_2 \right] +
\left[\mathcal{X}_3, \mathcal{X}_4 \right] \big)
-\left[\mathcal{X}_5, \mathcal{X}_6 \right] \; ,
\end{align}
\end{subequations}
where
$s:= \tfrac{1}{4} \,\e^{-4\tau} \in \R_{>0}$ and
$\lambda(s)=\big(\,\tfrac{1}{4s}\, \big)^{\frac{5}{4}}$. The
equations \eqref{eqn:flow_rescaled_X-matrices} are automatically satisfied in the
temporal gauge $X_\tau=0$ by taking \emph{constant} scalar fields
$X_\mu$ for $\mu=1,\dots,5$ satisfying the Higgs branch BPS equations
\eqref{eqn:BPSequations} of the quiver gauge theory.

Changing to a complex basis as before and defining
\begin{equation}
  \mathcal{Y}_j= \tfrac{1}{2}\, \left( \mathcal{X}_{2j-1} - \im \,
\mathcal{X}_{2j}  \right) \and  \overline{\mathcal{Y}}_{{j}}= \tfrac{1}{2} \,
\left( \mathcal{X}_{2j-1} + \im\, \mathcal{X}_{2j} \right) \for j=1,2,3\; ,
\end{equation}
the resulting holomorphicity conditions are
\begin{subequations}
\label{eqn:instanton_reformulated}
\begin{align}
  \big[ \mathcal{Y}_{{1}} , \mathcal{Y}_{{2}} \big]= 0 & \and
\big[ \, \overline{\mathcal{Y}}_{1} , \overline{\mathcal{Y}}_{2} \big] = 0 \; , 
\label{eqn:inst_commutators} \\[4pt]
\frac{\diff\mathcal{Y}_{{i}}}{ \diff s} =-2 \, \im \, \big[ 
\mathcal{Y}_{{i}} , \mathcal{Y}_{{3}} \big] & \and
\frac{\diff\overline{\mathcal{Y}}_{i}}{ \diff s} = 2\, \im \, \big[ \, \overline{\mathcal{Y}}_{i} , \overline{\mathcal{Y}}_{3} \big] \for i=1,2 \; , \label{eqn:inst_Nahm}
\end{align}
while the stability condition yields
\begin{equation}
 \frac{\diff\mathcal{Y}_3}{\diff s} + \frac{\diff\overline{\mathcal{Y}}_{{3}}}{\diff s} =
2\, \im \, \big[ \mathcal{Y}_{3} , \overline{\mathcal{Y}}_{{3}} \big] +2\,\im \,
\lambda(s) \,\Big( \big[ \mathcal{Y}_{1} , 
\overline{\mathcal{Y}}_{{1}} \big] + 
\big[ \mathcal{Y}_{2} , \overline{\mathcal{Y}}_{{2}} \big]\Big) \; . 
\label{eqn:inst_stability}
\end{equation}
\end{subequations}
Analogously to the generic situation, we define the subspace
\begin{equation}
 \mathbb{A}^{1,1}\big(\mathcal{E}^{k,l}\big) =  \Big\{ 
\big(\{\mathcal{Y}_j\},\{\, \overline{\mathcal{Y}}_{{j}} \} \big)\in \mathbb{A}\big(\mathcal{E}^{k,l}\big) \, : \,
\text{\eqref{eqn:inst_commutators} and \eqref{eqn:inst_Nahm} hold} \Big\} \; .
\end{equation}
\paragraph{Real gauge group}
On the space $ \mathbb{A}^{1,1}(\mathcal{E}^{k,l})$ there is an action of the 
gauge group 
\begin{equation}
 \widehat{\Gcal}\,^{k,l} := \Omega^0\big(\R_{>0},\Gcal^{k,l}\big) \ ,
\end{equation}
with $\Gcal^{k,l}\hookrightarrow\urm(p)$, given by\footnote{We assume that the paths $g(s):(0,\infty) \to\Gcal^{k,l}$ are sufficiently smooth.}
\begin{subequations}
\label{eqn:instanton_gauge_transf}
\begin{align}
\overline{\mathcal{Y}}_{{i}} \longmapsto \mathrm{Ad}(g) \overline{\mathcal{Y}}_{{i}} \for 
i=1,2 \and
\overline{\mathcal{Y}}_{{3}} \longmapsto \mathrm{Ad}(g) \overline{\mathcal{Y}}_{{3}} +
\frac{\im}{2} \, \Big(\, \frac{\diff g}{\diff s} \, \Big) \, g^{-1}  \; ,
\end{align}
\end{subequations}
for $g\in  \widehat{\Gcal}\,^{k,l}$.
One readily checks that the full set of 
equations~\eqref{eqn:instanton_reformulated} is invariant under these ``real'' 
gauge transformations. Moreover, one can always find a gauge transformation 
$g\in \widehat{\Gcal}\,^{k,l}$ such that $g\cdot\mathcal{X}_6=0$ or
equivalently $g\cdot\overline{\mathcal{Y}}_{{3}}= g\cdot \mathcal{Y}_{3}$.
\paragraph{Complex gauge group}
The space $ \mathbb{A}^{1,1}(\mathcal{E}^{k,l})$ also admits an action of the 
complex gauge group
\begin{equation}
 \big(\widehat{\Gcal}^{\,k,l}\big)^\C := \Omega^0\big(\R_{>0},(\Gcal^{k,l})^\C\big)  \; ,
\end{equation}
with $\big(\Gcal^{k,l}\big)^\C \hookrightarrow \glrm (p,\C)$. 
However, only the equations \eqref{eqn:inst_commutators} and \eqref{eqn:inst_Nahm} are 
invariant under the ``complex'' gauge transformations given by 
\begin{subequations}
\label{eqn:instanton_cplx_gauge_transf}
\begin{align}
\overline{\mathcal{Y}}_{{i}} \longmapsto \mathrm{Ad}(g) \overline{\mathcal{Y}}_{{i}} & \and
\mathcal{Y}_{i} \longmapsto \mathrm{Ad}(g^{*-1}) \mathcal{Y}_{i} 
\for i=1,2 \; ,\\[4pt]
\overline{\mathcal{Y}}_{{3}} \longmapsto \mathrm{Ad}(g) \overline{\mathcal{Y}}_{{3}} +
\frac{\im}{2} \, \Big(\, \frac{\diff g}{\diff s} \,\Big) \, g^{-1}   
& \and
\mathcal{Y}_{3} \longmapsto \mathrm{Ad}(g^{*-1}) \mathcal{Y}_{3} +
\frac{\im}{2} \,g^{*-1} \, \Big(\, \frac{\diff g^\dagger}{\diff s} \,
  \Big) \ ,
\end{align}
\end{subequations}
where $g\in \big(\widehat{\Gcal}^{\,k,l}\big)^\C $ and $g^{*-1} = (g^{-1})^\dagger$. 
\paragraph{Kähler structure}
Following the construction of
Section~\ref{subsec:moduli_generic}, the next step is to define a 
Kähler structure on $\mathbb{A}^{1,1}(\mathcal{E}^{k,l})$. The tangent space $T_{\Acal} 
\mathbb{A}(\mathcal{E}^{k,l})$ at point $\Acal$ is $\Omega^1\big( 
C(S^5/\ZK), \End_\uo(V^{k,l}) \big)$, so a tangent vector $\boldsymbol{x} = 
\boldsymbol{x}_{j} \, \theta^j + \overline{\boldsymbol{x}}_{{j}} \, \overline{\theta}{}^{{j}}$ over $\mathbb{A}^{1,1}(\mathcal{E}^{k,l})$ is 
defined by linearisation of the holomorphicity equations~\eqref{eqn:inst_commutators} and \eqref{eqn:inst_Nahm} for paths  
$\overline{\boldsymbol{x}}_{{j}}(s) : (0,\infty) \to \End_\uo (V^{k,l})^\C $. The gauge 
transformations are given by $\overline{\boldsymbol{x}}_{{j}}\mapsto \mathrm{Ad}(g) 
\overline{\boldsymbol{x}}_{{j}} $ for $j=1,2,3$.

A metric on $\mathbb{A}^{1,1}(\mathcal{E}^{k,l})$ can be defined from~\eqref{eqn:metric_connections} as
\begin{equation}
 \boldsymbol{g}_{|\Acal}(\boldsymbol{x},\boldsymbol{y}):= \frac{1}{2} \,
\int_{0^+}^\infty\, \diff s \ \sum_{j=1}^3\, \tr 
\big(\boldsymbol{x}_{j}^\dagger \,
\boldsymbol{y}_{j}^{\phantom{\dagger}}  + 
\boldsymbol{x}_j^{\phantom{\dagger}} \, \boldsymbol{y}_j^\dagger \big) \; ,
\end{equation}
where the integral over $S^5/\ZK$ drops out here as the tangent 
vectors at equivariant connections are independent of the coordinates of $S^5/ \ZK$. A 
symplectic form on $\mathbb{A}^{1,1}(\mathcal{E}^{k,l})$ can likewise
be defined from~\eqref{eqn:symplectic_form_connections} as
\begin{equation}
  \boldsymbol{\omega}_{|\Acal}(\boldsymbol{x},\boldsymbol{y}):=
\frac{\im}{2} \,\int_{0^+}^\infty\, \diff s
\ \sum_{j=1}^3 \,
\tr  \big( \boldsymbol{x}_{j}^\dagger \,
\boldsymbol{y}_{j}^{\phantom{\dagger}}  - 
\boldsymbol{x}_j^{\phantom{\dagger}} \, \boldsymbol{y}_j^\dagger \big) \; .
\end{equation}
Both $\boldsymbol{g} $ and 
$\boldsymbol{\omega} $  are gauge-invariant by construction.
Moreover, we immediately see that for the choice 
of complex structure\footnote{We essentially use the complex structure $J$ of 
$\C^3$.} $J (\,\overline{\boldsymbol{x}}_{{j}}) = \im \, \overline{\boldsymbol{x}}_{{j}}$ the 
symplectic form $\boldsymbol{\omega}$ and the metric $\boldsymbol{g}$ are compatible, i.e. $\boldsymbol{g}(\cdot, J \,
\cdot ) = \boldsymbol{\omega}(\cdot, \cdot)$.
\paragraph{Moment map}
On the Kähler manifold $\mathbb{A}^{1,1}(\mathcal{E}^{k,l})$ we define a moment 
map by
\begin{align}
 \mu : \mathbb{A}^{1,1}\big(\mathcal{E}^{k,l}\big) &\longrightarrow \widehat{\End}_\uo\big(V^{k,l}\big) \notag \\
\big(\{\mathcal{Y}_j\},\{\, \overline{\mathcal{Y}}_{{j}} \} \big) &\longmapsto  \frac{\diff\mathcal{Y}_3}{\diff s} + \frac{\diff\overline{\mathcal{Y}}_3}{\diff s} -
2\,\im\, \big[ \mathcal{Y}_{3} , \overline{\mathcal{Y}}_{{3}} \big] - 2\,\im \,
\lambda(s) \,\Big( \big[ \mathcal{Y}_{1} , 
\overline{\mathcal{Y}}_{{1}} \big] + 
\big[ \mathcal{Y}_{2} , \overline{\mathcal{Y}}_{{2}} \big]\Big) \; ,
\end{align}
which readily gives us the Kähler quotient for the instanton moduli space
\begin{equation}
 \mathcal{M}^{\sut}_{k,l} = \mu^{-1} (0) \, \big\slash \, \widehat{\Gcal}\,^{k,l} \; .
\end{equation}
\paragraph{Stable points}
We can alternatively describe the moduli space $\mathcal{M}_{k,l}^{\sut} $ via the stable points
\begin{equation}
\mathbb{A}^{1,1}_{\rm st}\big(\mathcal{E}^{k,l}\big) := \Big\{ 
\big(\{\mathcal{Y}_j\},\{\, \overline{\mathcal{Y}}_{{j}} \} \big) \in 
\mathbb{A}^{1,1}\big(\mathcal{E}^{k,l} \big) \, : \;
\big(\widehat{\Gcal}\,^{k,l}\big)^\C_{(\{\mathcal{Y}_j\},\{\, \overline{\mathcal{Y}}_{{j}} \} )} \cap 
\mu^{-1}(0)\neq \emptyset \Big\} \ ,
\end{equation}
and by taking the GIT quotient as before to get
\begin{equation}
\mathcal{M}_{k,l}^\sut \cong \mathbb{A}^{1,1}_{\rm st}\big(\mathcal{E}^{k,l}\big) \, \big\slash \, \big(\widehat{\Gcal}\,^{k,l}\big)^\C \; .
\end{equation}
We show below that it is sufficient to solve the 
holomorphicity equations (subject to certain boundary conditions), as the solution to the stability equation
then follows automatically by a 
complex gauge transformation. More precisely, for 
every point in $\mathbb{A}^{1,1}(\mathcal{E}^{k,l})$ there exists a unique 
point in its complex gauge orbit which satisfies the stability equation, i.e. every point in $\mathbb{A}^{1,1}(\mathcal{E}^{k,l})$ is stable.
\paragraph{Solutions of the holomorphicity equations}
Following~\cite{Donaldson:1984} one can regard the  
holomorphicity equations as being locally trivial. For this, 
we use a complex gauge transformation~\eqref{eqn:instanton_cplx_gauge_transf} to eliminate $\overline{\mathcal{Y}}_3$ via
\begin{equation}
 \widetilde{\overline{\mathcal{Y}}}_{{3}} = \mathrm{Ad}(g) 
\overline{\mathcal{Y}}_{{3}} + \frac{\im}{2} \, \Big(\, \frac{\diff g}{\diff s} \, \Big) \,
g^{-1} = 0 \; .
\end{equation}
From the holomorphicity equations \eqref{eqn:inst_Nahm} and \eqref{eqn:inst_commutators} one obtains in this gauge
\begin{equation}
\frac{\diff\widetilde{\overline{\mathcal{Y}}}_{{i}}}{\diff s}  =0 \and \widetilde{\overline{\mathcal{Y}}}_{{i}}=\overline{\mathcal{T}}_{{i}} \with
\big[\,\overline{\mathcal{T}}_{{1}},\overline{\mathcal{T}}_{{2}} \big]=0 \ ,
\end{equation}
where $\overline{\mathcal{T}}_{{i}}$ are \emph{constant} for $i=1,2$. Consequently the general local solution 
of the holomorphicity equations~\eqref{eqn:inst_commutators} and~\eqref{eqn:inst_Nahm} 
is
\begin{equation}
\overline{\mathcal{Y}}_{{i}} = 
\mathrm{Ad}(g^{-1}) \overline{\mathcal{T}}_{{i}} \with
\big[\, \overline{\mathcal{T}}_{{1}},\overline{\mathcal{T}}_{{2}} \big]=0 \and 
\overline{\mathcal{Y}}_{{3}}=- \frac{\im}{2} \, g^{-1}\, \frac{\diff g}{\diff s} \; ,
\label{eqn:solution_complex}
\end{equation}
with $g \in \big(\widehat{\Gcal}\,^{k,l}\big)^\C$. A solution to the commutator constraint chooses $\overline{\mathcal{T}}_{{i}}$ for 
$i=1,2$ as elements of the Cartan subalgebra of the complex Lie algebra 
$\End_\uo (V^{k,l})^\C $ of the structure
group~\eqref{eqn:gauge_group_fibre}. 
\paragraph{Solutions of the stability equation}
We also need to 
solve the stability equation~\eqref{eqn:inst_stability}, for which we
follow again~\cite{Donaldson:1984}. Recall that the complete 
set of instanton equations~\eqref{eqn:instanton_reformulated} is $\widehat{\Gcal}\,^{k,l}$-invariant, 
and for each $ g \in 
(\widehat{\Gcal}\,^{k,l})^\C$ define 
\begin{equation}
 h = h(g) = g\, g^\dagger : (0,\infty) \longrightarrow \big(\Gcal^{k,l}\big)^\C \, \big\slash \,\Gcal\,^{k,l}
\ \hookrightarrow \ \glrm (p,\C) \, \big\slash\, \urm(p) \ .
\end{equation}
Fix a $6$-tuple
$\big\{\,\overline{\mathcal{Y}}_{{j}},\mathcal{Y}_{j}\big\}_{j=1,2,3}$ and
define the gauge transformed $6$-tuple 
$\big\{\,
\widetilde{\overline{\mathcal{Y}}}_{{j}},\widetilde{\mathcal{Y}}_{j} \big\}_{j=1,2,3}$. We
will study the critical points of the functional
\begin{equation}
 \mathcal{L}_{\epsilon}[g] = \frac{1}{2} \,
\int_{{\epsilon}}^{\frac{1}\epsilon} \,\diff s \ \tr \Big( \, \big| 
\widetilde{\mathcal{Y}}_3 +  \widetilde{\overline{\mathcal{Y}}}_{{3}} \big|^2 + 2 
\lambda(s) \, \sum_{i=1}^{2} \, \big| \,
\widetilde{\overline{\mathcal{Y}}}_{{i}} \, \big|^2 \,
\Big) \for 0<\epsilon<1 \; . \label{eqn:Lagrange_functional}
\end{equation}

As the instanton equations are invariant under
$\urm(p)$-valued gauge transformations, we can restrict $g$ to take values 
in the quotient $\glrm(p,\C)\,\slash\, \urm(p)$ which may be identified with the set 
of positive Hermitian $p\times p$ matrices~\cite{Donaldson:1984}.
Hence it is sufficient to 
consider variations with $\delta g = \delta g^\dagger$ around $g=\mathds{1}_p$ (and with $\delta g \neq 0$). 
Then the gauge transformations~\eqref{eqn:instanton_cplx_gauge_transf} imply 
that
\begin{equation}
 \delta \overline{\mathcal{Y}}_{{3}} = \big[ \delta g , \overline{\mathcal{Y}}_{{3}} 
\big] + \frac{\im}{2} \, \frac{\diff\delta g}{\diff s}  \and \delta 
\overline{\mathcal{Y}}_{i} = \big[ \delta g , \overline{\mathcal{Y}}_{i} 
\big] \for i=1,2 \; .
\end{equation}
The variation then leads to
\begin{equation}
 \delta \mathcal{L}_{\epsilon}[g] = - \im \,
\int_{\epsilon}^{\frac1\epsilon} \,
 \diff s \ \tr \Big( \mu\big(\{\mathcal{Y}_j\},\{\, \overline{\mathcal{Y}}_{{j}} \} \big) \ \delta g 
\Big) \; ,
\end{equation}
i.e. the critical points of~\eqref{eqn:Lagrange_functional} form the zero-level 
set of the moment map.

Now we use the solution~\eqref{eqn:solution_complex} as an initial evaluation of $\mathcal{L}_{\epsilon}$. Then we obtain
the functional of $h$ given by
\begin{align}
 \mathcal{L}_{\epsilon} [h] =  \frac{1}{2} \,
\int_{{\epsilon}}^{\frac1\epsilon} \, \diff s \ \bigg(\, \frac{1}{4} \, \tr 
\Big(\, h^{-1} \,
\frac{\diff h}{\diff s} \, \Big)^2  + V(h) \bigg) \ ,
\end{align}
where the potential $V(h)= 2 \lambda(s)\, \sum_{i=1}^2 \,
\tr\big( h^{-1} \, \overline{\mathcal{T}}_{i}\, h \, \overline{\mathcal{T}}_{i}{}^\dagger 
\big)$ 
is positive. This implies that for any boundary values $h_\pm \in (\Gcal^{k,l})^\C 
\slash \Gcal^{k,l}$ there exists a continuous path\footnote{See for instance the note 
under~\cite[Corollary 2.13]{Donaldson:1984}: Since $\glrm(p,\C) \slash \urm (p)$ 
satisfies all necessary conditions for the existence of a unique stationary path 
between any two points, the quotient $(\Gcal^{k,l})^\C \slash \Gcal^{k,l} \cong \prod_{(n,m)}\,
\glrm(p_{(n,m)},\C) \slash \urm (p_{(n,m)}) \times \glrm(n+1,\C) \slash \urm (n+1) 
$ inherits these properties.}
\begin{equation}
 h_\epsilon : \left[\epsilon,\tfrac{1}{\epsilon} \right] \longrightarrow \big(\Gcal^{k,l}\big)^\C \, \big\slash \,
\Gcal^{k,l} \with h(\epsilon) =h_- \and h(\tfrac{1}{\epsilon}) = h_+ \; ,
\end{equation}
which is smooth on $\big(\epsilon,\tfrac{1}{\epsilon} \big)$ and 
minimises the functional $\Lcal_\epsilon$. Hence for any choice of complex gauge transformation $g$ 
such that $g \, g^\dagger =h_\epsilon$, the triple $g\cdot\big( \{ \overline{\mathcal{T}}_{i} 
\}_{i=1,2},0 \big) = \big( \{ \mathrm{Ad}(g)\overline{\mathcal{T}}_{i}  
\}_{i=1,2},\tfrac{\im}{2} \,(\tfrac{\diff g}{\diff s})\, g^{-1} \big) $ satisfies 
the stability equation $\mu\big(\{\mathcal{Y}_j\},\{\, \overline{\mathcal{Y}}_{{j}} \} \big) =0$ on 
$\big(\epsilon, \tfrac{1}{\epsilon}\big)$ for any $0<\epsilon <1$.

The uniqueness of the solution $h_\epsilon$ and its extension to the 
limit $\epsilon \to 0$ follows from~\cite{Donaldson:1984} similarly to the 
proof of~\cite[Lemma 3.17]{Kronheimer:1989}.\footnote{We omit
  a description of the required differential inequality as well as a
  treatment of potential pole contributions from $\lambda(s)$;
  see~\cite[Section~3]{Sperling:2015sra} for a general discussion of
  these issues.} The gauge transformation 
$g_\infty = (h_\infty)^{\frac12}$ is obtained from $h_\infty = \lim_{\epsilon \to 
0} \, h_\epsilon$.
However, the corresponding complex gauge transformation $g=g(h_\epsilon)$ 
is not unique. Similarly to~\cite{Donaldson:1984,Kronheimer:1989}, given a 
solution $\{\,\overline{\mathcal{Y}}_{{j}}  \}_{j=1,2,3}$ of the holomorphicity 
equations one can define two solutions $\{\,\overline{\mathcal{Y}}_{{j}}{}^{\prime} \}_{j=1,2,3}= 
\{g_1 \cdot \overline{\mathcal{Y}}_{{j}} \}_{j=1,2,3}$ and $\{\, \overline{\mathcal{Y}}_{{j}}{}^{\prime\prime} 
\}_{j=1,2,3} = \{g_2 \cdot \overline{\mathcal{Y}}_{{j}} \}_{j=1,2,3}$ of the stability 
equation for any $g_1,g_2\in(\widehat{\Gcal}\,^{k,l})^\C$. By uniqueness one has $g^{\phantom{\dag}}_1 \, g_1^\dagger = g^{\phantom{\dag}}_2 \, g_2^\dagger$; 
thus there exists $\tilde g \in \widehat{\Gcal}\,^{k,l}$ such that $g_1 (s) =  g_2 (s)\, \tilde g(s)$. This ambiguity in the choice of $g=g(h_\epsilon)$ can be 
removed as follows: The complete set of instanton equations is 
invariant under $\widehat{\Gcal}\,^{k,l}$ and a $\widehat{\Gcal}\,^{k,l}$ 
gauge transformation is sufficient to eliminate $\mathcal{X}_6$. Hence one can 
demand that the gauge transformation $\{\,\overline{\mathcal{Y}}_{{j}}{}^{\prime} \}_{j=1,2,3}= 
\{g \cdot \overline{\mathcal{Y}}_{{j}} \}_{j=1,2,3}$ of a solution 
$\{\,\overline{\mathcal{Y}}_{{j}} \}_{j=1,2,3}$ satisfies $\overline{\mathcal{Y}}_{{3}}{}^{\prime}= \mathcal{Y}_{{3}}{}^{\prime}$. This fixes $g=g(h_\epsilon)$ uniquely.
\paragraph{Boundary conditions} 
A trivial solution of~\eqref{eqn:flow_rescaled_X-matrices} is given by
\begin{equation}
 \mathcal{X}_6(s)=0 \and  \mathcal{X}_\mu (s)= T_\mu \with [T_\mu,T_\nu]=0 \for 
\mu,\nu=1,\ldots,5 \; ,
\end{equation}
where $T_\mu$ are constant elements in the Cartan subalgebra $\uoL^p$ of 
$\End_\uo(V^{k,l})$. From the rescaling~\eqref{eqn:rescale_X-matrices}
we then see that the original scalar fields $X_\mu$ are singular at the origin $r= 0$ (corresponding to $\tau \to -\infty 
$ or $ s \to \infty$).
Following~\cite{Kronheimer:1989,Hitchin:1991}, in the generic case we choose 
boundary conditions for 
$X_\mu$ such that\footnote{From now on we will no longer deal with the
  scalar field $X_6$ 
as it can always be gauged away.} $X_\mu(\tau) \to 0$ as $\tau\to+\infty$ for $\mu=1,\dots,5$. Arguing as in~\cite{Kronheimer:1989}, this implies the existence of the 
limit of $\mathcal{X}_{\mu}(s)$ for $s\to 0$ and hence the solutions extend over the 
half-closed interval $\R_{\geq0}$. Since~\eqref{eqn:flow_rescaled_X-matrices} is a system of first order ordinary differential equations, it suffices
to impose one additional boundary condition for the matrices $\mathcal{X}_\mu(s)$ on $[0,\infty)$ which we take to be
\begin{equation}
\label{eqn:boundary_cond}
\lim_{s\to 
\infty}\, \mathcal{X}_{\mu}(s) =\mathrm{Ad}(g_0) T_{\mu} \; ,
\end{equation}
for suitable $g_0 \in \Gcal^{k,l}$ ensuring compatibility with the $\sut$-equivariant structure from~\eqref{eqn:explicit_form_X-matrices} (cf.~Section~\ref{sec:SU(3)-equivariant_cone} for explicit examples). Then the value of $\mathcal{X}_\mu(s)$ at 
$s=0$ is completely determined by the solution.

From~\eqref{eqn:inst_Nahm} it follows that the paths $\overline{\mathcal{Y}}_{{i}}(s)$ for $i=1,2$ each lie respectively in the same 
adjoint orbits $\Ocal_i$ of the complex Lie algebra
$\End_\uo(V^{k,l})^\C$ for all $s\in[0,\infty)$.
Let $\overline{\mathcal{T}}_{i} = \tfrac{1}{2}\, \left( T_{2i-1} + \im \,
T_{2i} \right)$ for $i=1,2$, and denote by $\Ocal_{\overline{\mathcal{T}}_{i}}$ the adjoint orbit of $\overline{\mathcal{T}}_{i}$ in $\End_\uo(V^{k,l})^\C$. Then the boundary conditions~\eqref{eqn:boundary_cond} imply that the closures $\overline{\Ocal_{\overline{\mathcal{T}}_{i}}}$ contain $\Ocal_i$ for $i=1,2$. If the quintuple $\{T_\mu\}_{\mu=1,\dots,5}$ is \emph{regular} in the Cartan subalgebra 
of $\End_\uo(V^{k,l})$, i.e. the joint centraliser of $T_\mu$ in $\Gcal^{k,l}$ is the maximal torus $\uo^p$, then $\overline{\Ocal_{\overline{\mathcal{T}}_{i}}}=\Ocal_{\overline{\mathcal{T}}_{i}}$ are regular orbits and hence $\Ocal_{\overline{\mathcal{T}}_{i}}=\Ocal_i$~\cite{Kronheimer:1989}. By our previous results, there exists a unique complex gauge transformation $g$, which is bounded and framed, such that $\{g\cdot\overline{\mathcal{Y}}_{{j}}\}_{j=1,2,3}$ 
satisfies~\eqref{eqn:inst_stability} and $g\cdot
\overline{\mathcal{Y}}_{{3}}$ is
skew-Hermitian. Employing~\eqref{eqn:inst_commutators}, it follows
that in this case there is a map
\begin{align}
\mathcal{M}_{k,l}^\sut&\longrightarrow \Ocal_{\overline{\mathcal{T}}_{1}}\times
\Ocal_{\overline{\mathcal{T}}_{2}} \notag \\
\big(\{\mathcal{Y}_j(\tau)\}_{j=1,2,3} \,,\,\{\,
  \overline{\mathcal{Y}}_j(\tau)\}_{{j=1,2,3}} \} \big) &\longmapsto
                                                         \big(\,
                                                         \overline{\mathcal{Y}}_1(0),
                                                         \overline{\mathcal{Y}}_2(0)
                                                         \big)
\end{align}
from the
moduli space of solutions satisfying the boundary conditions
\eqref{eqn:boundary_cond} together with the equivariance
condition imposed by our construction. Arguing as in~\cite{Kronheimer:1989}, by our construction of local
solutions to the complex equations, and the existence of a unique
solution to the real equation within the complex gauge orbit of these
elements, this map is a bijection which moreover preserves the
holomorphic symplectic structure. This space is naturally a
complex symplectic manifold of (complex) dimension
$2 \, \dim(\Gcal^{k,l})^\C-\sum_{i=1}^2\,
\dim(\Zcal_{\overline{\mathcal{T}}_{i}})$ with the product of the
standard Kirillov-Kostant-Souriau symplectic forms on the orbits,
where
$\Zcal_{\overline{\mathcal{T}}_{i}}\subset(\Gcal^{k,l})^\C$ is
the subgroup that commutes with $\overline{\mathcal{T}}_{i} $ for
$i=1,2$. By our general constructions it is a K\"ahler manifold. In
the cases that $\sut$-equivariance forces $\overline{\mathcal{T}}_{i}
=0$ for some $i\in\{1,2\}$, the corresponding orbit closure
$\overline{\Ocal_{\overline{\mathcal{T}}_{i}}}$ should be replaced by
the nilpotent cone $\Ncal$ of dimension
$\dim(\Gcal^{k,l})^\C-p$ which consists of all nilpotent elements of
$\End_\uo(V^{k,l})^\C$. The variety $\Ncal$ has singularities
corresponding to non-regular nilpotent orbits, and in particular it contains the locus of Kleinian singularities $\C^2/\Z_p$ in complex codimension~$2$; see~\cite{Lechtenfeld:2014fza} for further
details. Thus in this case the moduli space is \emph{singular} and by
$\sut$-equivariance we expect that it contains the singular subvariety $\C^3/\Z_p$. 
% 
%%%%%%%%%%%%%%%%%%%%%%%%%%%%%%%%%%%%%%%%%%%%%%%%%%%%%%%%%%%%%%%%%%%%%%%%%%%%%%%%
%%%%%%%%%%%%%%%%%%%%%%%%%%%%%%%%%%%%%%%%%%%%%%%%%%%%%%%%%%%%%%%%%%%%%%%%%%%%%%%%
% 
\subsubsection{$\C^3$-invariance}
\label{subsec:trans-inv_case}
Now we turn our attention to the space of translationally-invariant connections 
$\mathbb{A}(\mathfrak{E}^{k,l})$ on the bundle~\eqref{eqn:bundle_TransInv}. The 
structure group $\mathfrak{G}^{k,l}$ of~\eqref{eqn:bundle_TransInv}
(which in this case coincides with the gauge group) is given
by~\eqref{eqn:gauge_group_transinv} and its Lie algebra $\gfrak^{k,l}$
by~\eqref{eqn:gauge_alg_transinv}. A generic element of the tangent space $T_\Acal 
\mathbb{A}(\mathfrak{E}^{k,l})$ at a point $\Acal \in 
\mathbb{A}(\mathfrak{E}^{k,l})$ is given by 
\begin{equation}
 W= W_\a \, \diff z^\a  +  \overline{W}_{{\a}} \, \diff
 \bar{z}^{{\a}} \ \in \ 
\Omega^{1}\big( \C^3/\ZK, \gfrak^{k,l} \big) \; ,
\end{equation}
with constant $W_\a,\overline{W}_{{\a}}$ for $\a=1,2,3$.
As before, let us define a metric $\boldsymbol{g}$ on 
$\mathbb{A}(\mathfrak{E}^{k,l})$. Gauge transformations of tangent vectors $\boldsymbol{w} 
=\boldsymbol{w}_{\a} \, \diff z^\a + \overline{\boldsymbol{w}}_{{\a}}
\, \diff \bar{z}^\a $ are given by
$\overline{\boldsymbol{w}}_{{\a}} \mapsto \mathrm{Ad}(g)\overline{\boldsymbol{w}}_{{\a}} $ 
for $\a = 1,2,3$. We deduce the metric to be
\begin{equation}
 \boldsymbol{g}_{|\Acal}(\boldsymbol{w},\boldsymbol{v}):= \frac{1}{2} \,
\sum_{\a=1}^3 \,\tr 
\big(\boldsymbol{w}_{\a}^\dagger \, \boldsymbol{v}_{\a}^{\phantom{\dagger}}  + 
\boldsymbol{w}_{\a}^{\phantom{\dagger}}\, \boldsymbol{v}_{\a}^\dagger
\big)  \ ,
\end{equation}
and a symplectic form via
\begin{equation}
   \boldsymbol{\omega}_{|\Acal}(\boldsymbol{w},\boldsymbol{v}) := 
\frac{\im}{2} \, \sum_{\a=1}^3 \, \tr 
\big( \boldsymbol{w}_{\a}^\dagger \, \boldsymbol{v}_{\a}^{\phantom{\dagger}}  - 
\boldsymbol{w}_{\a}^{\phantom{\dagger}} \, \boldsymbol{v}_{\a}^\dagger \big) \; .
\end{equation}
These definitions follow directly from the translationally-invariant limit
of~\eqref{eqn:metric_connections} and \eqref{eqn:symplectic_form_connections} 
(and agree with those of~\cite{SardoInfirri:1996ga}). Evidently 
the metric and symplectic structure are gauge-invariant.

Define the subspace of invariant connections that satisfy the 
holomorphicity conditions~\eqref{eqn:gen_instanton_1} as
\begin{equation}
 \mathbb{A}^{1,1}\big(\mathfrak{E}^{k,l} \big) = \Big\{
 \big(\{W_\a\},\{\, \overline{W}_{{\a}}\}\big) \in 
\mathbb{A}\big(\mathfrak{E}^{k,l}\big) \, : \, \big[\, \overline{W}_{{\alpha}}, \overline{W}_{{\beta}} 
\big] =0 \for \a,\b = 1,2,3 \Big\} \; ,
\end{equation}
which is a finite-dimensional Kähler space by the general considerations of 
Section~\ref{subsec:moduli_generic}.
\paragraph{Moment map}
The corresponding moment map can be introduced as before via
\begin{equation}
\begin{split}
 \mu : \mathbb{A}^{1,1}\big(\mathfrak{E}^{k,l}\big) &\longrightarrow
 \mathfrak{g}^{k,l} \\
\big(\{W_\a\},\{\, \overline{W}_{{\a}}\}\big) &\longmapsto
\im\,\sum_{\a=1}^3\, \big[W_\a , \overline{W}_{\a} \big] \; ,
\end{split}
\label{eqn:moment_map_transinv}
\end{equation}
but in this case it is possible to choose various gauge-invariant levels 
$\Xi$ from \eqref{eqn:levels} and consequently define different moduli spaces
\begin{equation}
 \mathcal{M}_{k,l}^{\C^3}(\Xi) = \mu^{-1}(\Xi) \, \big\slash\,
 \mathfrak{G}^{k,l} \; .
\end{equation}
\paragraph{Real gauge group}
The complete set of instanton equations~\eqref{eqn:gen_instanton_1} 
and~\eqref{eqn:gen_instanton_2} is invariant under the action of the gauge 
group~\eqref{eqn:gauge_group_transinv} with the usual transformations
\begin{equation}
 \overline{W}_{{\alpha}} \longmapsto \mathrm{Ad}(g)\overline{W}_{{\alpha}} \for \a=1,2,3
\end{equation}
for $g\in \mathfrak{G}^{k,l}
\hookrightarrow \urm(p)$.
\paragraph{Complex gauge group}
Recalling that the holomorphicity conditions allow for the introduction of a 
$(\mathfrak{G}^{k,l})^\C$-bundle, we find that the corresponding equations are 
invariant under $(\mathfrak{G}^{k,l})^\C$ gauge transformations. Again the stability
equation is not invariant under the action of the complex gauge group.
\paragraph{Stable points}
The set of stable points is defined as before to be
\begin{equation}
\mathbb{A}^{1,1}_{\rm st}\big(\mathfrak{E}^{k,l};\Xi \big) := \Big\{ 
\big(\{W_\alpha\},\{\,\overline{W}_{{\alpha}}\} \big)\in 
\mathbb{A}^{1,1}\big(\mathfrak{E}^{k,l}\big) \, : \,
\big(\mathfrak{G}^{k,l} \big)^\C_{(\{W_\alpha\},\{\,\overline{W}_{{\alpha}} \})} \cap 
\mu^{-1}(\Xi)\neq \emptyset  \Big\}  \; ,
\end{equation}
and by taking the GIT quotient one obtains the $\Xi$-dependent moduli spaces\footnote{This description is analogous to the quiver GIT quotients used by~\cite{Cirafici:2010bd,SardoInfirri:1996gb} to describe instanton moduli on $\C^3/\Z_{q+1}$ as representation moduli of the McKay quiver.}
\begin{equation}
\mathcal{M}_{k,l}^{\C^3}(\Xi) \cong \mathbb{A}^{1,1}_{\rm
  st}\big(\mathfrak{E}^{k,l};\Xi \big) \, \big\slash \,
\big(\mathfrak{G}^{k,l} \big)^\C \ .
\end{equation}
The moment map~\eqref{eqn:moment_map_transinv} 
transforms under $g\in (\mathfrak{G}^{k,l})^\C$ as
\begin{equation}
\mu \big(\{W_\alpha\},\{\,\overline{W}_{{\alpha}} \} \big) =\im\, \sum_{\alpha=1}^3\, \big[W_\alpha , 
\overline{W}_{{\alpha}} \big] \longmapsto \im\, \mathrm{Ad}(g)
\sum_{\alpha=1}^3 \, \big[ h^{-1} \,
W_\alpha \, h , \overline{W}_{{\alpha}} \big] \; ,
\end{equation}
where we introduced $h=h(g) = g^\dagger\, g \in (\mathfrak{G}^{k,l})^\C \slash 
\mathfrak{G}^{k,l}$. Similarly to before, $h$ can be
identified with a positive Hermitian $p\times p$
matrix. Moreover, $\mathrm{Ad}(g'\,) \Xi = \Xi$ for any $g' \in 
\mathfrak{G}^{k,l}$. By the embedding $\mathfrak{G}^{k,l} \hookrightarrow \urm(p)$ and the 
polar decomposition of an element $g \in (\mathfrak{G}^{k,l})^\C$ into $g= h'\,
\exp(\im \, X)$ for Hermitian $h' \in \mathfrak{G}^{k,l}$ and skew-adjoint $X \in 
\mathfrak{g}^{k,l}$, we have
\begin{equation}
 \mathrm{Ad} (g) \Xi = \mathrm{Ad}(h'\,) \big( 
\mathrm{Ad}(\exp{(\im \, X)}) \Xi \big)= \mathrm{Ad}(h'\, ) \big( \Xi + \im \,
\left[X,\Xi \right] \big) =\mathrm{Ad}(h'\,) \Xi = \Xi \; ,
\end{equation}
where we used the Baker-Campbell-Hausdorff formula and 
the fact that $\Xi$ is central in $\mathfrak{g}^{k,l}$. It follows that 
$\mathrm{Center}\big(\gfrak^{k,l} \big) \subset
\mathrm{Center}\big((\gfrak^{k,l})^\C \big)$. Hence a point 
$\big(\{W_\alpha\},\{\, \overline{W}_{{\alpha}}\} \big)\in \mathbb{A}^{1,1}(\mathfrak{E}^{k,l}) $ is 
stable if and only if there exists a positive Hermitian matrix $h$ (modulo unitary transformations) 
satisfying the equation
\begin{equation}
  \sum_{\alpha=1}^3 \, \big[ h^{-1} \, W_\alpha \, h ,
  \overline{W}_{{\alpha}} \big]  = -\im\, \Xi \ .
\end{equation}

By our general constructions the moduli spaces
$\Mcal_{k,l}^{\C^3}(\Xi)$ are K\"ahler spaces, which however are
generically \emph{not} smooth manifolds but have a complicated scheme
structure with branches of varying dimension that should be analysed within the context of a perfect
obstruction theory; such an analysis is beyond the scope of the
present paper. Generally, the canonical map $\Mcal_{k,l}^{\C^3}(\Xi)\to \Mcal_{k,l}^{\C^3}(0)$ is a partial resolution of singularities for generic $\Xi$. For example, in the case $p_{(n,m)}=1$ for all $(n,m)\in Q_0(k,l)$ (so that $V^{k,l}\cong\RepSu{k}{l}$ and $p=p_0$), for generic levels
$\Xi\neq0$ the moduli spaces $\Mcal_{k,l}^{\C^3}(\Xi)$ are schemes akin to the $\Z_p$-Hilbert scheme of $p=\dim(\,\RepSu{k}{l}\,)$
points on $\C^3$ for the $\Z_p$-action given
by~\eqref{eqn:ZK-action} (with $q=p-1$), which
are partial resolutions of the singular spaces $\Mcal_{k,l}^{\C^3}(0)$
that correspond to configurations of $p$ points
of $\C^3$ given as unions of $\Z_p$-orbits
(cf.~\cite{Cirafici:2010bd,SardoInfirri:1996ga} for the case of the
McKay quiver);\footnote{In the special case $q=1$ the $\Z_2$-Hilbert
  scheme is the product $M_1\times\C$ where $M_1$ is the total space
  of the canonical line bundle $\Ocal_{\C P^1}(-2)\to\C P^1$ (Eguchi-Hanson space), whereas for $q=2$ the $\Z_3$-Hilbert
scheme is the total space of the canonical line bundle $\Ocal_{\CP}
(−3)\to\CP$ (local del~Pezzo surface of degree~$0$).} these are precisely the same types of singularities encountered in the moduli spaces $\Mcal_{k,l}^\sut$ above.
%%%%%%%%%%%%%%%%%%%%%%%%%%%%%%%%%%%%%%%%%%%%%%%%%%%%%%%%%%%%%%%%%%%%%%%%%%%%%%%%
% 
\bigskip \section*{Acknowledgements}

\noindent
This work was partially supported by the Grant 
LE 838/13 from the Deutsche Forschungsgemeinschaft (DFG, Germany), 
by the
Consolidated Grant ST/L000334/1 from the UK Science and Technology
Facilities Council (STFC), and by the Action MP1405 QSPACE from the European Cooperation in Science and Technology
(COST).
% 
%%%%%%%%%%%%%%%%%%%%%%%%%%%%%%%%%%%%%%%%%%%%%%%%%%%%%%%%%%%%%%%%%%%%%%%%%%%%%%%%
%   
\begin{appendix}
\bigskip \section{Bundles on $\CP$}
\label{subsec:geometry_CP}
\subsection{Geometry of $\CP$}
\paragraph{$\sut$-equivariant 1-forms}
Consider the row vector $\beta^\top = ( \beta^1, \beta^2)$. The 
relations~\eqref{eqn:definitions} and~\eqref{eqn:identities} dictate the explicit 
form of the $1$-forms $\beta^i$ and their exterior derivatives as
\begin{subequations}
\label{eqn:diffs_1-forms}
\begin{alignat}{2}
\beta^i &= \frac{1}{\gamma} \, \diff y^i - \frac{1}{\gamma^2\, (\gamma 
+ 1)}\, y^i \, \sum_{j=1}^2\, \bar{y}^j \, \diff y^j \; , &\quad 
 \bar{\beta}^i &= \frac{1}{\gamma} \, \diff \bar{y}^i -
 \frac{1}{\gamma^2 \, (\gamma 
+ 1)} \, \bar{y}^i \, \sum_{j=1}^2\, y^j \,\diff \bar{y}^j \; ,\\[4pt]
 \diff \beta^1 &= -\beta^1 \wedge \big(B_{11} + \tfrac{3}{2}\, a \big) + 
\beta^2 \wedge \bar{B}_{12} \; , &\quad 
\diff \beta^2 &= - \beta^1 \wedge B_{12} + \beta^2  \wedge \big( B_{11} - 
\tfrac{3}{2} \,a \big) \; ,\\[4pt]
 \diff \bar{\beta}^1 &= - \big( B_{11} + \tfrac{3}{2} \, a \big) \wedge 
\bar{\beta}^1  -  B_{12} \wedge \bar{\beta}^2 \; , &\quad 
\diff \beta^2 &= \bar{B}_{12} \wedge \bar{\beta}^1  + \big( B_{11} - 
\tfrac{3}{2} \, a \big) \wedge \bar{\beta}^2 \; .
\end{alignat}
\end{subequations}
One can regard $\beta^i$ as a basis for the $(1,0)$-forms and $\bar{\beta}^i$ 
as a basis for the $(0,1)$-forms of the complex cotangent bundle
over the patch $\mathcal{U}_0$ of 
$\mathbb{C}P^2$ with respect to an almost complex structure $J$. The canonical $1$-forms $\diff 
y^i$ and $\diff \bar{y}^i$ could equally well be used for a
holomorphic decomposition 
with respect to $J$, but the forms $\beta^i$ and $\bar{\beta}^i$ are 
$\sut$-equivariant.

\paragraph{Hermitian Yang-Mills equations}
The canonical Kähler $2$-form on the patch $\mathcal{U}_0$ is given by
\begin{equation}
 \omega_{\mathbb{C}P^2} = -\im \, R^2 \, \beta^\top \wedge \bar{\beta} = 
\im \, R^2 \, \left( \beta^1\wedge \bar{\beta}^1 + \beta^2 \wedge 
\bar{\beta}^2 \right) \; ,\label{eqn:canonical_Kaehler_CP2}
\end{equation}
where $R$ is the radius of the linearly embedded projective line $\C P^1\subset \CP$.
The $1$-form $B_{(1)}$ is then an 
instanton connection by the following argument: Locally, one can define a $(2,0)$-form 
$\Omega$ proportional to $ \beta^1 \wedge \beta^2$. The Hermitian Yang-Mills 
equations for a curvature 
$2$-form $F$ are 
  \begin{equation}
   \Omega \wedge F =0 \and \omega_{\CP}\, \lrcorner\, F=0 \; ,
  \end{equation}
which translate to $F = F^{1,1} $ being a $(1,1)$-form for which 
$\tr(F^{1,1})=0$; here the contraction $\lrcorner$ between two forms $\eta$ 
and $\eta'$ is defined as $\eta\, \lrcorner\, \eta' := \star \left(\eta \wedge \star\, \eta'\, \right)$. 
The curvature $F_B= \diff B + B \wedge B = \bar{\beta} \wedge \beta^\top 
$ is a $(1,1)$-form which is
$\utwoL$-valued, i.e. $\tr(F_B)=2a \neq 0$. However
$F_a= 
\diff a =  \beta^\dagger \wedge \beta$ is also a $(1,1)$-form. Thus the curvature of the connection $B_{(1)} = B - 
\frac{1}{2}\, a\, \mathds{1}_2$ given 
by $F_{B_{(1)}}= F_B - \frac{1}{2}\, F_a\, \mathds{1}_2$ is a 
$(1,1)$-form and by construction traceless; hence $B_{(1)}$ is an
$\mathfrak{su}(2)$-valued 
connection satisfying the Hermitian Yang-Mills equations, i.e. it is
an instanton connection.
\subsection{Hopf fibration and associated bundles}
Consider the principal $\uo $-bundle $S^5 = \sut / \su \rightarrow 
\CP$. One can associate to it a complex vector 
bundle whose fibres carry any representation of the structure group $\uo$,
i.e. a complex vector space $V$ together with a group homomorphism $\rho : \uo  \to \glrm(V)$. 
Then the associated vector bundle $E$ is given as $E \coloneqq S^5 \times_\rho 
V \to \CP$. In particular, one can choose $V = \Reps{m}$
to be the one-dimensional irreducible representation of highest weight
$m\in\Z$. Following~\cite{Dolan:2009nz}, one then generates associated complex line bundles 
$L_{\frac m2} \coloneqq ( L^{\otimes m} )^{\frac{1}{2}}$. 
\paragraph{Chern classes and monopole charges}
Using the conventions of~\cite{Dolan:2009nz} for $\CP$, there is a
normalised volume form
\begin{equation}
\beta_{\rm vol} \coloneqq 
\frac{1}{2 \pi^2} \, \beta^1 \wedge \bar\beta^1 \wedge \beta^2 \wedge 
\bar\beta^2 \with \int_{\mathbb{C}P^2} \, \beta_{\rm vol} = 1 \ ,
\end{equation} 
and the canonical Kähler $2$-form~\eqref{eqn:canonical_Kaehler_CP2} with 
\begin{equation}
\omega_{\CP} \wedge 
\omega_{\CP} = - \big(2 \pi \, R^2 \big)^2 \, \beta_{\rm vol} \ . 
\label{eqn:omegaCPsquared}\end{equation}
Consider the connection $a$ from~\eqref{eqn:def_monopole_conn} on the line bundle 
$L$ associated to the Hopf bundle $S^5\to\CP$ and the fundamental representation. Since its curvature is $F_a= \tfrac{\im}{R^2} \, \omega_{\CP}$, the total Chern character of the monopole bundle $L$ is
\begin{equation}
 \mathrm{ch}(L) = \exp{\big( \tfrac{\im}{2 \pi} \, F_a
\big) } = \exp(\xi)
\end{equation}
where $\xi:= -\frac1{2 \pi \, R^2}\, \omega_{\CP}$. Then one immediately
reads off the first Chern class
\begin{equation}
 c_1(L) = \xi  \with \int_{\CP} \, \xi\wedge\xi = -1 \ .
\end{equation}
Since $[\xi]=[c_1(L)]$ generates $H^2(\mathbb{C}P^2,\mathbb{Z})
\cong\Z$~\cite{Lechtenfeld:2008nh}, this identifies the first Chern
number of $L$ as $-1$. Thus
$L\equiv L_1$ exists globally, and the dual bundle $L_{-1}$ has first Chern class 
$c_1(L_{-1})=-c_1(L)$ and hence first Chern number $+1$. For all other bundles $L_{\frac{m}{2}}$ one takes the connection to 
be $\frac{m}{2} \, a$, which changes the first Chern class
accordingly to
\begin{equation}
 c_1\big(L_{\frac m2} \big) = \tfrac{m}{2}\, \xi \ ,
\end{equation}
and the first Chern number to $-\frac m2$. 
Hence only for even values of $m$ do the line 
bundles $L_{\frac m2}$ exist globally in the sense of conventional bundles.
On the other hand, for odd values of $m$ the line bundles $L_{\frac
  m2}$ (and also the instanton bundles $I_n$ for odd values of the
isospin $n$~\cite{Dolan:2009nz}) are examples of 
\emph{twisted bundles}. The obstruction to the global existence of
these bundles is the failure of 
the cocycle condition for transition functions on triple overlaps of patches, which is given by a non-trivial integral $3$-cocycle representing the Dixmier-Douady class of an abelian gerbe; see for
example~\cite{Murray:2003pk} for more details. As argued
in~\cite{Dolan:2009nz}, the Chern number $\frac m2$ of the line bundle
$L_{-\frac m2}$ 
should be taken as the monopole charge rather than the $H_{\alpha_2}$-eigenvalue $m$ 
in the Biedenharn basis. 
\bigskip \section{Representations}
\subsection{Biedenharn basis}
\label{sec:Rep_Theory_SU3}
Let us summarise the relevant details we need concerning the Biedenharn 
basis~\cite{Biedenharn:1962haa,Baird:1963wv,Mukunda:1965}, which is defined as the
basis of eigenvectors according to~\eqref{eqn:def_Biederharn}; we 
follow~\cite{Lechtenfeld:2008nh,Dolan:2009nz} for the presentation and notation.
\paragraph{Generators}
The remaining generators of $\sutL$ act on this eigenvector basis as
\begin{subequations}
\begin{align}
  E_{\pm \, \alpha_{1}} \ket{n}{q}{m} &= \frac{1}{2} \, \sqrt{(n \mp q)\,
                                     (n \pm q +2) } \,
\ket{n}{q\pm2}{m} \; , \\[4pt]
  E_{\alpha_{2}} \ket{n}{q}{m} &= \sqrt{ \tfrac{n-q-2}{2(n+1)} } \,
\LaPlus{n,m} \, \ket{n+1}{q-1}{m+3} + \sqrt{ \tfrac{n+q}{2(n+1)} } \,
\LaMinus{n,m} \, \ket{n-1}{q-1}{m+3} \; ,\\[4pt]
 E_{\alpha_1 +\alpha_{2}} \ket{n}{q}{m} &= \sqrt{ \tfrac{n+q+2}{2(n+1)} } \,
\LaPlus{n,m} \, \ket{n+1}{q+1}{m+3} + \sqrt{ \tfrac{n-q}{2(n+1)} } \,
\LaMinus{n,m} \, \ket{n-1}{q+1}{m+3} \; ,
\end{align}
\end{subequations}
with $E_{\alpha_{2}}^\dagger = E_{\alpha_{2}}^\top =
E_{-\alpha_{2}}$ and 
$E_{\alpha_1 + \alpha_{2}}^\dagger = E_{\alpha_1 + \alpha_{2}}^\top =  
E_{-(\alpha_1 + \alpha_{2})}$. It is convenient to express the generators as
\begin{subequations}%
\label{eqn:matrix_elements_E1_E1+2}%
\begin{align}%
 E_{\alpha_1 + \alpha_2}^{+\, (n,m)} &=  
 \sum_{q \in Q_n} \, \sqrt{ \tfrac{n+q+2}{2(n+1)} } \,
\LaPlus{n,m} \, \ket{n+1}{q+1}{m+3} \bra{n}{q}{m} \; ,
 \\[4pt]
 E_{\alpha_1 + \alpha_2}^{-\,(n,m)} &= 
\sum_{q \in Q_n} \, \sqrt{ \tfrac{n-q}{2(n+1)} } \,
\LaMinus{n,m} \, \ket{n-1}{q+1}{m+3} \bra{n}{q}{m} \; , \\[4pt]
E_{\alpha_2}^{+\, (n,m)} &= 
\sum_{q \in Q_n} \, \sqrt{ \tfrac{n-q-2}{2(n+1)} } \,
\LaPlus{n,m} \, \ket{n+1}{q-1}{m+3} \bra{n}{q}{m} \; ,\\[4pt]
 E_{\alpha_2}^{-\,(n,m)} &=
 \sum_{q \in Q_n} \, \sqrt{ \tfrac{n+q}{2(n+1)} } \,
\LaMinus{n,m} \, \ket{n-1}{q-1}{m+3} \bra{n}{q}{m} \; ,
\end{align}
\end{subequations}
where $Q_n := \{ -n, -n+2, \ldots, n-2,n \}$ and
\begin{subequations}%
\label{eqn:def_Lambda}%
\begin{align}
 \LaPlus{n,m} &= \frac{1}{\sqrt{n+2}} \, \sqrt{\left(\tfrac{k+2l}{3} + 
\tfrac{n}{2} + \tfrac{m}{6} + 2 \right) \, \left(\tfrac{k-l}{3} + \tfrac{n}{2} + 
\tfrac{m}{6} + 1 \right) \, \left(\tfrac{2k+l}{3} - \tfrac{n}{2} - 
\tfrac{m}{6} \right)   } \; , \\[4pt]
 \LaMinus{n,m} &= \frac{1}{\sqrt{n}} \, \sqrt{\left(\tfrac{k+2l}{3} - 
\tfrac{n}{2} + \tfrac{m}{6} + 1 \right) \, \left(\tfrac{l-k}{3} + \tfrac{n}{2} - 
\tfrac{m}{6} \right) \, \left(\tfrac{2k+l}{3} + \tfrac{n}{2} - 
\tfrac{m}{6} +1\right)   } \; ,
\end{align}
\end{subequations}
with $\LaMinus{0,m}:=0$~\cite{Dolan:2009nz}.
The identity operator $\UniEnd{n,m}$ of the representation
$\underline{(n,m)}$ is given by 
\begin{equation}
 \UniEnd{n,m} = \sum_{q \in Q_n} \, \ket{n}{q}{m} \bra{n}{q}{m} \; . %
 \label{eqn:def_UniEnd}
\end{equation}
\paragraph{Fields}
The $1$-instanton connection~\eqref{eqn:def_B-one} is 
represented in the Biedenharn basis by
\begin{equation}
\begin{split}
 B_{(1)} &= B_{11} \, H_{\alpha_1}  + B_{12} \, E_{\alpha_{1}} - \left( 
B_{12} \,E_{ \alpha_{1}} \right)^\dagger \\[4pt]
&= \sum_{n,q,m} \, \Big(\, B_{11} \, q \, \ket{n}{q}{m} \bra{n}{q}{m} + \frac{1}{2} \,
B_{12} \, \sqrt{(n-q)\, (n+q+2)} \, \ket{n}{q+2}{m} \bra{n}{q}{m} \\
 &\phantom{ = \sum_{n,q,m} \, \Big(\, B_{11} \, q \, \ket{n}{q}{m} \bra{n}{q}{m} + 
} - \frac{1}{2}\, \bar{B}_{12} \, \sqrt{(n+q)\, (n-q+2)} \, \ket{n}{q-2}{m} 
\bra{n}{q}{m} \, \Big) \\
&\equiv \bigoplus_{(n,m)\in Q_0(k,l)}\,  B_{{(n,m)}} \; ,
\end{split}
\end{equation}
where $B_{{(n,m)}} \in \Omega^{1}\big(\suL 
,\End\big(\,\Reps{(n,m)} \, \big) \big)$.
One further introduces matrix-valued $1$-forms given by
\begin{subequations}%
\label{eqn:def_beta-matrix_forms}%
\begin{equation}
 \barbetaZK = \barbetaZK^1\,  E_{\alpha_1 + \alpha_2} + \barbetaZK^2 \,
E_{\alpha_2} \equiv \bigoplus_{(n,m)\in Q_0(k,l)} \, \left( 
\BarBetaPlusi{n,m} + 
\BarBetaMinusi{n,m} \right) \; ,
\end{equation}
with the morphism-valued $1$-forms
\begin{align}
 \BarBetaPMi{n,m} \in 
\Omega^1\Big(S^5 \slash \ZK\,,\,\Hom\big(\, \Reps{(n,m)}\,,\,
  \Reps{(n\pm1,m+3)}\, \big) 
\Big) \; , 
\end{align}
and the corresponding adjoint maps 
\begin{align}
 \BetaPMi{n,m} \in 
\Omega^1\Big(S^5 \slash 
\ZK\,,\,\Hom\big(\, \Reps{(n\pm 1,m+3)}\,,\,\Reps{(n,m)}\,\big) \Big) \; .
\end{align}
\end{subequations}
They have the explicit form
\begin{equation}
\begin{split} 
\BarBetaPMi{n,m} = \frac{\LaPM{n,m}}{\sqrt{2(n+1)}} \ \sum_{q \in Q_n} \,
\Big(  &\sqrt{n\pm q + 1 \pm 1 } \ \bar{\beta}^1_{q+1}  \,
\ket{n\pm 1}{q+1}{m+3}\bra{n}{q}{m} 
 \\*
&+ \sqrt{n\mp q + 1 \pm 1 } \ \bar{\beta}^2_{q+1} \, \ket{n\pm1}{q-1}{m+3}\bra{n}{q}{m} 
\Big) \; .
\end{split}
\end{equation}
\paragraph{Skew-Hermitian basis}
Similarly to~\cite{Popov:2010rf}, for a given representation 
$\RepSu{k}{l}$ of the generators $I_i$ and $I_\mu$ defined 
in~\eqref{eqn:alternative_basis_SU(3)} the decomposition into the Biedenharn 
 basis yields
% % 
% 
\begin{subequations}
\begin{alignat}{2}
 I_1 &= \bigoplus_{(n,m)} \, \Geni{1}{(n,m)}& &= \bigoplus_{\pm\,,\,(n,m)} \,
\left( E_{\alpha_1 + 
\alpha_2}^{\pm\, (n,m)} - E_{-\alpha_1 - 
\alpha_2}^{\pm\, (n,m)} \right) \; , \\[4pt]
I_2 &= \bigoplus_{(n,m)} \, \Geni{2}{(n,m)} & &= - \im \, \bigoplus_{\pm\,,\,(n,m)} \,
\left( E_{\alpha_1 + \alpha_2}^{\pm\,(n,m)} + E_{-\alpha_1 - 
\alpha_2}^{\pm\,(n,m)}  \right) \; ,\\[4pt]
 I_3 &= \bigoplus_{(n,m)} \, \Geni{3}{(n,m)}& &= \bigoplus_{\pm\,,\,
   (n,m)} \, \left( 
E_{\alpha_2}^{\pm\,(n,m)} 
- E_{ - \alpha_2}^{\pm\,(n,m)}  \right) \; , \\[4pt]
I_4 &= \bigoplus_{(n,m)} \, \Geni{4}{(n,m)}& &= - \im \, \bigoplus_{\pm\,,\,(n,m)} \,
\left( E_{\alpha_2}^{\pm\,(n,m)} + 
E_{ - \alpha_2}^{\pm\,(n,m)}  \right)  \; ,\\[4pt]
I_5 &= \bigoplus_{(n,m)}\, I_5^{(n,m)}& &= -\frac{\im}{2} \, 
\bigoplus_{(n,m)} \, H_{\alpha_2}^{(n,m)} \; .
\end{alignat}
\end{subequations}
The commutation relations $\left[I_i, I_a \right] = f_{ia}^{\ \ b} \, I_b $
and $\left[I_i, I_5 \right] = 0$ induced
by~\eqref{eqs:structure_constants} respectively imply relations among the
components given by
\begin{subequations}
\label{eqn:comm_relations}
 \begin{align}
I_i^{(n',m')} \, I_a^{(n,m)} &=
I_a^{(n,m)} \, I_i^{(n,m)} + f_{i a}^{\ \ b} \,
I_b^{(n,m)} \; , \label{eqn:comm_relation1}\\[4pt] 
I_i^{(n,m)} \, I_5^{(n,m) } &= 
I_5^{(n,m) } \, I_i^{(n,m)} \; , \label{eqn:comm_relation2}
 \end{align}
\end{subequations}
where $i \in \{ 6,7,8\}$, $a \in \{ 1,2,3,4\}$, $I_i = \bigoplus_{(n,m)} \,
I_i^{(n,m)} $ and $(n',m'\,)=(n\pm1,m+3)$.
\subsection{Flat connections}
\label{sec:flat_conn_appendix}
One can compute the matrix elements of $\Acal_0$ 
from~\eqref{eqn:ansatz_flat_conn_all} with respect to 
the Biedenharn basis. By choosing an $\sut$-representation 
$\RepSu{k}{l} $, which induces an $\su$-representation by restriction, one induces a connection $\Acal_0$ on the
vector V-bundle
\begin{equation}
 \widetilde{\mathcal{V}}_{\RepSu{k}{l}}  \xrightarrow{ 
\RepSu{k}{l} } \mathrm{G}\slash \widetilde{\mathrm{K}} \with 
\mathcal{V}_{\RepSu{k}{l}} \coloneqq \mathrm{G}  \times_{\mathrm{K}} 
\RepSu{k}{l}  
\label{eqn:associated_bundle_C(k,l)}
\end{equation}
associated to the principal V-bundle~\eqref{eqn:bundle_SU3_S5_ZK}. Then the 
connection $\Acal_0$ can be decomposed into morphism-valued $1$-forms
\begin{equation}
 \Acal_0 = \bigoplus_{(n,m)\in Q_0(k,l)}\, \Big( \Bfield{n,m} - 
\frac{\im \,m }{2} \, \eta \, \UniEnd{n,m} +
\BarBetaPlusi{n,m} + 
\BarBetaMinusi{n,m} - \BetaPlusi{n,m} - \BetaMinusi{n,m} \Big)
\end{equation}
with respect to this basis. The computation of the vanishing curvature 
$\Fcal_0=0$ yields 
relations between the different matrix elements given by
\begin{subequations}
\label{eqn:flat_connection_generic}
\begin{align}
&\begin{aligned}
   \diff \Bfield{n,m} + \Bfield{n,m} 
\wedge \Bfield{n,m} - \tfrac{\im\, m}{2} \, \diff \eta \, \UniEnd{n,m} 
=\; &\BarBetaPlusi{n-1,m-3} \wedge \BetaPlusi{n-1,m-3}+ \BarBetaMinusi{n+1,m-3} \wedge \BetaMinusi{n+1,m-3}  \\*
  &+ \BetaPlusi{n,m} \wedge \BarBetaPlusi{n,m} + \BetaMinusi{n,m} \wedge \BarBetaMinusi{n,m} \; ,
 \end{aligned}\\[4pt]
&0=  \diff  \BarBetaPMi{n,m} + \Bfield{n+1,m+3} \wedge 
\BarBetaPMi{n,m} + \BarBetaPMi{n,m} \wedge \Bfield{n,m} - \tfrac{3\,
  \im}{2} \, \eta \, \UniEnd{n\pm1,m+3} \wedge \BarBetaPMi{n,m} \; , \\[4pt]
 &0= \BarBetaPlusi{n,m} \wedge \BarBetaMinusi{n+1,m-3} + 
\BarBetaMinusi{n+2,m} \wedge \BarBetaPlusi{n+1,m-3} \; , \\[4pt]
 &0=\BarBetaPlusi{n,m} \wedge \BetaMinusi{n,m} + 
\BetaMinusi{n+1,m+3} \wedge \BarBetaPlusi{n-1,m+3}  \; ,\\[4pt]
 &0= \BarBetaPMi{n,m} \wedge \BarBetaPMi{n\mp 1,m-3} \; ,
\end{align}
\end{subequations}
plus their conjugate equations.
\subsection{Quiver connections}
\label{sec:Quiver_conn_appendix}
One can also compute the matrix elements of the curvature~\eqref{eqn:curv_quiver_matrix} in the Biedenharn basis. For this, the curvature 
$\Fcal= \diff \Acal + \Acal \wedge \Acal$ is arranged into components
\begin{equation}
 \CurvQui{n,m}{n',m'} \in \Omega^2\Big(\mathcal{E}^{k,l}\,,\, 
\End\big(E_{p_{(n,m)}},E_{p_{(n',m')}}\big)\otimes 
\End\big(\,\Reps{(n,m)}\,,\,\Reps{(n',m'\,)}\,\big)\Big) \; ,
\end{equation}
which can be simplified by using the 
relations~\eqref{eqn:flat_connection_generic}. 
We denote the curvature of the connection $\ConnQui{n,m}$ 
on the bundle~\eqref{eqn:def_hermitian_bundles} by
 \begin{subequations}
  \begin{equation}
   \FConnQui{n,m}\coloneqq \diff \ConnQui{n,m} + \ConnQui{n,m} \wedge 
\ConnQui{n,m}
  \end{equation}
and the bifundamental covariant derivatives of the Higgs fields as
  \begin{align}
   D \HomoPhiPMi{n,m} &\coloneqq \diff  \HomoPhiPMi{n,m} + 
\ConnQui{n\pm1,m+3} \, \HomoPhiPMi{n,m} 
- \HomoPhiPMi{n,m} \, \ConnQui{n,m} \; ,\\[4pt]
  D \EndPsiQui{n,m} &\coloneqq \diff \EndPsiQui{n,m} + \ConnQui{n,m} \,
\EndPsiQui{n,m} - \EndPsiQui{n,m} \, \ConnQui{n,m} \; .
  \end{align}
  \end{subequations}
Then the non-zero curvature components read as
\begin{subequations}%
\label{eqn:components_curv}%
\begin{align}
  \CurvQui{n,m}{n,m}=& \, \FConnQui{n,m} \otimes \UniEnd{n,m} - D \EndPsiQui{n,m}  
\wedge \tfrac{\im \, m }{2}\, \eta \, \UniEnd{n,m} \notag \\ & - \big( 
\EndPsiQui{n,m} - \mathds{1}_{p_{(n,m)}} \big) \otimes \tfrac{\im\, m }{2}\, \diff \eta \,
\UniEnd{n,m} \notag \\
&+ \big(\mathds{1}_{p_{(n,m)}} - \HomoPhiPlusi{n-1,m-3} \,
\HomoPhiAdjPlusi{n-1,m-3} \big)\otimes \BarBetaPlusi{n-1,m-3} \wedge 
\BetaPlusi{n-1,m-3} \notag \\
&+ \big(\mathds{1}_{p_{(n,m)}} - \HomoPhiMinusi{n+1,m-3} \,
\HomoPhiAdjMinusi{n+1,m-3} \big) \otimes \BarBetaMinusi{n+1,m-3} 
\wedge \BetaMinusi{n+1,m-3} \notag \\
&+ \big(\mathds{1}_{p_{(n,m)}} - \HomoPhiAdjPlusi{n,m}\,  
\HomoPhiPlusi{n,m} \big) \otimes \BetaPlusi{n,m} \wedge 
\BarBetaPlusi{n,m}\notag \\
&+ \big(\mathds{1}_{p_{(n,m)}} - \HomoPhiAdjMinusi{n,m}\,  
\HomoPhiMinusi{n,m} \big) \otimes \BetaMinusi{n,m} \wedge 
\BarBetaMinusi{n,m} \; ,\\[4pt]
% 
%%%%%%%%%%%%%%%%%%%%%%%%%%%%%%%%%%%%%%%%%%%%%%%%%%%%%%%%%%%%%%%%%%%%%%%%%%%%%%%%
%  
\CurvQui{n,m}{n\pm1,m+3} =& \, D \HomoPhiPMi{n,m} \wedge 
\BarBetaPMi{n,m} - \big( (m+3)\, \EndPsiQui{n\pm1,m+3} \, \HomoPhiPMi{n,m} \notag \\ &\phantom{\, D \HomoPhiPMi{n,m} \wedge 
\BarBetaPMi{n,m}} - m \,
\HomoPhiPMi{n,m} \, \EndPsiQui{n,m} - 3 \HomoPhiPMi{n,m} 
\big) \otimes \tfrac{\im}{2}\, \eta \, \UniEnd{n\pm1,m+3} \wedge 
\BarBetaPMi{n,m} \; , \\[4pt]
% 
%%%%%%%%%%%%%%%%%%%%%%%%%%%%%%%%%%%%%%%%%%%%%%%%%%%%%%%%%%%%%%%%%%%%%%%%%%%%%%%%
%
\CurvQui{n+1,m-3}{n+1,m+3}= & \, \big( \HomoPhiPlusi{n,m}\,  
\HomoPhiMinusi{n+1,m-3} - \HomoPhiMinusi{n+2,m}\, 
\HomoPhiPlusi{n+1,m-3} \big)  \otimes \BarBetaPlusi{n,m} \wedge 
\BarBetaMinusi{n+1,m-3} \; , \\[4pt]
% 
%%%%%%%%%%%%%%%%%%%%%%%%%%%%%%%%%%%%%%%%%%%%%%%%%%%%%%%%%%%%%%%%%%%%%%%%%%%%%%%%
% 
\CurvQui{n-1,m+3}{n+1,m+3}= & \, -\big(\HomoPhiPlusi{n,m}\,  
\HomoPhiAdjMinusi{n,m} - \HomoPhiAdjMinusi{n+1,m+3} \,
\HomoPhiPlusi{n-1,m+3} \big) \otimes \BarBetaPlusi{n,m} 
\wedge \BetaMinusi{n,m} \; ,
\end{align}
which are accompanied by the anti-Hermiticity conditions
\begin{align}
\CurvQui{n',m'}{n,m} = - \left(\CurvQui{n,m}{n',m'}\right)^\dagger \; .
\end{align}
\end{subequations}
By setting $\psi_{(n,m)}=\mathds{1}_{p_{(n,m)}}$ for all $(n,m)\in Q_0(k,l)$, these curvature matrix elements correctly reproduce those computed in~\cite{Dolan:2009nz} for equivariant dimensional reduction over $\CP$.
\bigskip \section{Quiver bundle examples}
\label{sec:examples_quiver}
\paragraph{$\RepSuHead{1}{0}$-quiver }
Consider the fundamental $3$-dimensional representation $\RepSu{1}{0}$ of 
$\mathrm{G}=\sut $. Its decomposition into irreducible
$\su$-representations is given by
\begin{equation}
 \RepSu{1}{0} \big|_{\su } = \Reps{(0,-2)} \ \oplus \ \Reps{(1,1)} \; , %
 \label{eqn:split_C(1,0)}
\end{equation}
wherein $\Reps{(0,-2)}$ is the $\su $-singlet and $\Reps{(1,1)}$ is the 
$\su $-doublet. Using the general quiver construction of 
Section~\ref{subsec:construction_quiver}, the $\G $-action 
dictates the existence of bundle morphisms
\begin{subequations}
 \begin{alignat}{2}
  \phi  &:= \phi_{(0,-2)}^+ \in 
 \Hom\big(E_{p_{(0,-2)}},E_{p_{(1,1)}}\big) \; , \quad  & \phi^\dagger  
&:= 
 \big( \phi^+ \big)_{(0,-2)}^\dagger \in 
 \Hom\big(E_{p_{(1,1)}},E_{p_{(0,-2)}}\big)  \; , \\[4pt]
 \psi_0 &:= \psi_{(0,-2)} \in \End\big(E_{p_{(0,-2)}} \big) \; , \quad & 
\psi_1 
&:= \psi_{(1,1)} \in \End\big(E_{p_{(1,1)}} \big) \; .
 \end{alignat}
\end{subequations}
\paragraph{$\RepSuHead{2}{0}$-quiver}
The $6$-dimensional representation $\RepSu{2}{0}$ of $\sut $ splits under 
restriction to $\su $ as
\begin{equation}
 \RepSu{2}{0} \big|_{\su } = \Reps{(2,2)}  \ \oplus \ \Reps{(1,-1)} \
 \oplus \ 
\Reps{(0,-4)} \; .%
 \label{eqn:split_C(2,0)}
\end{equation}
The $\sut $-action intertwines the irreducible $\su $-modules 
and the corresponding bundles. The actions of $E_{\alpha_1 + \alpha_2}$ and 
$E_{\alpha_2}$ respectively yield Higgs fields
\begin{subequations}
\begin{equation}
\phi_0:= \phi^+_{(0,-4)} \in
\Hom\big(E_{p_{(0,-4)}},E_{p_{(1,-1)}} \big) 
\; , \quad \phi_1:= \phi^+_{(1,-1)} \in 
\Hom\big(E_{p_{(1,-1)}},E_{p_{(2,2)}} \big) \; .
\end{equation}
Due to the non-zero restrictions of $H_{ \alpha_2}$ to its eigenspaces $\Reps{(0,-4)}$, $\Reps{(1,-1)}$ and 
$\Reps{(2,2)}$, one further has three bundle endomorphisms 
\begin{equation}
 \psi_0:= \EndoPsi{(0,-4)} \in \End\big(E_{p_{(0,-4)}} \big) \; ,
 \quad  \psi_1:= \EndoPsi{(1,-1)} \in 
\End\big(E_{p_{(1,-1)}}\big)   \; 
, \quad \psi_2:= \EndoPsi{(2,2)} \in \End\big(E_{p_{(2,2)}}\big) \; .
\end{equation}
\end{subequations}
\paragraph{$\RepSuHead{1}{1}$-quiver}
The $8$-dimensional adjoint representation of $\sut $ splits under restriction to $\su $ as
\begin{equation}
 \RepSu{1}{1} \big|_{\su } =\Reps{(1,3)} \ \oplus \ \Reps{(0,0)} \
 \oplus \ 
\Reps{(2,0)} \ \oplus \ \Reps{(1,-3)} \; .%
\label{eqn:split_C(1,1)}
\end{equation}
The action of $\sut $ implies the existence of the following bundle morphisms:
The actions of $E_{\alpha_1 + \alpha_2}$ and $E_{\alpha_2}$ 
translate into the Higgs fields
\begin{subequations}
\begin{alignat}{2}
\phi_1^+:= \phi^+_{(1,-3)} &\in \Hom\big(E_{p_{(1,-3)}},E_{p_{(2,0)}}\big)\; , 
& \quad   
\phi_1^-:= \phi^-_{(1,-3)} &\in \Hom\big(E_{p_{(1,-3)}},E_{p_{(0,0)}}\big) \; 
, \\[4pt]
\phi_0^+:= \phi^+_{(0,0)} &\in \Hom\big(E_{p_{(0,0)}},E_{p_{(1,3)}}\big) \; 
, & \quad
 \phi_0^-:= \phi^-_{(2,0)} &\in \Hom\big(E_{p_{(2,0)}},E_{p_{(1,3)}}\big) \; ,
\end{alignat}
whereas the action of $H_{\alpha_2}$ generates
\begin{equation}
 \psi^\pm:= \EndoPsi{(1,\pm\, 3)} \in \End\big(E_{p_{(1,\pm3)}}\big) \; .
\end{equation}
\end{subequations}
Note that $H_{ \alpha_2}$ neither introduces endomorphisms on 
$\Reps{(0,0)}$ and $\Reps{(2,0)}$ nor does it intertwine these $\su
$-multiplets. This follows from the fact that these representations are 
subspaces of the kernel of $H_{ \alpha_2}$, and that $H_{ 
\alpha_2}$ commutes with the entire Lie algebra $\suL$.
\bigskip \section{Equivariant dimensional reduction details}
\subsection{1-form products on $\CP$}
\label{sec:reduction_CP2}
The metric on $\mfd{d} \times \CP$ is given as 
\begin{equation}
 \diff s^2 = \diff s^2_{M^d} + \diff s_{\CP}^2 \; , \label{eqn:metric_red_CP2}
\end{equation}
where
\begin{equation}
\diff s^2_{M^d} = G_{\mu' \nu'}\, \diff 
x^{\mu'} \otimes \diff x^{\nu'}
\end{equation}
with $(x^{\mu'})$ local real coordinates on the manifold $M^d$ and $\mu' ,\nu',\ldots = 1, \ldots ,d $. The metric on $\CP$ is written as 
\begin{equation}
 g_{\CP}:= \diff s^2_{\CP} = R^2 \, \left( \beta^1 \otimes \bar{\beta}^1 + 
\bar{\beta}^1 \otimes \beta^1 + \beta^2 \otimes \bar{\beta}^2 +\bar{\beta}^2 
\otimes \beta^2 \right) \; .
\end{equation}
This metric is compatible with the Kähler 
form~\eqref{eqn:canonical_Kaehler_CP2}, and by defining the complex 
structure $J$ via $ \omega_{\CP} (\cdot, \cdot) = g_{\CP}( \cdot , J\, \cdot 
)$ on the cotangent bundle of $\CP$ one obtains $J \beta^i = \im \, \beta^i$ 
and $J \bar{\beta}^i = - \im \, \bar{\beta}^i$ for $i=1,2$. The corresponding Hodge operator is denoted $\starCP$, with the non-vanishing $1$-form products
\begin{subequations}
\begin{align}
 \starCP 1 &= R^4 \, \beta^1 \wedge \bar{\beta}^1 \wedge \beta^2 \wedge 
\bar{\beta}^2 = 2 \big(\pi\, R^2\big)^2 \, \beta_{\mathrm{vol}} \; ,\\[4pt]
\bar{\beta}^1 \wedge \starCP \beta^1 &= \bar{\beta}^2 \wedge \starCP \beta^2 = 
\beta^1 \wedge \starCP \bar{\beta}^1 = \beta^2 \wedge \starCP \bar{\beta}^2 = 2 
\pi^2 \, R^2 \, \beta_{\mathrm{vol}} \label{eqn:star_oneform} \; , \\[4pt]
\starCP \bar{\beta}^1 \wedge \beta^1 &= \beta^2 \wedge \bar{\beta}^2 \; , 
\qquad \starCP \bar{\beta}^2 \wedge \beta^2 = \beta^1 \wedge \bar{\beta}^1 \; 
,\\[4pt]
\starCP \bar{\beta}^1 \wedge \beta^2 &= \bar{\beta}^1 \wedge \beta^2 \; , 
\qquad \starCP \bar{\beta}^2 \wedge \beta^1 = \bar{\beta}^2 \wedge \beta^1 \; .
\end{align}
\end{subequations}
For later use we shall also need to compute various products involving matrix-valued $1$-forms. Firstly, we have\footnote{The expressions~\eqref{eqn:trace_results_CP_old} correct the trace formulas from~\cite[eq.~(B.7)]{Dolan:2009nz}.}
\begin{subequations}
 \begin{align}
\tr \, \frac{\BetaPMi{n,m}
\wedge 
\starCP \BarBetaPMi{n,m}}{\LaPM{n,m}^2}
&= 2 \pi^2 \, R^2 \, (n+1 \pm 1 ) \, \betavol \; , \\[4pt]
% 
%%%%%%%%%%%%%%%%%%%%%%%%%%%%%%%%%%%%%%%%%%%%%%%%%%%%%%%%%%%%%%%%%%%%%%%%%%%%%%%%
%  
\tr \, \frac{\BetaPMi{n,m} \wedge 
\BarBetaPMi{n,m} \wedge \starCP \big( \BetaPMi{n,m} \wedge 
\BarBetaPMi{n,m} \big)^\dagger  }{ \LaPM{n,m}^4 }  
&= 2 \pi^2 \, (n+1 \pm 1)\, \betavol \; , \\[4pt]
% 
%%%%%%%%%%%%%%%%%%%%%%%%%%%%%%%%%%%%%%%%%%%%%%%%%%%%%%%%%%%%%%%%%%%%%%%%%%%%%%%%
% 
\tr \, \frac{ \BarBetaPMi{n,m} \wedge 
\BetaPMi{n,m} \wedge \starCP  \big( \BarBetaPMi{n,m} 
\wedge \BetaPMi{n,m} \big)^\dagger  }{ \LaPM{n,m}^4} 
&=2 \pi^2 \, \frac{(n+ 1 \pm1 )^2}{n+1} \, \betavol \label{eqn:trace-result-wrong} \; 
,\\[4pt]
% 
%%%%%%%%%%%%%%%%%%%%%%%%%%%%%%%%%%%%%%%%%%%%%%%%%%%%%%%%%%%%%%%%%%%%%%%%%%%%%%%%
% 
\tr \, \frac{ \BarBetaPlusi{n,m} \wedge 
\BarBetaMinusi{n+1,m-3} \wedge \starCP \big( \BarBetaPlusi{n,m} 
\wedge \BarBetaMinusi{n+1,m-3} \big)^\dagger }{ \LaPlus{n,m}^2 \, 
\LaMinus{n+1,m-3}^2 } 
&= 2 \pi^2\, \frac{n+1}{3} \, \betavol \; ,\\[4pt]
% 
%%%%%%%%%%%%%%%%%%%%%%%%%%%%%%%%%%%%%%%%%%%%%%%%%%%%%%%%%%%%%%%%%%%%%%%%%%%%%%%%
% 
\tr \, \frac{\BarBetaPlusi{n,m} \wedge 
\BetaMinusi{n,m} \wedge \starCP \big( \BarBetaPlusi{n,m} 
\wedge \BetaMinusi{n,m} \big)^\dagger }{ \LaPlus{n,m}^2 \,
\LaMinus{n,m}^2} 
&= 2 \pi^2 \, \frac{n\,(n+2)}{n+1} \, \betavol \; .
\end{align}
\label{eqn:trace_results_CP_old}
\end{subequations}
The trace formulas~\eqref{eqn:trace_results_CP_old} will have to be 
supplemented by
\begin{subequations}
\begin{align}
\tr \, \frac{\BetaPlusi{n,m} \wedge 
\BarBetaPlusi{n,m} 
\wedge \starCP \big( \BetaMinusi{n,m} \wedge 
\BarBetaMinusi{n,m} \big)^\dagger }{\LaPlus{n,m}^2\, \LaMinus{n,m}^2} 
&= 2 \pi^2 \, \frac{2n \, (n+2)}{3(n+1)} \; \betavol \ , \\[4pt]
 \tr \, \frac{ \BarBetaPlusi{n-1,m-3} \wedge 
\BetaPlusi{n-1,m-3} 
\wedge \starCP \big( \BarBetaMinusi{n+1,m-3} \wedge 
\BetaMinusi{n+1,m-3} \big)^\dagger  }{\LaPlus{n-1,m-3}^2\,
\LaMinus{n+1,m-3}^2} 
&= 2 \pi^2 \, \frac{2(n+1)}{3} \; \betavol \ , \\[4pt]
\tr \, \frac{\BetaPMi{n,m} \wedge 
\BarBetaPMi{n,m} 
\wedge \starCP \big( \BarBetaPMi{n\mp1,m-3} \wedge \BetaPMi{n\mp1,m-3} 
\big)^\dagger }{\LaPM{n,m}^2\, \LaPM{n\mp1,m-3}^2 } 
&=-2 \pi^2 \, \frac{n\, (n+2)}{n+1\mp1} \; \betavol \ , \\[4pt]
\tr \, \frac{\BetaPMi{n,m} \wedge 
\BarBetaPMi{n,m} 
\wedge \starCP \big( \BarBetaMPi{n\pm1,m-3} \wedge \BetaMPi{n\pm1,m-3} 
\big)^\dagger }{ \LaPM{n,m}^2\, \LaMP{n\pm1,m-3}^2 } 
&= 2 \pi^2 \, \Big(\, \frac{n\, (n+2)}{3(n+1\pm1)} - (n+1) \, \Big) \; \betavol
\end{align}
\label{eqn:trace_results_CP_new}
\end{subequations}
and one additionally needs the traces
\begin{subequations}
\begin{align}
\tr \, \frac{\BetaPMi{n,m} \wedge 
\BarBetaPMi{n,m} }{\Lambda^{\pm}_{k,l}(n,m)^2} &= -\frac{\im 
}{2 R^2} \, (n + 1 \pm 1) \, \omega_{\CP} =
\starCP \, \tr \, \frac{\big(\BetaPMi{n,m} 
\wedge 
\BarBetaPMi{n,m} \big)^\dagger}{\Lambda^{\pm}_{k,l}(n,m)^2} \ , \\[4pt]
\tr \, \frac{\BarBetaPMi{n\mp1,m-3} \wedge 
\BetaPMi{n\mp1,m-3} 
}{\Lambda^{\pm}_{k,l}(n\mp1,m-3)^2} &=\frac{\im }{2 R^2}\, (n + 1) \,
\omega_{\CP} =
\starCP \, \tr \, \frac{\big(\BarBetaPMi{n\mp1,m-3}
\wedge 
\BetaPMi{n\mp1,m-3}\big)^\dagger }{\Lambda^{\pm}_{k,l}(n\mp1,m-3)^2} \ .
\end{align}
\end{subequations}
% 
% % 
% %%%%%%%%%%%%%%%%%%%%%%%%%%%%%%%%%%%%%%%%%%%%%%%%%%%%%%%%%%%%%%%%%%%%%%%%%%%%%%%%
% %%%%%%%%%%%%%%%%%%%%%%%%%%%%%%%%%%%%%%%%%%%%%%%%%%%%%%%%%%%%%%%%%%%%%%%%%%%%%%%%
% % 
\subsection{1-form products on $S^5$}
\label{subsec:reduction_S5}
Let us write the metric~\eqref{eqn:metric_S5} in the forms
\begin{equation}
 \diff s_{S^5}^2 = 
 g_{ij}\, \left( \betaphi^i 
\otimes \barbetaphi^j + \barbetaphi^j \otimes \betaphi^i  \right) +
g_{55} \, \eta 
\otimes \eta = 2 R^2\, \delta_{ab} \,
e^a \otimes e^b + r^2 \, e^5 \otimes e^5 \ ,
\end{equation}
for $i,j=1,2$ and $a,b=1,2,3,4$, where $r$ is the radius of the $S^1$-fibre of the Hopf bundle $S^5\to \CP$; the corresponding Hodge operator is denoted $\starS$. 
Define the normalised volume form $\etavol$ on $S^5$ as
\begin{align}
 \starS 1 &= - (2 \pi)^3 \, r \, R^4 \ \etavol \with \betavol \wedge \eta =- 4 \pi \
\etavol= -\frac2{\pi^2}\, e^{12345} \and \int_{S^5}\, \etavol = 1 \; .
\end{align}
In the computation of the reduced action~\eqref{eqn:YM-action_matrix} we use the identities
\begin{subequations}
\begin{align} 
 e^\mu \wedge \starS e^\nu &= \sqrt{g}\, g^{\mu \nu}\, e^{12345} = 
\left\{ \begin{matrix} 4 \pi^3\, r\, R^2 \ \etavol \; , & \mu=\nu=a \ , \\ 
\frac{(2\pi)^3\, R^4}{r} \ \etavol \; , & \mu=\nu =5 \\ 0 \; , & \mu \neq \nu \ ,
\end{matrix} \right. \; ,\\[4pt]
e^{\mu \nu} \wedge \starS e^{\rho \sigma} &= \left\{ \begin{matrix} \sqrt{g} \,
g^{\mu \rho} \, g^{\nu \sigma}\, e^{12345} \; , & \mu=\rho , \ \nu=\sigma \; , \\ - \sqrt{g} \,
g^{\mu \sigma} \, g^{\nu \rho}\, e^{12345} \; , & \mu= \sigma , \ \nu = \rho  \\ 0 \; 
, & \mathrm{otherwise} \ , \end{matrix} \right. \; ,\\[4pt]
e^{ab} \wedge \starS e^{ab} &= 2 \pi^3\, r \ \etavol \and  e^{a5} \wedge 
\starS e^{a5} = \tfrac{4 \pi^3 \, R^2}{r} \ \etavol \; .
\end{align}
\end{subequations}
We can reduce the action of the Hodge operator in $5$ dimensions to the action of $\star_{\CP}$ from Appendix~\ref{sec:reduction_CP2} to get
\begin{subequations}
 \begin{alignat}{2}
  \starS \betaphi^i &= r \, \left(\starCP \betaphi^i \right) \wedge \eta  \; ,
& \quad \starS \barbetaphi^i &= r\, \left(\starCP \barbetaphi^i \right)  
\wedge \eta \; , \\[4pt]
 \starS \left( \betaphi^i \wedge \barbetaphi^j \right) &= r\, \left( \starCP 
\betaphi^i \wedge \barbetaphi^j  \right) \wedge \eta \; , & \quad 
\starS \left( \betaphi^i \wedge \betaphi^j \right) &= r\, \left( \starCP 
\betaphi^i \wedge \betaphi^j  \right) \wedge \eta  \; , \\[4pt]
\starS \left(  \eta \wedge \betaphi^i \right) &= \tfrac{1}{r}\, \starCP 
\betaphi^i \; , & \starS \left( \eta \wedge \barbetaphi^i \right) &= 
\tfrac{1}{r} \starCP \barbetaphi^i  \; , \\[4pt]
 \starS \eta &=\tfrac{2(\pi\, R^2)^2}{r} \ \betavol \; , &  \eta \wedge \starS 
\eta &= - \tfrac{(2 \pi)^3\, R^4}{r} \ \etavol \; .
 \end{alignat}
\end{subequations}
We can additionally compute
\begin{subequations}
 \begin{align}
  \diff \eta &= -2 \omega_3 = \im \, \left(\betaphi^1 \wedge \barbetaphi^1 + 
\betaphi^2 \wedge \barbetaphi^2 \right) = -\tfrac{1}{R^2} \, \omega_{\CP} \; , \\[4pt]
  \starS \diff \eta &= 
-\tfrac{1}{R^2}\, \starS \omega_{\CP}  = -\tfrac{r}{R^2}\, (\starCP 
\omega_{\CP} )\wedge \eta =\tfrac{r}{R^2} \, \omega_{\CP} \wedge \eta \; , \\[4pt]
\diff \eta \wedge \starS \diff \eta &= - 2 (2 \pi)^3\, r \ \etavol \ ,
 \end{align}
\end{subequations}
wherein we used $\starCP \omega_{\CP}= - \omega_{\CP}$ and~\eqref{eqn:omegaCPsquared}.
Note that due to the structure of the extension from $\CP$ to $S^5$, the matrices 
accompanying contributions from $\eta$ or $\diff\eta$ are always proportional to the identity operators $\UniEnd{n,m}$; thus their inclusion does not alter the 
trace formulas of Appendix~\ref{sec:reduction_CP2}.
% 
% % 
% %%%%%%%%%%%%%%%%%%%%%%%%%%%%%%%%%%%%%%%%%%%%%%%%%%%%%%%%%%%%%%%%%%%%%%%%%%%%%%%%
% %%%%%%%%%%%%%%%%%%%%%%%%%%%%%%%%%%%%%%%%%%%%%%%%%%%%%%%%%%%%%%%%%%%%%%%%%%%%%%%%
% %
\subsection{Yang-Mills action}
\label{sec:details_red_S5_appendix}
The reduction of~\eqref{eqn:YM-action_MxS5} proceeds by writing
\begin{equation}
 \tr \, \mathcal{F} \wedge \star\, \mathcal{F} = -\sum_{(n,m)\in Q_0(k,l)} \,
\tr\big(\mathcal{F} \wedge \star\,
\mathcal{F}^\dagger\, \big)_{(n,m),(n,m)} \; .
\end{equation}
We insert the 
explicit non-vanishing components~\eqref{eqn:components_curv}, rescale the horizontal Higgs fields
\begin{equation}
 \HomoPhiPMi{n,m} \longrightarrow \tfrac{1 }{\LaPM{n,m}} \, \HomoPhiPMi{n,m}
\end{equation}
as in~\cite{Dolan:2009nz} (but \emph{not} the vertical Higgs fields $\psi_{(n,m)}$), and  evaluate the traces over the representation spaces $\Reps{(n,m)} $ 
for each weight $(n,m)\in Q_0(k,l)$ using the matrix products from
Appendix~\ref{sec:reduction_CP2} and the relations of Appendix~\ref{subsec:reduction_S5}. Finally,
one then integrates over $S^5$ using the unit volume form $\etavol$ introduced in Appendix~\ref{subsec:reduction_S5}.
The dimensionally reduced Yang-Mills action on $\mfd{d}$ 
then reads as\footnote{By setting
  $\psi_{(n,m)}=\mathds{1}_{p_{(n,m)}}$ for all $(n,m)\in Q_0(k,l)$ and $r=\frac1{4\pi}$
  in~\eqref{eqn:YM-action_S5} we obtain the quiver gauge theory action
  for equivariant dimensional reduction over the complex projective
  plane $\CP$; this reduction eliminates the last nine lines of~\eqref{eqn:YM-action_S5} and the resulting expression corrects~\cite[eq.~(3.5)]{Dolan:2009nz}.}
\begin{align}
 S=&\, \frac{2\pi^3\,r\, R^4}{\tilde{g}^2} \, \int_{\mfd{d}} 
\, \diff^d x \ \sqrt{G} \ 
\sum_{(n,m)\in Q_0(k,l)} \, \tr \bigg( (n+1)\,\big(F_{(n,m)}\big)_{\mu' \nu'}^\dagger \,
\big(F_{(n,m)}\big)^{\mu' \nu'} \notag \\
%
%%%%%%%%%%%%%%%%%%%%%%%%%%%%%%%%%%%%%%%%%%%%%%%%%%%%%%%%%%%%%%%%%%%%%%%%%%%%%%%%
% 
&\, +\frac{n+2}{R^2} \, \big(D_{\mu'} 
\HomoPhiPlusi{n,m}\big)^\dagger \, D^{\mu'} \HomoPhiPlusi{n,m}
+  \frac{n+1}{R^2} \, D_{\mu'} \HomoPhiPlusi{n-1,m-3}\, \big(D^{\mu'} 
\HomoPhiPlusi{n-1,m-3}\big )^\dagger \notag \\
%
%%%%%%%%%%%%%%%%%%%%%%%%%%%%%%%%%%%%%%%%%%%%%%%%%%%%%%%%%%%%%%%%%%%%%%%%%%%%%%%%
% 
&\, +\frac{n}{R^2} \, \big(D_{\mu'} 
\HomoPhiMinusi{n,m}\big)^\dagger \, D^{\mu'} \HomoPhiMinusi{n,m}
+ \frac{n+1}{R^2} \, D_{\mu'} \HomoPhiMinusi{n+1,m-3} \, \big( D^{\mu'} 
\HomoPhiMinusi{n+1,m-3}\big)^\dagger \notag \\
%
%%%%%%%%%%%%%%%%%%%%%%%%%%%%%%%%%%%%%%%%%%%%%%%%%%%%%%%%%%%%%%%%%%%%%%%%%%%%%%%%
%
&\, + \frac{n+2}{R^4} \, \Big( \LaPlus{n,m}^2 \, \mathds{1}_{p_{(n,m)}} -  \HomoPhiAdjPlusi{n,m} \,
\HomoPhiPlusi{n,m} \Big)^2 
\notag\\
%
%%%%%%%%%%%%%%%%%%%%%%%%%%%%%%%%%%%%%%%%%%%%%%%%%%%%%%%%%%%%%%%%%%%%%%%%%%%%%%%%
%
&\, + \frac{n}{R^4}\, \Big( \LaMinus{n,m}^2 \,\mathds{1}_{p_{(n,m)}}
- \HomoPhiAdjMinusi{n,m} \, \HomoPhiMinusi{n,m} \Big)^2 
\notag  \\
%
%%%%%%%%%%%%%%%%%%%%%%%%%%%%%%%%%%%%%%%%%%%%%%%%%%%%%%%%%%%%%%%%%%%%%%%%%%%%%%%%
% 
&\, + \frac{n+1}{n\, R^4}\, \Big( \LaPlus{n-1,m-3}^2 \,\mathds{1}_{p_{(n,m)}}
- \HomoPhiPlusi{n-1,m-3}\,  
\HomoPhiAdjPlusi{n-1,m-3}  \Big)^2 \notag \\
%
%%%%%%%%%%%%%%%%%%%%%%%%%%%%%%%%%%%%%%%%%%%%%%%%%%%%%%%%%%%%%%%%%%%%%%%%%%%%%%%%
% 
&\, + \frac{(n+1)^2}{(n+2)\, R^4}\,
\Big( \LaMinus{n+1,m-3}^2 \,\mathds{1}_{p_{(n,m)}} - \HomoPhiMinusi{n+1,m-3}\, 
 \HomoPhiAdjMinusi{n+1,m-3}  \Big)^2 \notag \\
%
%%%%%%%%%%%%%%%%%%%%%%%%%%%%%%%%%%%%%%%%%%%%%%%%%%%%%%%%%%%%%%%%%%%%%%%%%%%%%%%%
% 
&\, + \frac{2(n+3)}{3R^4} \, \Big|  
\HomoPhiPlusi{n,m}\, \HomoPhiMinusi{n+1,m-3} 
- \frac{ \LaPlus{n,m}\, \LaMinus{n+1,m-3} }{\LaPlus{n+1,m-3} \,
\LaMinus{n+2,m}} \, \HomoPhiMinusi{n+2,m}\,  
\HomoPhiPlusi{n+1,m-3} \Big|^2 \notag \\
%
%%%%%%%%%%%%%%%%%%%%%%%%%%%%%%%%%%%%%%%%%%%%%%%%%%%%%%%%%%%%%%%%%%%%%%%%%%%%%%%%
% 
&\, + \frac{2n\, (n+2)}{(n+1)\, R^4} \, \Big|  
\HomoPhiPlusi{n,m}\, \HomoPhiAdjMinusi{n,m} \notag\\ &\, \qquad \qquad \qquad \qquad \qquad 
 - \frac{ \LaPlus{n,m}\, \LaMinus{n,m}}{\LaPlus{n-1,m+3}\, \LaMinus{n+1,m+3}} \,
\HomoPhiAdjMinusi{n+1,m+3} \, \HomoPhiPlusi{n-1,m+3} \Big|^2 
\notag \\
%
%%%%%%%%%%%%%%%%%%%%%%%%%%%%%%%%%%%%%%%%%%%%%%%%%%%%%%%%%%%%%%%%%%%%%%%%%%%%%%%%
% 
&\, + \frac{4n \, (n+2)}{3(n+1)\, R^4} \,\Big( 
\big(\LaPlus{n,m}^2 \,\mathds{1}_{p_{(n,m)}} - \HomoPhiAdjPlusi{n,m} \,
\HomoPhiPlusi{n,m} \big) \notag\\ &\, \qquad \qquad \qquad \qquad \qquad \times \ \big(\LaMinus{n,m}^2 \,\mathds{1}_{p_{(n,m)}} - \HomoPhiAdjMinusi{n,m} \,
 \HomoPhiMinusi{n,m} \big) \Big) \notag \\
%
%%%%%%%%%%%%%%%%%%%%%%%%%%%%%%%%%%%%%%%%%%%%%%%%%%%%%%%%%%%%%%%%%%%%%%%%%%%%%%%%
% 
&\, -\frac{2(n+2)}{R^4} \, \Big( \big(\LaPlus{n,m}^2 \,\mathds{1}_{p_{(n,m)}}
- \HomoPhiAdjPlusi{n,m} \,
\HomoPhiPlusi{n,m} \big) \notag\\ &\, \qquad \qquad \qquad \qquad \qquad \times \ \big(\LaPlus{n-1,m-3}^2 \,\mathds{1}_{p_{(n,m)}} - 
\HomoPhiPlusi{n-1,m-3}\, \HomoPhiAdjPlusi{n-1,m-3} \big)  
\Big) \notag \\
%
%%%%%%%%%%%%%%%%%%%%%%%%%%%%%%%%%%%%%%%%%%%%%%%%%%%%%%%%%%%%%%%%%%%%%%%%%%%%%%%%
%
&\, +\frac{2}{R^4}\,\Big(\, \frac{n}{3} -n-1 \,\Big) \,
\Big( \big(\LaPlus{n,m}^2 \,\mathds{1}_{p_{(n,m)}} - \HomoPhiAdjPlusi{n,m} \,
 \HomoPhiPlusi{n,m} \big) \notag\\ &\, \qquad \qquad \qquad \qquad \qquad \times \ \big(\LaMinus{n+1,m-3}^2 \, \,\mathds{1}_{p_{(n,m)}} - 
\HomoPhiMinusi{n+1,m-3}\, \HomoPhiAdjMinusi{n+1,m-3} \big)  
\Big) \notag \\
%
%%%%%%%%%%%%%%%%%%%%%%%%%%%%%%%%%%%%%%%%%%%%%%%%%%%%%%%%%%%%%%%%%%%%%%%%%%%%%%%%
%  
&\, + \frac{2}{R^4} \, \Big(\, \frac{n+2}{3} - n-1
\,\Big) \, \Big( \big(\LaMinus{n,m}^2 \,\mathds{1}_{p_{(n,m)}} - 
\HomoPhiAdjMinusi{n,m} \,
 \HomoPhiMinusi{n,m} \big) \notag\\ &\, \qquad \qquad \qquad \qquad \qquad \times \ \big( \LaPlus{n-1,m-3}^2 \,\mathds{1}_{p_{(n,m)}} - 
\HomoPhiPlusi{n-1,m-3}\, \HomoPhiAdjPlusi{n-1,m-3} \big) \Big) 
\notag \\
%
%%%%%%%%%%%%%%%%%%%%%%%%%%%%%%%%%%%%%%%%%%%%%%%%%%%%%%%%%%%%%%%%%%%%%%%%%%%%%%%%
%  
&\, - \frac{2n}{R^4}\, \Big( \big(\LaMinus{n,m}^2 \,\mathds{1}_{p_{(n,m)}}
-\HomoPhiAdjMinusi{n,m} \, \HomoPhiMinusi{n,m} \big) \notag\\ &\, \qquad \qquad \qquad \qquad \qquad \times \ \big(\LaMinus{n+1,m-3}^2\,\mathds{1}_{p_{(n,m)}} - 
\HomoPhiMinusi{n+1,m-3}\, \HomoPhiAdjMinusi{n+1,m-3} \big) 
\Big)\notag \\
%
%%%%%%%%%%%%%%%%%%%%%%%%%%%%%%%%%%%%%%%%%%%%%%%%%%%%%%%%%%%%%%%%%%%%%%%%%%%%%%%%
% 
&\, + \frac{4 (n+1)}{3 R^4}\, \Big( 
\big(\LaPlus{n-1,m-3}^2 \,\mathds{1}_{p_{(n,m)}}
 - \HomoPhiPlusi{n-1,m-3}\,
\HomoPhiAdjPlusi{n-1,m-3} \big) \notag\\ &\, \qquad \qquad \qquad \qquad \qquad  \times \ \big( \LaMinus{n+1,m-3}^2 \,\mathds{1}_{p_{(n,m)}} - 
\HomoPhiMinusi{n+1,m-3}\, \HomoPhiAdjMinusi{n+1,m-3} \big)  
\Big)\notag  \\
%
%%%%%%%%%%%%%%%%%%%%%%%%%%%%%%%%%%%%%%%%%%%%%%%%%%%%%%%%%%%%%%%%%%%%%%%%%%%%%%%%
%  
&\, + \frac{(n+1)\, m^2}{4r^2} \, D_{\mu'} \EndPsiQui{n,m} \, \big(D^{\mu'} 
\EndPsiQui{n,m} \big)^\dagger 
+ \frac{2 (n+1)\, m^2}{R^4}\, \Big( \EndPsiQui{n,m} - \mathds{1}_{p_{(n,m)}} \Big)^2 
\notag \\
%
%%%%%%%%%%%%%%%%%%%%%%%%%%%%%%%%%%%%%%%%%%%%%%%%%%%%%%%%%%%%%%%%%%%%%%%%%%%%%%%%
%  
&\, -\frac{m\, (n+2)}{R^4}\, \Big( \big( \LaPlus{n,m}^2 \, \mathds{1}_{p_{(n,m)}}
- \HomoPhiAdjPlusi{n,m} \,
\HomoPhiPlusi{n,m} \big) \, \big(\EndPsiQui{n,m} - 
\mathds{1}_{p_{(n,m)}} \big) \Big) \notag \\
%
%%%%%%%%%%%%%%%%%%%%%%%%%%%%%%%%%%%%%%%%%%%%%%%%%%%%%%%%%%%%%%%%%%%%%%%%%%%%%%%%
%  
&\, -\frac{m\, n}{R^4} \, \Big( \big( \LaMinus{n,m}^2 \, \mathds{1}_{p_{(n,m)}}
- \HomoPhiAdjMinusi{n,m} \, 
\HomoPhiMinusi{n,m} \big) \, \big(\EndPsiQui{n,m} - 
\mathds{1}_{p_{(n,m)}} \big) \Big) \notag \\
%
%%%%%%%%%%%%%%%%%%%%%%%%%%%%%%%%%%%%%%%%%%%%%%%%%%%%%%%%%%%%%%%%%%%%%%%%%%%%%%%%
%  
&\, + \frac{m \, (n+1 )}{ R^4} \, \Big( \big( \LaPlus{n-1,m-3}^2\, \mathds{1}_{p_{(n,m)}}
- \HomoPhiPlusi{n-1,m-3}\,   
\HomoPhiAdjPlusi{n-1,m-3} \big) \, \big(\EndPsiQui{n,m} - 
\mathds{1}_{p_{(n,m)}} \big) \Big) \notag \\
%
%%%%%%%%%%%%%%%%%%%%%%%%%%%%%%%%%%%%%%%%%%%%%%%%%%%%%%%%%%%%%%%%%%%%%%%%%%%%%%%%
%  
&\, + \frac{m \, (n+1 )}{ R^4} \, \Big( \big( \LaMinus{n+1,m-3}^2 \, \mathds{1}_{p_{(n,m)}}
- \HomoPhiMinusi{n+1,m-3}\,  
\HomoPhiAdjMinusi{n+1,m-3} \big) \, \big(\EndPsiQui{n,m} - 
\mathds{1}_{p_{(n,m)}} \big) \Big) \notag \\
%
%%%%%%%%%%%%%%%%%%%%%%%%%%%%%%%%%%%%%%%%%%%%%%%%%%%%%%%%%%%%%%%%%%%%%%%%%%%%%%%%
%  
&\, +\frac{n +1}{4 R^2 \, r^2} \, \Big| m \, \EndPsiQui{n,m} \,
\HomoPhiPlusi{n-1,m-3} - (m-3)\, \HomoPhiPlusi{n-1,m-3}\,  
\EndPsiQui{n-1,m-3} - 3 \HomoPhiPlusi{n-1,m-3} \Big|^2 \notag \\
%
%%%%%%%%%%%%%%%%%%%%%%%%%%%%%%%%%%%%%%%%%%%%%%%%%%%%%%%%%%%%%%%%%%%%%%%%%%%%%%%%
%  
&\, +\frac{n +1}{4 R^2 \, r^2} \, \Big| m \, \EndPsiQui{n,m} \,
\HomoPhiMinusi{n+1,m-3} - (m-3)\, \HomoPhiMinusi{n+1,m-3}\,  
\EndPsiQui{n+1,m-3} - 3 \HomoPhiMinusi{n+1,m-3} \Big|^2 \notag \\
%
%%%%%%%%%%%%%%%%%%%%%%%%%%%%%%%%%%%%%%%%%%%%%%%%%%%%%%%%%%%%%%%%%%%%%%%%%%%%%%%%
%  
&\, +\frac{n +2}{4 R^2 \, r^2} \, \Big| (m+3) \, \EndPsiQui{n+1,m+3} \,
\HomoPhiPlusi{n,m} - m \, \HomoPhiPlusi{n,m}  
\EndPsiQui{n,m} - 3 \HomoPhiPlusi{n,m} \Big|^2 \notag \\
%
%%%%%%%%%%%%%%%%%%%%%%%%%%%%%%%%%%%%%%%%%%%%%%%%%%%%%%%%%%%%%%%%%%%%%%%%%%%%%%%%
%  
\label{eqn:YM-action_S5} &\, +\frac{n}{4 R^2 \, r^2} \, \Big| (m+3) \, \EndPsiQui{n-1,m+3} \,
\HomoPhiMinusi{n,m} - m \, \HomoPhiMinusi{n,m}\, 
\EndPsiQui{n,m} - 3 \HomoPhiMinusi{n,m} \Big|^2 \bigg) \ . 
\end{align}
Note that while the trace in~\eqref{eqn:YM-action_matrix} is
taken over the full fibre space $V^{k,l}$ of the equivariant vector
bundle~\eqref{eqn:decomp_E(k,l)}, in~\eqref{eqn:YM-action_S5} the
trace over the $\su\times\uo$-representations $\Reps{(n,m)}$ has
already been evaluated.
\end{appendix}
\bigskip \bibliographystyle{JHEP}    
 {\footnotesize{\bibliography{references}}}

\providecommand{\href}[2]{#2}\begingroup\raggedright\begin{thebibliography}{10}

\bibitem{Kachru:1998ys}
S.~Kachru and E.~Silverstein, {\it {$4D$ conformal theories and strings on
  orbifolds}},  {\em Phys. Rev. Lett.} {\bf 80} (1998) 4855--4858,
  [\href{http://arxiv.org/abs/hep-th/9802183}{{\tt hep-th/9802183}}].

\bibitem{Lawrence:1998ja}
A.~E. Lawrence, N.~A. Nekrasov, and C.~Vafa, {\it {On conformal field theories
  in $4$-dimensions}},  {\em Nucl. Phys. B} {\bf 533} (1998) 199--209,
  [\href{http://arxiv.org/abs/hep-th/9803015}{{\tt hep-th/9803015}}].

\bibitem{Kehagias:1998gn}
A.~Kehagias, {\it {New type IIB vacua and their F-theory interpretation}},
  {\em Phys. Lett. B} {\bf 435} (1998) 337--342,
  [\href{http://arxiv.org/abs/hep-th/9805131}{{\tt hep-th/9805131}}].

\bibitem{Klebanov:1998hh}
I.~R. Klebanov and E.~Witten, {\it {Superconformal field theory on $3$-branes
  at a Calabi-Yau singularity}},  {\em Nucl. Phys. B} {\bf 536} (1998)
  199--218, [\href{http://arxiv.org/abs/hep-th/9807080}{{\tt hep-th/9807080}}].

\bibitem{Morrison:1998cs}
D.~R. Morrison and M.~R. Plesser, {\it {Nonspherical horizons 1}},  {\em Adv.
  Theor. Math. Phys.} {\bf 3} (1999) 1--81,
  [\href{http://arxiv.org/abs/hep-th/9810201}{{\tt hep-th/9810201}}].

\bibitem{Martelli:2004wu}
D.~Martelli and J.~Sparks, {\it {Toric geometry, Sasaki-Einstein manifolds and
  a new infinite class of AdS/CFT duals}},  {\em Commun. Math. Phys.} {\bf 262}
  (2006) 51--89, [\href{http://arxiv.org/abs/hep-th/0411238}{{\tt
  hep-th/0411238}}].

\bibitem{Kallen:2012cs}
J.~Källén and M.~Zabzine, {\it {Twisted supersymmetric $5D$ Yang-Mills theory
  and contact geometry}},  {\em JHEP} {\bf 1205} (2012) 125,
  [\href{http://arxiv.org/abs/1202.1956}{{\tt arXiv:1202.1956}}].

\bibitem{Qiu:2013pta}
J.~Qiu and M.~Zabzine, {\it {$5D$ super Yang-Mills on $Y^{p,q}$ Sasaki-Einstein
  manifolds}},  {\em Commun. Math. Phys.} {\bf 333} (2015) 861--904,
  [\href{http://arxiv.org/abs/1307.3149}{{\tt arXiv:1307.3149}}].

\bibitem{Gauntlett:2004yd}
J.~P. Gauntlett, D.~Martelli, J.~F. Sparks, and D.~Waldram, {\it
  {Sasaki-Einstein metrics on $S^2\times S^3$}},  {\em Adv. Theor. Math. Phys.}
  {\bf 8} (2004) 711--734, [\href{http://arxiv.org/abs/hep-th/0403002}{{\tt
  hep-th/0403002}}].

\bibitem{Gauntlett:2004hh}
J.~P. Gauntlett, D.~Martelli, J.~F. Sparks, and D.~Waldram, {\it {A new
  infinite class of Sasaki-Einstein manifolds}},  {\em Adv. Theor. Math. Phys.}
  {\bf 8} (2006) 987--1000, [\href{http://arxiv.org/abs/hep-th/0403038}{{\tt
  hep-th/0403038}}].

\bibitem{Cvetic:2005ft}
M.~Cvetic, H.~Lu, D.~N. Page, and C.~Pope, {\it {New Einstein-Sasaki spaces in
  $5$ and higher dimensions}},  {\em Phys. Rev. Lett.} {\bf 95} (2005) 071101,
  [\href{http://arxiv.org/abs/hep-th/0504225}{{\tt hep-th/0504225}}].

\bibitem{Ivanova:2013mea}
T.~A. Ivanova, O.~Lechtenfeld, A.~D. Popov, and R.~J. Szabo, {\it {Orbifold
  instantons, moment maps and Yang-Mills theory with sources}},  {\em Phys.
  Rev. D} {\bf 88} (2013) 105026, [\href{http://arxiv.org/abs/1310.3028}{{\tt
  arXiv:1310.3028}}].

\bibitem{Sperling:2015sra}
M.~Sperling, {\it {Instantons on Calabi-Yau cones}},
  \href{http://arxiv.org/abs/1505.01755}{{\tt arXiv:1505.01755}}.

\bibitem{Candelas:1985en}
P.~Candelas, G.~T. Horowitz, A.~Strominger, and E.~Witten, {\it {Vacuum
  configurations for superstrings}},  {\em Nucl. Phys. B} {\bf 258} (1985)
  46--74.

\bibitem{Lechtenfeld:2014fza}
O.~Lechtenfeld, A.~D. Popov, and R.~J. Szabo, {\it {Sasakian quiver gauge
  theories and instantons on Calabi-Yau cones}},
  \href{http://arxiv.org/abs/1412.4409}{{\tt arXiv:1412.4409}}.

\bibitem{IR}
Y.~Ito and M.~Reid, {\it {The McKay correspondence for finite subgroups of
  $\sltc$}},  in {\em Higher Dimensional Complex Varieties} (M.~Andreatta and
  T.~Peternell, eds.), pp.~220--240, de Gruyter, 1996.
\newblock \href{http://arxiv.org/abs/alg-geom/9411010}{{\tt alg-geom/9411010}}.

\bibitem{IN}
Y.~Ito and H.~Nakajima, {\it {McKay correspondence and Hilbert schemes in
  dimension $3$}},  {\em Topology} {\bf 39} (2000) 1155--1191,
  [\href{http://arxiv.org/abs/math/9803120}{{\tt math/9803120}}].

\bibitem{DGM}
M.~R. Douglas, B.~R. Greene, and D.~R. Morrison, {\it {Orbifold resolution by
  D-branes}},  {\em Nucl. Phys. B} {\bf 506} (1997) 84--106,
  [\href{http://arxiv.org/abs/hep-th/9704151}{{\tt hep-th/9704151}}].

\bibitem{Douglas:2000qw}
M.~R. Douglas, B.~Fiol, and C.~Romelsberger, {\it {The spectrum of BPS branes
  on a noncompact Calabi-Yau}},  {\em JHEP} {\bf 0509} (2005) 057,
  [\href{http://arxiv.org/abs/hep-th/0003263}{{\tt hep-th/0003263}}].

\bibitem{SardoInfirri:1996ga}
A.~V. Sardo-Infirri, {\it {Partial resolutions of orbifold singularities via
  moduli spaces of HYM-type bundles}},
  \href{http://arxiv.org/abs/alg-geom/9610004}{{\tt alg-geom/9610004}}.

\bibitem{Cirafici:2010bd}
M.~Cirafici, A.~Sinkovics, and R.~J. Szabo, {\it {Instantons, quivers and
  noncommutative Donaldson-Thomas theory}},  {\em Nucl. Phys. B} {\bf 853}
  (2011) 508--605, [\href{http://arxiv.org/abs/1012.2725}{{\tt
  arXiv:1012.2725}}].

\bibitem{Ivanova:2012vz}
T.~A. Ivanova and A.~D. Popov, {\it {Instantons on special holonomy
  manifolds}},  {\em Phys. Rev. D} {\bf 85} (2012) 105012,
  [\href{http://arxiv.org/abs/1203.2657}{{\tt arXiv:1203.2657}}].

\bibitem{Lechtenfeld:2008nh}
O.~Lechtenfeld, A.~D. Popov, and R.~J. Szabo, {\it {$\sut$-equivariant quiver
  gauge theories and nonabelian vortices}},  {\em JHEP} {\bf 0808} (2008) 093,
  [\href{http://arxiv.org/abs/0806.2791}{{\tt arXiv:0806.2791}}].

\bibitem{Boyer:2008}
C.~P. Boyer and K.~Galicki, {\em {Sasakian Geometry}}.
\newblock Oxford University Press, Oxford, 2008.

\bibitem{Dolan:2009nz}
B.~P. Dolan and R.~J. Szabo, {\it {Dimensional reduction and vacuum structure
  of quiver gauge theory}},  {\em JHEP} {\bf 0908} (2009) 038,
  [\href{http://arxiv.org/abs/0905.4899}{{\tt arXiv:0905.4899}}].

\bibitem{Conti:2005}
D.~Conti and S.~Salamon, {\it {Generalized Killing spinors in dimension $5$}},
  {\em Trans. Amer. Math. Soc.} {\bf 359} (2007) 5319--5343,
  [\href{http://arxiv.org/abs/math/0508375}{{\tt math/0508375}}].

\bibitem{Fernandez:2006ux}
M.~Fernandez, S.~Ivanov, V.~Munoz, and L.~Ugarte, {\it {Nearly hypo structures
  and compact nearly Kähler $6$-manifolds with conical singularities}},  {\em
  J. London Math. Soc.} {\bf 78} (2008) 580--604,
  [\href{http://arxiv.org/abs/math/0602160}{{\tt math/0602160}}].

\bibitem{Biedenharn:1962haa}
L.~C. Biedenharn, {\it {Invariant operators of the Casimir type}},  {\em Phys.
  Lett.} {\bf 3} (1962) 69--70.

\bibitem{Baird:1963wv}
G.~E. Baird and L.~C. Biedenharn, {\it {On the representations of semisimple
  Lie groups. 2}},  {\em J. Math. Phys.} {\bf 4} (1963) 1449--1466.

\bibitem{Mukunda:1965}
N.~Mukunda and L.~K. Pandit, {\it {Tensor methods and a unified representation
  theory of {${\rm SU}(3)$}}},  {\em J. Math. Phys.} {\bf 6} (1965) 746--765.

\bibitem{AlvarezConsul:2001uk}
L.~Alvarez-C\'onsul and O.~Garc\'ia-Prada, {\it {Dimensional reduction and
  quiver bundles}},  {\em J. Reine Angew. Math.} {\bf 556} (2003) 1--46,
  [\href{http://arxiv.org/abs/math/0112160}{{\tt math/0112160}}].

\bibitem{Bunk:2014coa}
S.~Bunk, O.~Lechtenfeld, A.~D. Popov, and M.~Sperling, {\it {Instantons on
  conical half-flat 6-manifolds}},  {\em JHEP} {\bf 1501} (2015) 030,
  [\href{http://arxiv.org/abs/1409.0030}{{\tt arXiv:1409.0030}}].

\bibitem{Donaldson:1985}
S.~K. Donaldson, {\it {Anti-self-dual Yang-Mills connections over complex
  algebraic surfaces and stable vector bundles}},  {\em Proc. London Math.
  Soc.} {\bf 50} (1985) 1--26.

\bibitem{Uhlenbeck:1986}
K.~Uhlenbeck and S.-T. Yau, {\it {On the existence of Hermitian Yang-Mills
  connections in stable vector bundles}},  {\em {Commun. Pure Appl. Math.}}
  {\bf 39} (1986) S257--S293.

\bibitem{Donaldson:1987}
S.~K. Donaldson, {\it {Infinite determinants, stable bundles and curvature}},
  {\em Duke Math. J.} {\bf 54} (1987) 231--247.

\bibitem{Atiyah:1982}
M.~F. Atiyah and R.~Bott, {\it {The Yang-Mills equations over Riemann
  surfaces}},  {\em Phil. Trans. Roy. Soc. London A} {\bf 308} (1982) 523--615.

\bibitem{SardoInfirri:1996gb}
A.~V. Sardo-Infirri, {\it {Resolutions of orbifold singularities and flows on
  the McKay quiver}},  \href{http://arxiv.org/abs/alg-geom/9610005}{{\tt
  alg-geom/9610005}}.

\bibitem{Cirafici:2011cd}
M.~Cirafici, A.~Sinkovics, and R.~J. Szabo, {\it {Instanton counting and
  wall-crossing for orbifold quivers}},  {\em Ann. Henri Poincar\'e} {\bf 14}
  (2013) 1001--1041, [\href{http://arxiv.org/abs/1108.3922}{{\tt
  arXiv:1108.3922}}].

\bibitem{Cirafici:2012qc}
M.~Cirafici and R.~J. Szabo, {\it {Curve counting, instantons and McKay
  correspondences}},  {\em J. Geom. Phys.} {\bf 72} (2013) 54--109,
  [\href{http://arxiv.org/abs/1209.1486}{{\tt arXiv:1209.1486}}].

\bibitem{Deser:2014}
A.~Deser, O.~Lechtenfeld, and A.~D. Popov, {\it {Sigma-model limit of
  Yang-Mills instantons in higher dimensions}},  {\em Nucl. Phys. B} {\bf 894}
  (2015) 361--373, [\href{http://arxiv.org/abs/1412.4258}{{\tt
  arXiv:1412.4258}}].

\bibitem{Thomas:2006}
R.~P. Thomas, {\it {Notes on {GIT} and symplectic reduction for bundles and
  varieties}},  {\em {Surv. Diff. Geom.}} {\bf 10} (2006) 221--273,
  [\href{http://arxiv.org/abs/math/0512411}{{\tt math/0512411}}].

\bibitem{Donaldson:1984}
S.~K. Donaldson, {\it {Nahm's equations and the classification of monopoles}},
  {\em Commun. Math. Phys.} {\bf 96} (1984) 387--407.

\bibitem{Kronheimer:1989}
P.~B. Kronheimer, {\it {A Hyper-Kählerian structure on coadjoint orbits of a
  semisimple complex group}},  {\em J. London Math. Soc.} {\bf 42} (1990)
  193--208.

\bibitem{Hitchin:1991}
N.~J. Hitchin, {\it {Hyper-Kähler manifolds}},  {\em Ast\'erisque} {\bf 206}
  (1992) 137--166.

\bibitem{Murray:2003pk}
M.~K. Murray and M.~A. Singer, {\it {Gerbes, Clifford modules and the index
  theorem}},  {\em Ann. Global Anal. Geom.} {\bf 26} (2004) 355--367,
  [\href{http://arxiv.org/abs/math/0302096}{{\tt math/0302096}}].

\bibitem{Popov:2010rf}
A.~D. Popov and R.~J. Szabo, {\it {Double quiver gauge theory and nearly
  Kähler flux compactifications}},  {\em JHEP} {\bf 1202} (2012) 033,
  [\href{http://arxiv.org/abs/1009.3208}{{\tt arXiv:1009.3208}}].

\end{thebibliography}\endgroup

\end{document}